\documentclass[]{interact}

\usepackage[dvips]{lscape}  

\usepackage[numbers,sort&compress,merge]{natbib}
\bibpunct[, ]{[}{]}{,}{n}{,}{,}
\renewcommand\bibfont{\fontsize{10}{12}\selectfont}

\usepackage{color}

\begin{document}

\title{Mutual information and mutual correlation: their spin-free formulations and comparison}

\author{
\name{Ji\v{r}\'{\i} Pittner\thanks{CONTACT J. Pittner. Email: jiri.pittner@jh-inst.cas.cz}}
\affil{J. Heyrovsk\'{y} Institute of Physical Chemistry, Academy of Sciences of the Czech \mbox{Republic, v.v.i.}, Dolej\v{s}kova 3, 18223 Prague 8, Czech Republic}
}

\maketitle

\begin{abstract}
Orbital entropies, pair entropies, and mutual information have become popular tools for analysis of strongly correlated wave functions.
They can quantitatively measure how strongly an orbital participates in the electron correlation 
and reveal the correlation pattern between different orbitals. 
However, this pattern can become rather complicated and sometimes difficult to interpret for large active spaces
and is not invariant with respect to the spin projection ($M_s$) component of the spin multiplet state. 
We introduce a modified spin-free orbital entropy, pair entropy, and mutual information, which  simplify the correlation analysis and are invariant with respect to $M_s$. 
By comparison of these quantities with their ``original'' spin-including counterparts
one can distinguish static correlation due to spin couplings from the ``genuine'' strong correlation  due to a multiconfigurational character of the wave function.
Recently, Evangelista introduced mutual correlation computed from the two-body cumulant as an alternative correlation measure,
which is easier to compute for some methods.
We present here a spin-free analogue of this quantity, which is $M_s$ invariant,
 and perform a comparison of the entropy-based and cumulant-based correlation measures in
both spin-free and original variants.
We illustrate the approaches on iron-sulfur bound complexes with one and two iron atoms.

\end{abstract}


\section{Introduction}

The problem of electron correlation has a central importance in quantum chemistry. 
The correlation effects\cite{sinanoglu1963} are usually divided into two or three categories, which are, however, somewhat fuzzy and can hardly be made rigorously disjunct.
The dynamic (also called weak) correlation occurs due to pairwise Coulomb repulsion of the electrons when they come close to each other
and manifests itself in many small contributions of excited Slater determinants to the wave function.
On the other hand, when there is no single dominant Slater determinant,  one speaks about strongly (or non-dynamically) correlated system.
Sometimes one distinguishes two sub-cases --- static correlation, where several Slater determinants with large coefficients are needed to
construct a state with a correct spin symmetry, and ``true'' strong correlation, which is related to quasidegenerate orbitals in molecules undergoing bond breaking, in transition metal complexes, etc.
For systems exhibiting dynamic (weak) correlation only, many single-reference methods exist, which are able to achieve a satisfactory ``chemical accuracy''.
For strongly correlated systems, the situation is considerably more difficult, since one has to accurately take into account both the strong (and static) as well as the dynamic correlation, which is always present.
Traditionally, the presence of the strong correlation has been judged based on the magnitude of configuration interaction (CI) coefficients or the size of coupled cluster (CC) amplitudes. 

New tools for this purpose have been developed based on the notions of entanglement and correlation, which are central to quantum information theory (QIT)  \cite{Ziesche-1995,Nagy-1996,Nalewajski-2000,legeza2021fermionic,schilling2020H2,schilling2021jctc,schilling2024qst}.
The entanglement and correlation analysis employs the orbital entropy, which is a measure of correlation of a given orbital with all other orbitals, 
and the mutual information, which measures the correlation of two orbitals with each other, regardless of other orbitals.
In this article we do not distinguish between quantum and classical correlation of the orbitals, which are both captured by the aforementioned quantities.
The QIT-based approaches have become more popular in recent two decades \cite{Legeza-2003b,Legeza-2004a,Legeza-2004b,Legeza-2006a,Rissler-2006,Barcza-2011,Boguslawski-2012b,Boguslawski-2013,Barcza-2014,Boguslawski-2014,boguslawski2016prb,boguslawski2021jcp,boguslawski2022pccp},
since they are particularly suitable for use with the DMRG method\cite{dmrg2023}, which also gained popularity in quantum chemistry, particularly
for studies of strongly correlated transition metal compounds \cite{Marti-2008a,Chan-2011,Chan-2012,Harris-2014,fe2s2chan2014,boguslawski2015cpl,gonzalez2015pccp,fe2s2veis2024,legeza2026femoco}.
They are, however, not restricted to any particular wave function ansatz and can be used with other strongly correlated methods as well, for example pCCD \cite{boguslawski2016prb,boguslawski2021jcp,boguslawski2022pccp}.

Among the important applications of the entropy-based measures belongs the optimization of orbital ordering, which is essential to
make the DMRG calculations efficient \cite{Legeza-2003b,Moritz-2005a}  and
the automated approach to active space selection  \cite{stein-reiher-2016}, which has the potential to turn the application of  multi-configurational methods into a black box approach.
Further development in this direction represents the Quantum Information-Assisted Complete Active Space Optimization \cite{schilling2023jpcl}, where besides the active space selection also an orbital optimization minimizing the out-of-CAS orbital correlation is performed.

The orbital entropies and mutual informations are not invariant with respect to orbital transformations,
so the correlation pattern depends on the form of the orbitals (e.g. localization) \cite{schilling2023qst,schilling2024faraday,angeli2024}
and a sensible (chemically intuitive) choice can be very important for its interpretation \cite{Brandejs2019}.
This non-invariance can actually be turned into an advantage, simplifying the  multireference character of the wave function by
a suitable orbital transformation.
Of course, the orbital correlation cannot be fully transformed away by the change of on-particle basis, there exists an orbital basis for which it has a minimum value called particle correlation \cite{schilling2024faraday}.
This idea dates back to L\"{o}wdin \cite{lowdin1}, who has shown that natural orbitals give rise to most compact CI expansion, but the relationship to quantum information theory was uncovered only recently.
Legeza et al. have shown that by performing suitable orbital transformations (called fermionic mode optimization in DMRG \cite{legeza2016prl}), it is possible to minimize the multireference character 
(i.e. the orbital correlation) in the wave function \cite{legeza2023jmc}.
The orbital transformations are not limited to DMRG, but can be exploited in other strongly correlated methods as well.
Li~Manni et al. \cite{limanni2023jctc,dobrautz2022prb,limanni2021pccp,limanni2020jctc}
introduced a ``Quantum Anamorphosis'' approach in the spin-adapted GUGA-FCIQMC context and were able to achieve a dramatic compression
of the multireference wave function by localization of the orbitals carrying the unpaired spins.
They applied it to studies of polynuclear transition metal clusters, including the iron-sulfur compounds
\cite{limanni2021jpca,dobrautz2021jctc}, 
and by MO localization and reordering were able to achieve remarkable sparsity of the FCIQMC Hamiltonian matrices
leading to improvements in computational efficiency as well as insight into the spin structure of the investigated states.
Along similar lines, Schilling et al. \cite{schilling2024jpcl} have employed orbital transformations minimizing the total orbital correlation to improve the ability of the tailored  CCSD method to capture the correlation energy.
Recently Li \cite{entanglement-minimized2025} presented an improved algorithm for computation of entanglement-minimized orbitals, which allowed
to obtain a much better initial state for quantum computation.

The importance of the reduced density matrices and their cumulants has been recognized in quantum chemistry long time ago \cite{lowdin1,mcweeny1967rdm,kutzelnigg1999cumulant}.
A lot of effort went into using the two-body reduced density matrix (2RDM) as a basic variable \cite{mazziotti2012review}, however, the key problem is to find 
the necessary and sufficient conditions for $N$-representability of the 2RDM.
The situation is similar to the density functional theory (DFT), where the exact density functional is not known.  
Approximate 2RDM $N$-representability conditions have been developed, which enable practicable calculations  \cite{mazziotti2012review}.
In hindsight, the difficulty of the 2RDM $N$-representability problem is not surprising, since it was proved that finding the ground state of a two-body Hamiltonian is a NP-complete problem (or QMA-complete on a quantum computer) \cite{kempe_2006}. 
There is thus not much hope to find an exact, closed and computationally (polynomially) efficient form of the 2RDM $N$-representability conditions,
as it would solve NP-complete problems.
Nevertheless, RDMs and their cumulants are very useful tools in the many-electron theory, where they enabled for example the generalization
of normal ordering and Wick theorem to a multideterminantal reference state \cite{kmgwt,mukherjeecpl,kutzelnigg2003cumulant,schaefer2020cumulant}.

Since for the totally uncorrelated Hartree-Fock wave function the cumulants vanish (and higher RDMs are thus functions of the 1RDM), it is intuitively clear that the cumulants must contain information about the correlation in the wave function.
Another approach to quantify the correlation in quantum many-body systems can thus be based on the cumulant of the two-body reduced density matrix\cite{mazziotti2006cumulant,kong2011valeev,evangelista2025mutual}.
In his recent work \cite{evangelista2025mutual} Evangelista introduced a quantity called mutual correlation, which can be computed from the two body cumulant, and has investigated its properties.
For some quantum chemical methods the two-body RDM is easier to compute than the (Fock-space) two-orbital RDM (or the corresponding subset of up to 4-body RDM elements \cite{Boguslawski-2014}), which makes this quantity practically useful.
However, similarly to the mutual information, the mutual correlation is not invariant with respect to components of a spin multiplet with different $M_s$. 
In this work we introduce a spin-free version of the mutual correlation of an orbital pair and a corresponding measure of orbital correlation analogous to the QIT orbital entropy.
We also perform numerical comparison of the original and spin-free variants of mutual correlation with mutual information on larger
and chemically relevant systems beyond the minimalistic H$_4$ example provided in \cite{evangelista2025mutual}.

Iron-sulfur clusters and complexes represent an important class of compounds highly relevant in biochemistry \cite{fe2s2chan2014,fecomplexgeom,fe2s2complexgeom}, which exhibit strong correlation due to the large number of unpaired electron in the iron 3d-shell.
They facilitate redox chemistry in living matter by carrying out electron transfer and catalytic  processes essential e.g. for nitrogen fixation, respiration, and photosynthesis. 
Their computational treatment thus represents a challenge, due to the complicated electronic structure and many close lying states of different spin multiplicity.
Although the high-spin state in these systems is dominated by a single configuration (but its lower-$M_s$ components are still multideterminantal),
the dynamic correlation can energetically favor the low-spin states, which typically have more complicated multiconfigurational character and require thus adequate methods \cite{fe2s2veis2024,NeeseROHF2024,limanni2021jpca,dobrautz2021jctc}.
We use these complexes as a nontrivial test cases for a comparison of the mutual information and mutual correlation in the original and the newly introduced spin-free variant
 and illustrate their application to the correlation analysis.

\section{Theory}

\subsection{A brief review of orbital entropy and mutual information}

In quantum chemistry, the quantum information theoretic measures are typically applied to molecular orbitals, which are considered 
as individual subsystems or ``sites'', corresponding to a Fock-space-type expansion of the normalized many-body wave function $|\Psi\rangle$
\begin{equation}
\label{wavefunction}
|\Psi\rangle = \sum_{n_1, n_2,\ldots,n_k=1}^4 C^{n_1,n_2,\ldots,n_k} |\psi_{n_1}\rangle \otimes|\psi_{n_2}\rangle \otimes  \ldots \otimes  |\psi_{n_k}\rangle
\;\;,
\end{equation}
where $k$ is the number of orbitals and
the orbital states $|\psi_{n_i}\rangle$ span the 4-dimensional basis 
\begin{equation}
\label{orbitalbasis}
{\cal B} = \{|\rangle, |\uparrow\rangle, |\downarrow\rangle, |\uparrow\downarrow\rangle\} \;\;\;.
\end{equation}
The tensor of expansions coefficients $C^{n_1,\ldots,n_k}$ is typically rather sparse, combining only 
the terms of the same number of electrons and $S_z$ projection (and possibly the molecular spatial symmetry).
Selecting an orbital number $i$ and considering the rest of the orbitals as an ``environment'' ${\cal E}_i= \{l: l=1,\ldots,k; l\neq i\}$ 
one can define the orbital RDM
\begin{equation}
\label{rho_i}
\rho_i = {\rm Tr}_{{\cal E}_i} |\Psi\rangle\langle\Psi|
\;\;,
\end{equation}
which is a $4\times4$ matrix and is diagonal due to the particle number and $S_z$ restrictions.
The orbital entropy can then be expressed as 
\begin{equation}
\label{orbitalentropy}
S_i \equiv S(\rho_i) = - {\rm Tr}_{\cal B} (\rho_i \ln \rho_i) =  - \sum_{p=1}^4 (\rho_i)_{pp}\, \ln (\rho_i)_{pp} \;\;
\;\;,
\end{equation}
taking advantage of the diagonality of $\rho_i$.
The sum of all orbital entropies (in the active space of $k$ orbitals) is often called total quantum information
\begin{equation}
\label{totalentropy}
S_{\rm tot} = \sum_{i=1}^k S_i
\;\;.
\end{equation}
Similarly, for an orbital pair $i,j$ one can trace over the ``environment'' ${\cal E}_{ij}= \{l: l=1,\ldots,k; l\neq i \wedge l\neq j\}$
to define the (orbital) pair RDM
\begin{equation}
\rho_{ij} = {\rm Tr}_{{\cal E}_{ij}} |\Psi\rangle\langle\Psi|
\;\;,
\end{equation}
which is a $16\times 16$ matrix, block-diagonal according to the particle numbers and $S_z$ values of the orbital pair states from ${\cal B}\otimes{\cal B}$.
The pair entropy is then analogously computed as
\begin{equation}
\label{pairentropy}
S_{ij} \equiv S(\rho_{ij}) = - {\rm Tr}_{{\cal B}\otimes{\cal B}} (\rho_{ij} \ln \rho_{ij}) =  - \sum_{p=1}^{16} \omega_p\, \ln \omega_p \;\;
\;\;,
\end{equation}
where $ \omega_p$ are the eigenvalues of $\rho_{ij}$ (we omit the indices $i,j$ at $\omega_p$ for simplicity).
The mutual information between orbitals $i$ and $j$ is then defined as
\begin{equation}
\label{pairmutual}
I_{ij} = S_i +S_j -S_{ij}
\;\;.
\end{equation}
Using these quantities, a weighted complete graph on $k$ vertices can be constructed, $S_i$ being vertex weights and $I_{ij}$ serving
as edge weights.
A picture of this graph with weight coded by color and/or line thickness then illustrates the correlation in the molecule (for a particular choice of orbitals).
As a measure of the overall correlation in a given state, the sum of pairwise mutual informations can be employed
\begin{equation}
\label{Itot}
I_{\rm tot} = \sum_{i<j} I_{ij} 
\;\;.
\end{equation}

The orbital and pair RDMs can be expressed as expectations values using the 16 possible orbital basis transition operators $O^{ij}$  defined as
\begin{equation}
\psi_i = O^{ij} \psi_j\;\; \forall \psi_i , \psi_j \in {\cal B}
\;\;.
\end{equation}
The orbital RDM matrix elements are expectation values of these operators, while the pair RDM matrix elements can be expressed as expectation values of products of two such operators \cite{Rissler-2006,Boguslawski-2013}.
Since these operators can in turn be expressed as linear combinations of products of at most four fermionic creation or annihilation operators, 
the orbital and pair RDMs can be obtained for any wave function ansatz for which the corresponding 1-body to 4-body ``traditional''  reduced density matrices can be computed \cite{Boguslawski-2014}.
In the DMRG method, a very efficient computation of the orbital and pair RDM based on contractions of the matrix product state wave functions is possible \cite{Schollwock-2011}.

\subsection{Spin-free orbital entropy and mutual information}

To define the spin-free orbital and pair entropy, we start with the orbital basis (\ref{orbitalbasis}) and consider the two singly occupied orbital states as a single indistinguishable ``microstate''
\begin{equation}
\label{orbitalbasis2}
\tilde{\cal B} = \{|\rangle, \{|\uparrow\rangle, |\downarrow\rangle\}, |\uparrow\downarrow\rangle\} \equiv  \{|0\rangle,  |1\rangle,  |2\rangle \}
\;\;,
\end{equation}
or, in another words, the orbital basis $\tilde{\cal B}$ only distinguishes the number of electrons in the orbital. 
The spin-free orbital  entropy is then computed as a trace of the spin-free orbital RDM $\tilde{\rho}_i$ over the spin-free orbital basis
\begin{equation}
\label{orbitalentropy2}
\tilde{S}_i = -{\rm Tr}_{\tilde{\cal B}} (\tilde{\rho_i} \ln \tilde{\rho_i}) = - \sum_{p=1}^3 (\tilde{\rho}_i)_{pp}  \ln (\tilde{\rho}_i)_{pp}
\;\;,
\end{equation}
where the nonzero diagonal elements of $\tilde{\rho}_i$ are
\begin{equation}
\label{rhotilde}
(\tilde{\rho}_i)_{11}= ({\rho}_i)_{11}, \;\; (\tilde{\rho}_i)_{22}= ({\rho}_i)_{22}+ ({\rho}_i)_{33},\;\; (\tilde{\rho}_i)_{33}= ({\rho}_i)_{44}
\;\;.
\end{equation}
It is easy to see that $\tilde{S}_i \leq S_i$ \cite{myspinfree2025},
which is consistent with the fact that when computing $\tilde{S}_i$ we discard some information.

The basis, over which the pair entropy expression is traced, must be the direct product of the two spin-free orbital bases $\tilde{\cal B}$ \cite{myspinfree2025}.
In analogy to (\ref{orbitalentropy2}), we thus define the spin-free pair entropy as
\begin{equation}
\label{pairentropy2}
\tilde{S}_{ij} = - \sum_{q=1}^9 \tilde{\omega}_q \ln \tilde{\omega}_q 
\;\;,
\end{equation}
where the summation runs over the nine spin-free basis states $\tilde{\cal B}_q$ from $\tilde{\cal B} \otimes \tilde{\cal B}$, 
i.e. $\tilde{\cal B}_q \in \{|00\rangle, |01\rangle,|02\rangle, |10\rangle,\ldots, |22\rangle \}$, where $|00\rangle$ abbreviates $|0\rangle\otimes |0\rangle$ etc.. 
The quantities $\tilde{\omega}_q$ in (\ref{pairentropy2}) are the ``spin-summed eigenvalues''
 of $\rho_{ij}$
\begin{equation}
\label{spinsum2}
 \tilde{\omega}_q  = \sum_{p=1}^{16} \omega_p \sum_{k \in {\cal B}_q} c_{pk}^2
\;\;.
\end{equation}
Here  $\omega_p$ are the ``original'' eigenvalues of  $\rho_{ij}$ and $c_{pk}$ are the corresponding normalized eigenvectors, while 
$ {\cal B}_q$ denotes the subset of the pair basis $\cal B  \otimes \cal B$ which corresponds to the $q$-th element of  $\tilde{\cal B} \otimes \tilde{\cal B}$.
For example, for $\tilde{\cal B}_q=|1\rangle \otimes |2\rangle$, $ {\cal B}_q = \{|\uparrow\rangle\otimes |\uparrow\downarrow\rangle, |\downarrow\rangle\otimes \uparrow\downarrow\rangle \}$.

In \cite{myspinfree2025} we have shown that  $0 \leq  \tilde{\omega}_q \leq 1$, i.e.  $\tilde{\omega}_q$ quantities can be interpreted as probabilities and employed in the entropy expression (\ref{pairentropy2}), corresponding to
a diagonal spin-free pair density matrix 
\begin{equation}
\label{tilderho2}
(\tilde{\rho}_{ij})_{pq} =  \tilde{\omega}_q \delta_{pq} \;\;,
\end{equation}
where the indices $p,q$ run over the 9 elements of the basis  $\tilde{\cal B}_i \otimes \tilde{\cal B}_j$. 
We have also shown that the spin-free pair entropy has the original pair entropy as an upper bound $\tilde{S}_{ij} \leq S_{ij}$,
again corresponding to the reduction of available information.
Moreover, for two  totally uncorrelated orbitals $i$ and $j$ we have seen that the spin-free pair entropy reduces to sum of the spin-free orbital entropies
$\tilde{S}_{ij} = \tilde{S}_i + \tilde{S}_j$,
which allows to define the spin-free mutual information analogously to (\ref{pairmutual})
\begin{equation}
\label{pairmutual2}
\tilde{I}_{ij} = \tilde{S}_i +\tilde{S}_j -\tilde{S}_{ij}
\;\;.
\end{equation}
Intuitively, it is clear that if we discard some
information by performing the spin summation, we cannot increase the amount of information correlating one orbital with another,
and in \cite{myspinfree2025} we have indeed shown that $\tilde{I}_{ij} \leq I_{ij}$.
The transition from the original mutual information (\ref{pairmutual}) to the spin-free one (\ref{pairmutual2})
 can thus only simplify the diagrams for the correlation analysis.

The main point of defining the spin-free correlation measures was, of course, to remove the dependence on the component of the 
spin-multiplets. 
This can be shown based on the assumption that a single set of molecular orbitals for both spins and for all components of the spin multiplet
have been employed. The theoretical argument as well as numerical confirmation was presented in \cite{myspinfree2025}.

\subsection{Mutual correlation}
The $n$-body reduced density matrix ($n$RDM) for a $N$-electron wave functions $\Psi$ has been introduced by L\"{o}wdin \cite{lowdin1}
in the first-quantized coordinate representation
\begin{equation}
\label{rdmoriginal}
\Gamma_n(x_1',x_2',\ldots,x_n'|x_1,x_2,\ldots,x_n) = {N \choose n} \int \Psi^*(x_1',x_2',\ldots,x_n',x_{n+1},\ldots,x_N) \Psi(x_1,\ldots,x_N) dx_{n+1}\cdots dx_N
\;\;\;,
\end{equation}
where $x_i \equiv (r_i \sigma_i)$ are spatial and spin coordinates of $i$-th electron.
They can be further (formally) integrated over spin variables to yield the spin-free RDMs
\begin{equation}
\label{rdmoriginalspinfree}
\Gamma_n(r_1',r_2',\ldots,r_n'|r_1,r_2,\ldots,r_n) = \int \Gamma_n(r_1'\sigma_1,r_2'\sigma_2,\ldots,r_n'\sigma_n|r_1\sigma_1,r_2\sigma_2,\ldots,r_n\sigma_n) d\sigma_1 \cdots  d\sigma_n
\;\;\;,
\end{equation}
The corresponding second quantized form of 1RDM and 2RDM in spinorbital representation
can be expressed as
\begin{equation}
\label{gamma1}
\gamma_R^P = \langle\Psi|a_P^\dagger a_R|\Psi\rangle
\end{equation}
and
\begin{equation}
\label{gamma2}
\Gamma_{RS}^{PQ} =  \langle\Psi|a_P^\dagger a_Q^\dagger a_S a_R|\Psi\rangle
\;\;\;,
\end{equation}
where $a_P^\dagger$ $(a_P)$ is a creation (annihilation) operator for an electron in spinorbital $P$ (we use capital indices $P,Q,\ldots$ for spinorbitals, while lowercase indices denote spatial orbitals).
The spin summation can be performed analogously, yielding the spin-free 1RDM and 2RDM $\gamma_r^p $ and $\Gamma_{rs}^{pq}$.
The eigenvectors of the 1RDM $\gamma$ are the natural (spin)orbitals and the corresponding eigenvalues can be interpreted as their occupation numbers $n_P$ \cite{lowdin1}.
Since in a Hartree-Fock wave functions the only spinorbital occupations numbers are 0 and 1, the size of the fractional part indicates the presence of correlation, and one can use the Von Neumann entropy of 1RDM 
\begin{equation}
\label{rdm1entropy}
S_{\rm VNE}(\gamma) = -\sum_P n_P \ln n_P
\end{equation}
as a correlation measure \cite{kais2005,mazziotti2006cumulant} alternative to the total quantum information (\ref{totalentropy}).
Actually, it is possible to use individual terms of this sum as a measure how each spinorbital is involved in the correlation
and sum the spin $\alpha$ and $\beta$ contribution of a given orbital to obtain an analogy of orbital entropy
\begin{equation}
\label{rdm1entropy2} 
S'_p = - n_{p} \ln n_{p} -  n_{\bar{p}} \ln n_{\bar{p}}
\;\;\;,
\end{equation}
where $p$ and $\bar{p}$ denote the $\alpha$ and $\beta$  spinorbitals of the spatial orbital $p$.

General-order cumulants of the RDMs can be defined via a generating function and represent the size-extensive part of the RDMs\cite{kutzelnigg2003cumulant,kong2011valeev}.
Here we will need only the two-body cumulant $\Lambda_{RS}^{PQ}$ which can be expresses as
\begin{equation}
\label{cumulant}
\Lambda_{RS}^{PQ} = \Gamma_{RS}^{PQ}  - (\gamma_R^P \gamma_S^Q - \gamma_S^P \gamma_R^Q)
\;\;\;.
\end{equation}
Together with the 1RDM it contains all information needed to compute expectation values of two-electron operators.
It vanishes for the Hartree-Fock wave function and the square of its Frobenius norm has thus been suggested as a measure of total correlation 
\cite{mazziotti2006cumulant}
\begin{equation}
\label{cumulantnorm}
C = \frac{1}{4} \sum_{PQRS} |\Lambda_{RS}^{PQ}|^2
\;\;\;.
\end{equation}
A few other correlation measures based on  $\Lambda$ which are less relevant for this work have been suggested; see Ref.~\cite{evangelista2025mutual} and references therein.
The main idea presented by Evangelista \cite{evangelista2025mutual} was to partition the correlation metrics (\ref{cumulantnorm}) to contributions of
subsystems (subsets of spinorbitals), which then allows to define a mutual correlation of an orbital pair in analogy to the mutual information.
Dividing the spinorbital basis to two mutually disjunct subsets $A$ and $B$, the correlation in subsystem $X$ = $A$ or $B$ is 
according to (\ref{cumulantnorm})
\begin{equation}
\label{cumulantnormX}
C_X = \frac{1}{4} \sum_{PQRS \in X} |\Lambda_{RS}^{PQ}|^2 \;\;\;.
\end{equation}
Evangelista then defined the mutual correlation between a pair of subsystems $A$ and $B$ as
\begin{equation}
\label{mutualcorrelation}
M_{AB} = C - C_A - C_B
\end{equation}
and suggested also a general partitioning to an arbitrary number of mutually disjunct subsystems,
where up to four-subsystem terms arise, as the two-body cumulant has four indices.
Combining (\ref{cumulantnorm}), (\ref{cumulantnormX}) and (\ref{mutualcorrelation}) and exploiting the hermiticity 
of the cumulant, the mutual correlation can be expressed as
\begin{equation}
\label{mutualcorrelationexplicit}
M_{AB} = \sum_{P \in A} \sum_{QRS \in B} |\Lambda_{RS}^{PQ}|^2 + \frac{1}{2} \sum_{PQ \in A}\sum_{RS \in B}  |\Lambda_{RS}^{PQ}|^2
	+  \sum_{PR \in A}\sum_{QS \in B}  |\Lambda_{RS}^{PQ}|^2 +  \sum_{PQR \in A} \sum_{S \in B} |\Lambda_{RS}^{PQ}|^2 
\;\;\;.
\end{equation}

Finally, to obtain the  mutual correlation of an orbital pair, the subsystems for the general partitioning are chosen
as the spinorbital pairs arising from all spatial orbitals
\begin{equation}
A_p = \{ \phi_p\alpha, \phi_p\beta \} \equiv  \{ \phi_p, \phi_{\bar{p}} \} 
\;\;\;.
\end{equation}
Using (\ref{mutualcorrelationexplicit}), the mutual correlation can be expressed as a sum of spin-summed quantities
\begin{equation}
\label{orbitalmutualcorrelation}
M_{pq} = \lambda_{qq}^{pq} + 1/2 \lambda_{qq}^{pp} + \lambda_{pq}^{pq} + \lambda_{qp}^{pp}
\;\;\;,
\end{equation}
where $\lambda_{rs}^{pq}$ is a sum of squares of the cumulant elements over all spin possibilities
for the orbitals $p,q,r,s$ that preserve $M_s$
\begin{equation}
\label{orbitalmutualcorrelationcontrib}
\lambda_{rs}^{pq} = |\Lambda_{rs}^{pq}|^2 + |\Lambda_{\bar{r}\bar{s}}^{\bar{p}\bar{q}}|^2 
	+ |\Lambda_{r\bar{s}}^{p\bar{q}}|^2 +  |\Lambda_{r\bar{s}}^{q\bar{p}}|^2
	+   |\Lambda_{s\bar{r}}^{p\bar{q}}|^2 +  |\Lambda_{s\bar{r}}^{q\bar{p}}|^2
\;\;\;,
\end{equation}
where $\beta$ spinorbitals are indicated by the over-bar.

\subsection{Spin-free mutual correlation}

Similarly to the orbital entropies and mutual information,
the orbital contribution to 1RDM  Von Neumann entropy  (\ref{rdm1entropy2}) and the mutual correlation (\ref{orbitalmutualcorrelation}) are not
invariant with respect to  the $M_s$ component of a spin multiplet.
In the following we shall develop invariant quantities which can be employed as correlation measures in an analogous manner.
First, we introduce the the spin-summed 1RDM $\tilde{\gamma}$
\begin{equation}
\label{gamma1ss}
\tilde{\gamma}_r^p = \gamma_r^p +  \gamma_{\bar{r}}^{\bar{p}}
\end{equation}
and 2RDM $\tilde{\Gamma}$
\begin{equation}
\label{gamma2ss}
\tilde{\Gamma}_{rs}^{pq} =  \Gamma_{rs}^{pq} +  \Gamma_{\bar{r}s}^{\bar{p}q} + \Gamma_{r\bar{s}}^{p\bar{q}} + \Gamma_{\bar{r}\bar{s}}^{\bar{p}\bar{q}} 
\;\;\;.
\end{equation}
The normalized $N$-electron wave function, from which the RDMs derive, can be expressed as a product of spatial and spin wave functions,
\begin{equation}
\label{spinspatial} 
\Psi = \Psi_{\rm P}(N,S) \Psi_{\rm S}(N,S,M_s) \;\;,
\end{equation}
which must belong to mutually adjoint irreducible representations of the permutation group ${\cal S}_N$ to maintain the total fermionic antisymmetry.
The spatial wave function $\Psi_{\rm P}(N,S)$ depends on the spin $S$, but is not dependent on the $M_s$ quantum number.
The key observation is that the spin-summed RDMs are $M_s$-invariant, as the $\Psi_S^* \Psi_S$ expression 
gets fully integrated  over the spin variables of all electrons in the two consecutive steps (\ref{rdmoriginal}) and (\ref{rdmoriginalspinfree}).
We should thus strive to express the cumulant-based expression for the orbital mutual correlation in terms of the spin-summed 2RDM and 1RDM.

Expressing the spin-summed cumulant as
\begin{equation}
\label{cumulantss}
\tilde{\Lambda}_{rs}^{pq} =  \Lambda_{rs}^{pq} +  \Lambda_{\bar{r}s}^{\bar{p}q} + \Lambda_{r\bar{s}}^{p\bar{q}} + \Lambda_{\bar{r}\bar{s}}^{\bar{p}\bar{q}}
\end{equation}
and using the definition of the cumulant (\ref{cumulant}) we obtain
\begin{eqnarray}
\tilde{\Lambda}_{rs}^{pq} &=& \Gamma_{rs}^{pq} +  \Gamma_{\bar{r}s}^{\bar{p}q} + \Gamma_{r\bar{s}}^{p\bar{q}} + \Gamma_{\bar{r}\bar{s}}^{\bar{p}\bar{q}}  \\
&-& \gamma_r^p \gamma_s^q +  \gamma_r^q \gamma_s^p -  \gamma_{\bar{r}}^{\bar{p}} \gamma_{\bar{s}}^{\bar{q}} +  \gamma_{\bar{r}}^{\bar{q}} \gamma_{\bar{s}}^{\bar{p}} - \gamma_r^p  \gamma_{\bar{s}}^{\bar{q}} -  \gamma_{\bar{r}}^{\bar{p}}  \gamma_s^q \;\;\;, \\
\end{eqnarray}
which can be rewritten in terms of the spin-summed RDMs as
\begin{equation}
\label{cumulantss2}
\tilde{\Lambda}_{rs}^{pq} = \tilde{\Gamma}_{rs}^{pq} - \tilde{\gamma}_r^p \tilde{\gamma}_s^q +  \tilde{\gamma}_r^q \tilde{\gamma}_s^p
 -\gamma_r^q \gamma_{\bar{s}}^{\bar{p}} -\gamma_{\bar{r}}^{\bar{q}} \gamma_s^p 
\end{equation}
except the last two terms, which thus spoil the $M_s$ invariance of the spin summed cumulant.
If we simply neglect these terms, the result would be nonzero for a closed shell restricted Hartree-Fock wave function (where $\gamma_r^p = \gamma_{\bar{r}}^{\bar{p}}$), which 
is clearly not acceptable if the results should serve as a basis for a correlation measure.
We thus propose to use for this purpose a ``pseudo-cumulant'' where the $  \tilde{\gamma}_r^q \tilde{\gamma}_s^p$ term is scaled down by 1/2 factor, and the last two terms (which for RHF equal to half of this term) are omitted
\begin{equation}
\label{pseudocumulantss}
{\Lambda'}_{rs}^{pq} = \tilde{\Gamma}_{rs}^{pq} - \tilde{\gamma}_r^p \tilde{\gamma}_s^q + \frac{1}{2}  \tilde{\gamma}_r^q \tilde{\gamma}_s^p
\;\;\;.
\end{equation}
This quantity is thus manifestly $M_s$ invariant and still vanishes for the closed shell RHF wave function.
It can be used in place of the original cumulant to perform the correlation analysis with a general partitioning \cite{evangelista2025mutual},
and, in particular, the spin-free mutual correlation of orbitals $p,q$ can be obtained in analogy to (\ref{orbitalmutualcorrelation}) as
\begin{equation}
\label{spinfreemutualcorrelation}
\tilde{M}_{pq} = |{\Lambda'}_{qq}^{pq}|^2 + 1/2 |{\Lambda'}_{qq}^{pp}|^2 + |{\Lambda'}_{pq}^{pq}|^2 + |{\Lambda'}_{qp}^{pp}|^2
\;\;\;.
\end{equation}

Another question is whether we can define some spin-free alternative of the orbital Von Neumann entropy (\ref{rdm1entropy2}),
which would be $M_s$ invariant.
Of course, provided the 2RDM, we can compute the orbital density matrix $(\rho_i)$ and the spin-free orbital entropy (\ref{orbitalentropy2}),
but we might still try to find another measures.
It is clearly not possible to use spatial orbital occupation numbers (scaled by 1/2) in the entropy-like formula (\ref{orbitalentropy2}),
since a high-spin open shell single-determinantal wave function would yield non-zero entropy of the open shell orbitals.
However, if we do not insist on an entropy-like expression, we can observe that the sum of mutual correlations for a given orbital with all other orbitals should indicate how much this orbital is involved in the correlation and define the measures accordingly
\begin{eqnarray}
M_p &=& \sum_{q\neq p} M_{pq} \\
\tilde{M}_p &=& \sum_{q\neq p} \tilde{M}_{pq} \\
\end{eqnarray}
in both spin-including and spin-free versions.
Actually, one could employ analogous partial sums of the mutual information $I_{pq}$ and $\tilde{I}_{pq}$  for the analysis in place of the orbital entropies,
but there is no need to do that, as the orbital entropy is easily available in both spin-dependent and spin-free version.
Finally, we can sum all elements of the mutual correlation like the mutual information (\ref{Itot}) to give  a measure of the overall correlation in a given state
\begin{equation}
\label{Mtot}
M_{\rm tot} = \sum_{p<q} M_{pq}  = \frac{1}{2} \sum_p M_p
\end{equation}
and analogously for the spin-free variant $\tilde{M}_{\rm tot}$.
Unfortunately an analogous inequality to  $\tilde{I}_{pq} \leq I_{pq}$ does not hold between $M_{pq}$ and $\tilde{M}_{pq}$, however, as we will see from the numerical results, 
the correlation diagrams get simplified when transitioning to the spin-free  $\tilde{M}_{pq}$.
In the rest of the paper, we will perform a numerical assessment of the newly proposed spin-free correlation measures.

\section{Computational details}

\begin{figure}[h]
\caption{
\label{fecomplex}
Structure of the  [Fe(SCH$_3$)$_4$]$^-$ and [Fe$_2$S$_2$(SCH$_3$)$_4$]$^{2-}$ complexes
}
\begin{center}
\includegraphics[width=6cm]{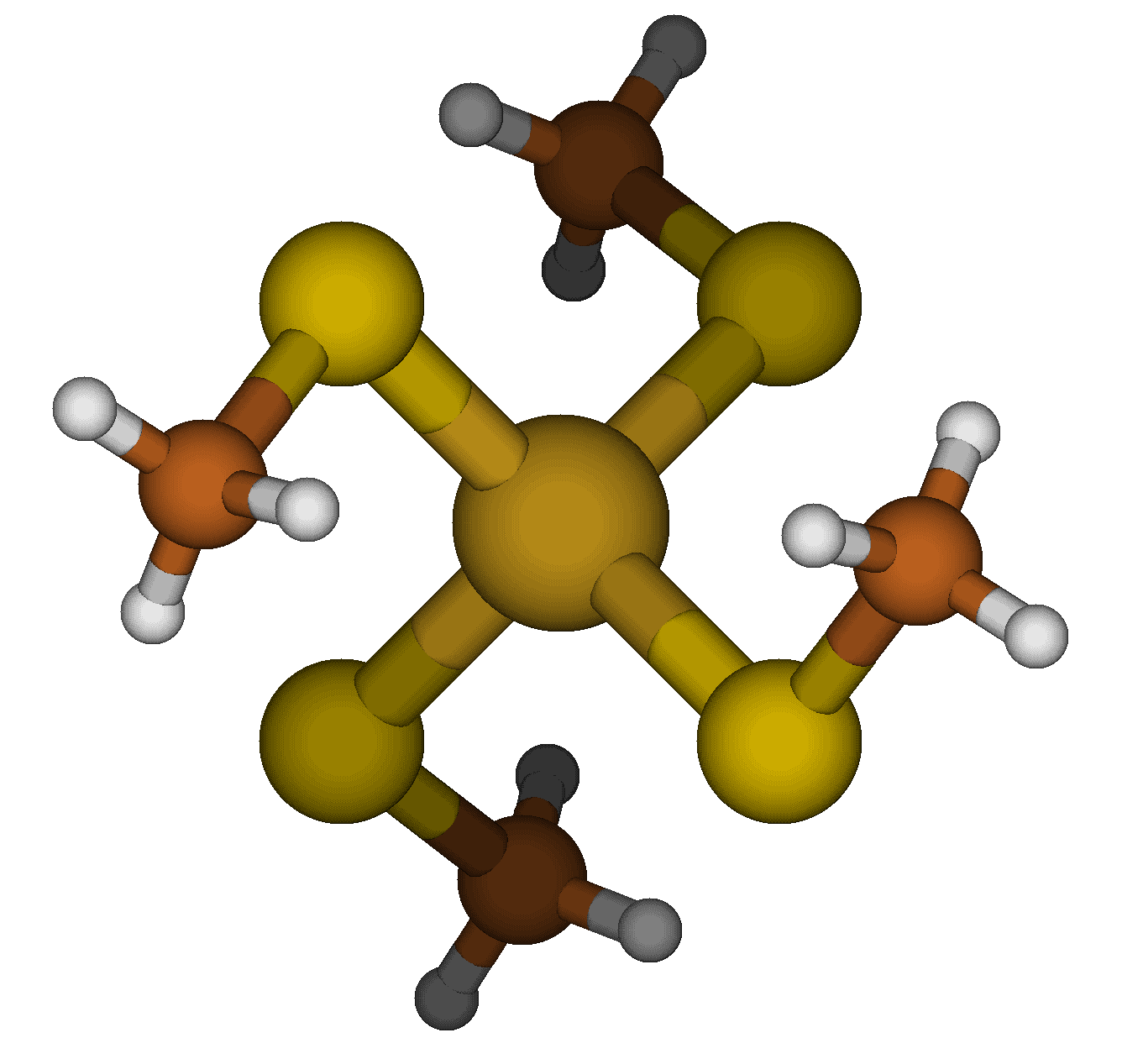} \\
\includegraphics[width=7cm]{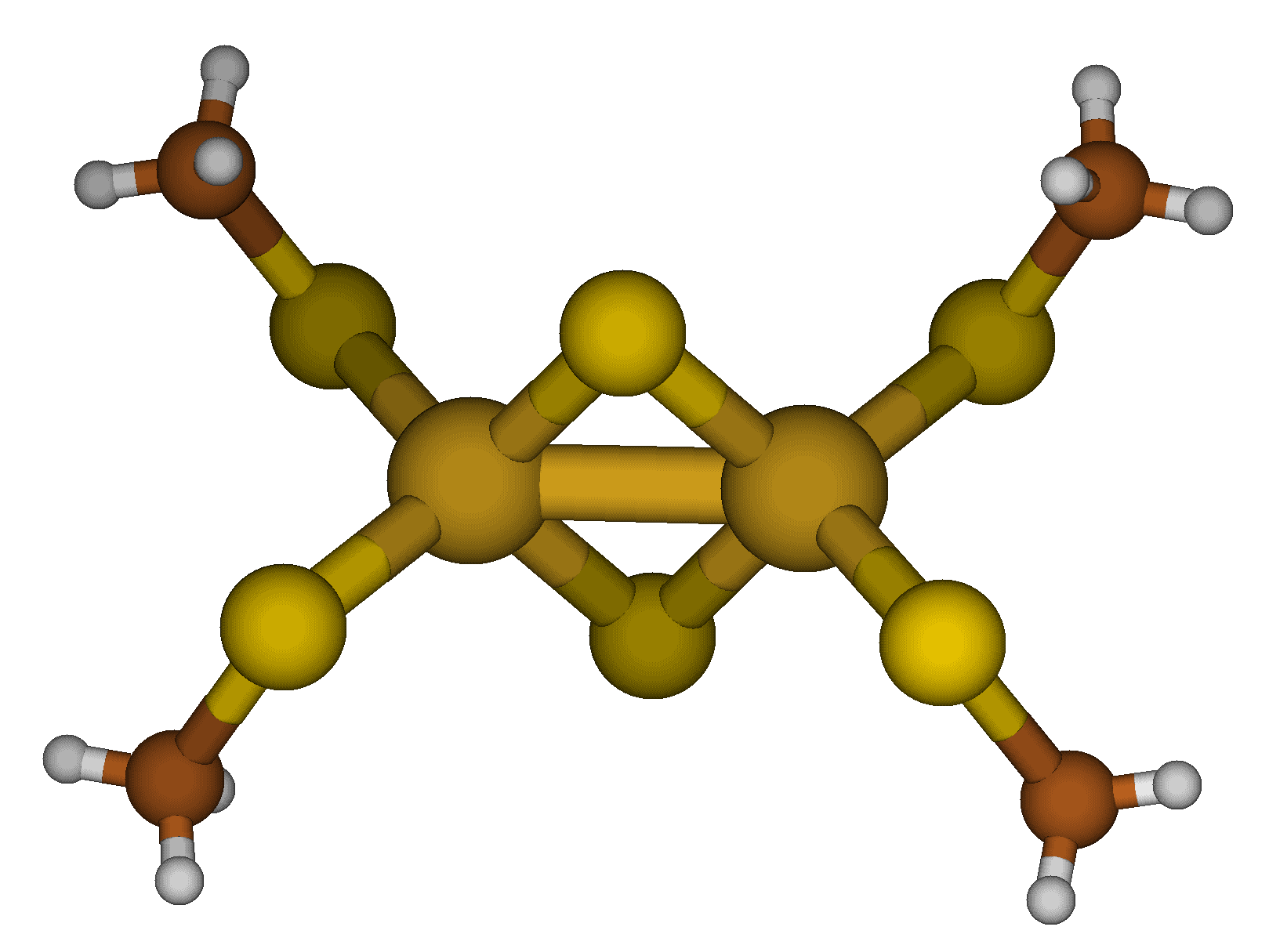}
\end{center}
\end{figure}

The iron complexes [Fe(SCH$_3$)$_4$]$^-$ and [Fe$_2$S$_2$(SCH$_3$)$_4$]$^{2-}$ (cf. Fig.~\ref{fecomplex}) 
represent  nontrivial prototypes of a strongly correlated molecule and can thus serve as an interesting
example for comparison of the mutual information, mutual correlation and their spin-free variants.
The geometries of the complexes were originally reported in Refs.~\cite{fecomplexgeom,fe2s2complexgeom} and  are also available in the Supplementary Information of  Ref.~\cite{NeeseROHF2024}.
We employed the Orca program \cite{Orca} and the Def2-TZVP basis set to perform a configuration-averaged ROHF SCF calculations and
integral dump. The MOLMPS code\cite{molmps2020} was used for the DMRG calculations of orbital density matrices,
while the BLOCK2 code\cite{block2code} was employed for the reduced density matrices (due to better numerical stability for these quantities).
We employed our own C++ code for the subsequent entropy and cumulant analysis. 

For the single-iron complex, the ROHF SCF was done configuration-averaged for 5 electrons in the 5 d-orbitals, yielding one orbital set. 
DMRG(13,14) calculations with bond dimension 512 were performed for states with multiplicities 2,4,6.
For the double-iron-sulfur complex, the configuration-averaged SCF was performed
for the 10 electrons in iron d-orbitals  and subsequently DMRG (22,21) calculations with bond dimension 2048 have been computed for the spin multiplicities 1,3,\ldots,11.
It turns out that in this complex the dynamic correlation is critical to get the correct energetic ordering of the states;
but in the (22,21) space the energy ordering is already qualitatively correct (singlet ground state), in agreement with the post-DMRG and NEVPT2 calculations \cite{fe2s2veis2024}.
We did not perform computations for several values of bond dimension and extrapolation to infinite bond dimension,
as we did not aim to compute the correlation energy with high accuracy, however, the active spaces and bond dimensions employed give a qualitatively correct energy ordering of the states and are sufficient for the qualitative correlation analysis.

\section{Results}

\subsection{A complex with one iron atom}

As a first realistic example of the presented spin-free correlation analysis we turn to  the  [Fe(SCH$_3$)$_4$]$^-$ complex,
which contains an Fe(III) center in a S$_4$ symmetry coordination environment.
We have computed the ground state and two lowest excited ones which are all dominated by Slater determinants generated by different occupation and spin coupling of the iron's d orbitals.
Their DMRG(13,14) energies are -3011.13544(6-8), -3011.0279(49-50), and  -3011.016680 a.u., respectively, where the digits in parenthesis indicate the numerical inaccuracy between individual $M_s$ components,
which ought to be degenerate.

An overview of the spin-free and spin-including total quantum information, total Von Neumann entropy, mutual information and mutual correlation for the three states of this complex are given in Tab.~\ref{fecomplextable}.
As can be seen, the spin-free quantities are invariant with respect to $M_s$, while the spin-including ones strongly depend on the spin projection,
reflecting the different static correlation due to spin coupling.
It is interesting to see that the total Von Neumann entropy computed from the 1RDM correlates well with the total  quantum information,
although it does not reach as high value for the $S=1/2$ state.
The spin-free total mutual correlation $\tilde{M}_{\rm tot}$ monotonously grows with decreasing $S$,
as the spin-free total mutual information $\tilde{I}_{\rm tot}$ does,  while $M_{\rm tot}$ has an opposite trend
than $I_{\rm tot}$ between the (3/2,1/2) and (1/2,1/2) states.
Notice also that mutual correlation attains considerably smaller values than mutual information, in agreement with the results of Evangelista \cite{evangelista2025mutual} for the $H_4$ model.

 The ground state follows the Hund's rule and turns to be a sextet, with each of the five MOs formed from 3d-orbitals of iron being singly occupied.
Reconstruction of the leading terms of the wave function from the MPS form confirmed that the $M_s=5/2$ component is single-determinantal,
while the $M_s=3/2$ component is dominated by 5 equally contributing Slater determinants, where four of the MOs corresponding to the iron 3d-orbitals have $\alpha$-spin and one has $\beta$ spin. For  $M_s=1/2$, ${5 \choose 2} = 10$ Slater determinants with three  $\alpha$ and two $\beta$ spins contribute equally. 
Clearly, different $M_s$ components much differ in the static correlation involved, increasing in the low spin  direction, which will strongly affect the
``traditional'' correlation diagrams, but shall be filtered out in their spin-free counterparts.

\begin{table}
\caption{
\label{fecomplextable}
Spin-free and spin-including total quantum informations $\tilde{S}_{\rm tot}$, $S_{\rm tot}$, total Von Neumann entropy from 1RDM  $S_{\rm VNE}$, mutual informations  $\tilde{I}_{\rm tot}$, $I_{\rm tot}$ and mutual correlations  $\tilde{M}_{\rm tot}$, $M_{\rm tot}$ for three states of the  [Fe(SCH$_3$)$_4$]$^-$ complex and their $M_s$ components, computed using configuration averaged ROHF orbitals.
}
\begin{tabular}{ccccccccc}
\hline
$S$ & $M_s$ &  $\tilde{S}_{\rm tot}$ & $S_{\rm tot}$ & $S_{\rm VNE}$ &  $\tilde{I}_{\rm tot}$ & $I_{\rm tot}$ &   $\tilde{M}_{\rm tot}$ & $M_{\rm tot}$ \\
\hline
5/2 & 5/2 & 0.085 & 0.087 & 0.045 & 0.044 & 0.082 & 0.001 & 0.000 \\
5/2 & 3/2 & 0.085 & 2.588 & 2.518 & 0.044 & 3.354 & 0.001 & 0.861 \\
5/2 & 1/2 & 0.085 & 3.451 & 3.378 & 0.044 & 4.556 & 0.001 & 1.937 \\
3/2 & 3/2 & 0.879 & 0.891 & 0.514 & 0.898 & 1.082 & 0.058 & 0.049 \\
3/2 & 1/2 & 0.879 & 2.792 & 2.255 & 0.898 & 2.960 & 0.058 & 0.809 \\
1/2 & 1/2 & 3.709 & 5.623 & 3.061 & 2.208 & 4.372 & 0.169 & 0.390 \\
\hline
\end{tabular}

{\small $\tilde{S}_{\rm tot}$,$S_{\rm tot}$,$\tilde{I}_{\rm tot}$, $I_{\rm tot}$ from Ref.~\cite{myspinfree2025}}
\end{table}

\begin{figure}[h]
\caption{
\label{spin6fecomplexanalysis}
Correlation analysis for the ground state ($S=5/2$) of the [Fe(SCH$_3$)$_4$]$^-$ complex, based on mutual information (left) and mutual correlation (right).
The bar graphs shows orbital entropies (left) or orbital sum of mutual information (right), while the color-coded size of the  mutual information (left) and mutual correlation (right) matrix elements is displayed below.
The weighted graphs combining the quantities are plotted as well.
The top row contains results for  $M_s=5/2$ , second row corresponds to  $M_s=3/2$  third row  $M_s=1/2$, and the bottom gives the spin-free results, which are identical for all $M_s$.
In the correlation graph, the orbitals in their ascending energy order are placed in the clockwise direction starting from the gray indicated ``noon''.
Since mutual correlation systematically attains smaller values than mutual information, the mutual correlations have been rescaled (see text for details).
Configuration-averaged ROHF orbitals have been employed.
}
\begin{center}
\begin{tabular}{ccc}
\includegraphics[width=8cm]{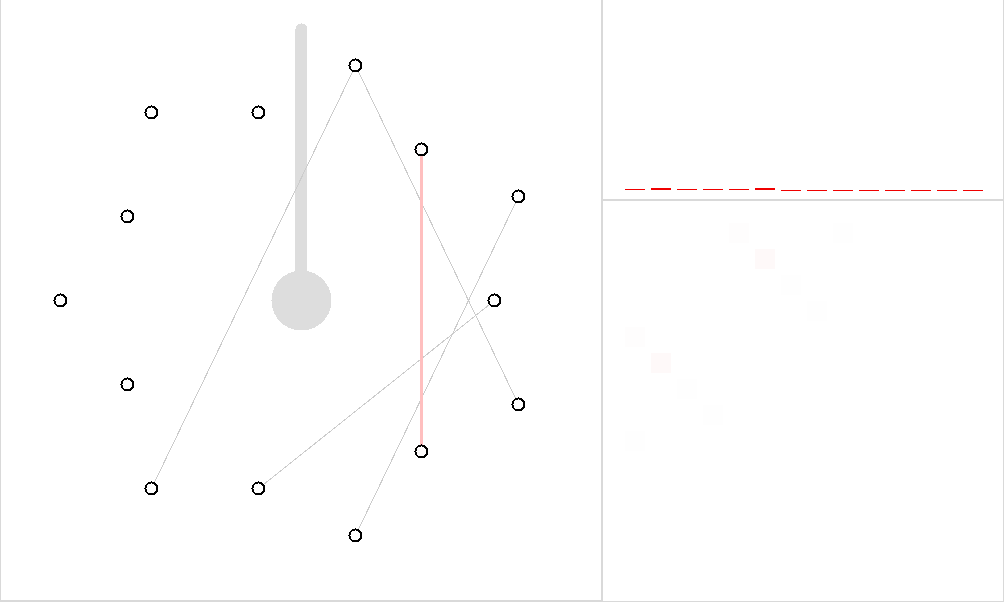} & $\;\;\;$ &\includegraphics[width=8cm]{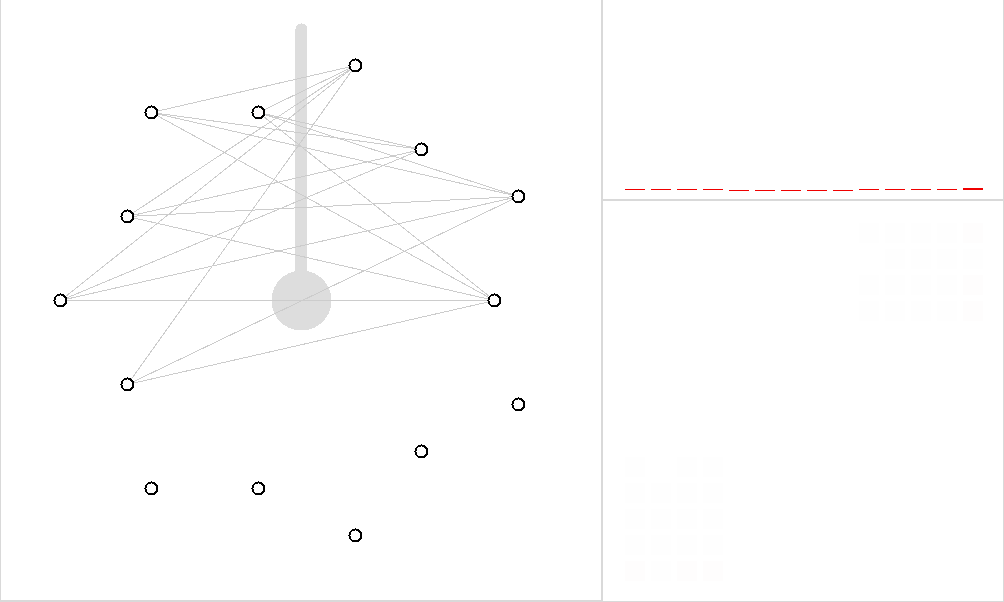}\\
\includegraphics[width=8cm]{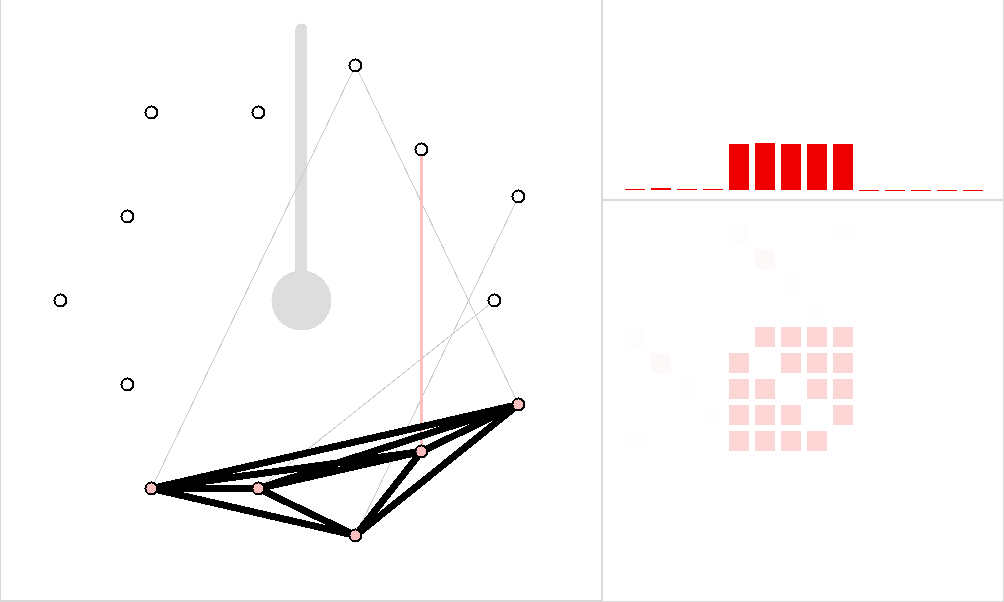} & $\;\;\;$ &\includegraphics[width=8cm]{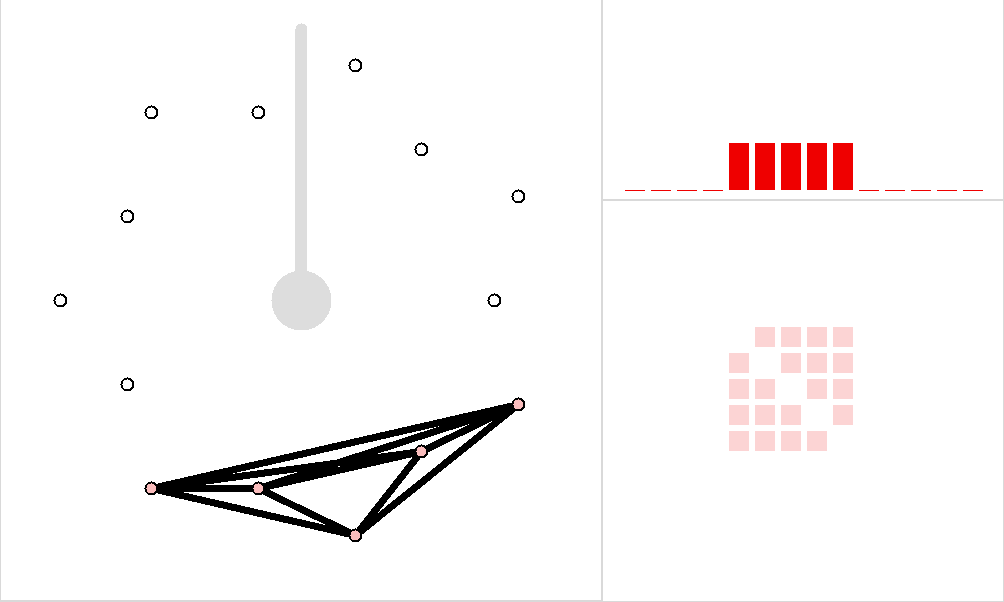}\\
\includegraphics[width=8cm]{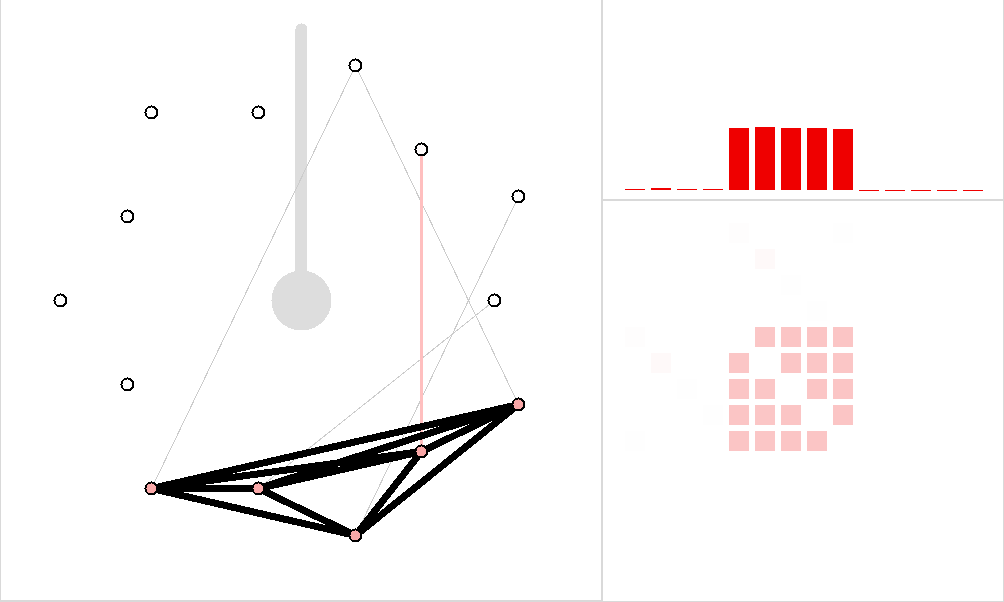} & $\;\;\;$ &\includegraphics[width=8cm]{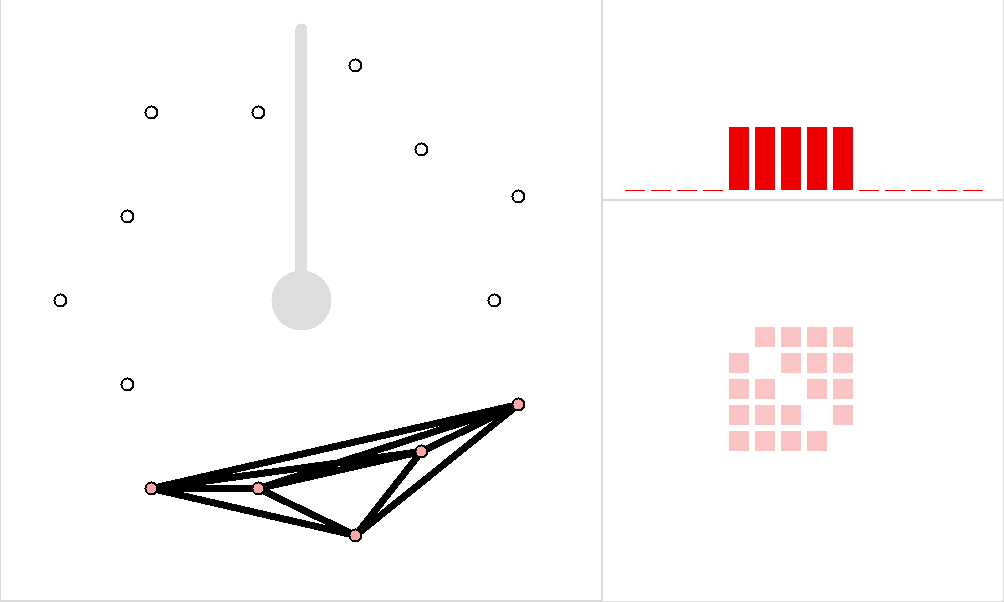}\\
\includegraphics[width=8cm]{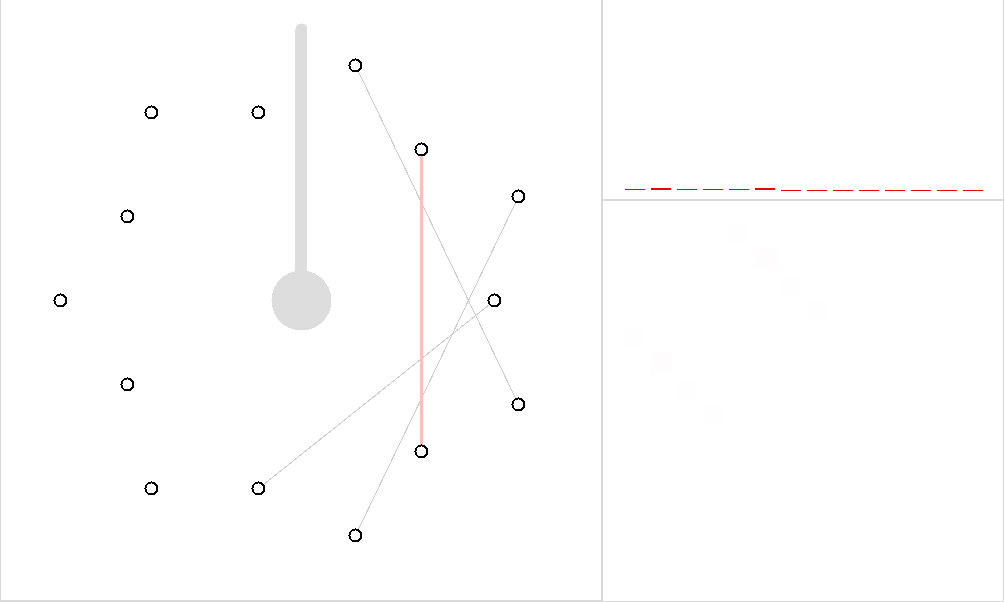} & $\;\;\;$ &\includegraphics[width=8cm]{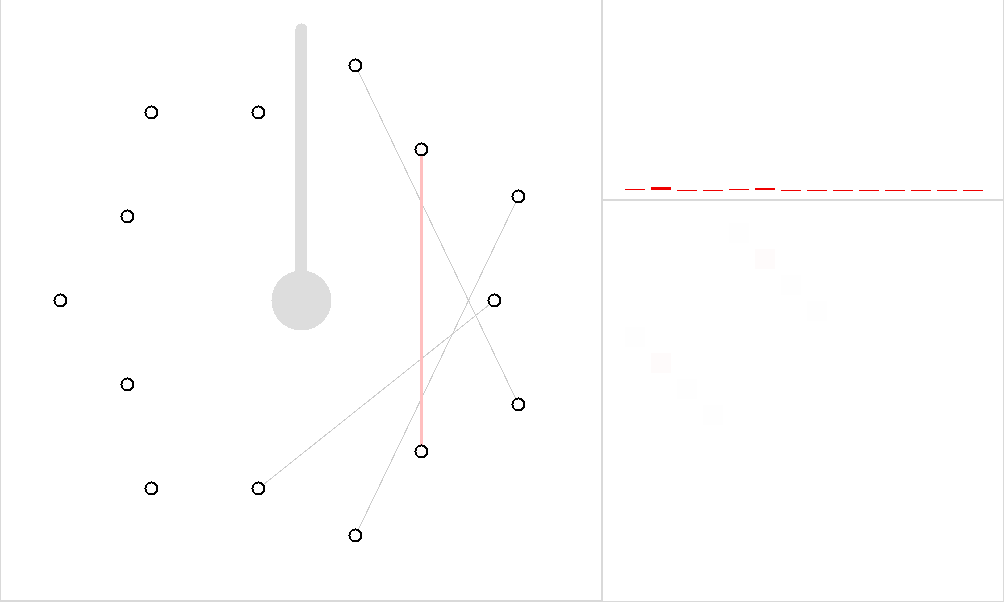}\\
\end{tabular}
\end{center}
\end{figure}

Figure~\ref{spin6fecomplexanalysis} shows the correlation analysis for the $S=5/2$ ground state for its three positive $M_s$ components in decreasing order, and the spin-free one at the bottom.
We have computed the  spin-free mutual information and mutual correlation from each $M_s$ component and verified that they are identical up to numerical noise (for mutual information the figures can be seen in Ref.~\cite{myspinfree2025}).
The left column shows analysis based on the orbital entropy and mutual information, while on the right the mutual correlation results are plotted, with partial sums of mutual correlation in place of the orbital entropies.
Since the mutual correlation attains a different absolute scale than mutual information, in this figure and henceforth the mutual correlations, interpreted as a vector for all pairs of orbitals $\{M_{pq}, \forall p<q\}$,
have been rescaled to have the same norm as analogous vectors of the mutual information.
As can be seen, the correlation diagrams based on mutual information and mutual correlation are in this case almost indistinguishable. 
The picture clearly shows the static correlation due to spin coupling in the $M_s=3/2$ component  and a slightly stronger one in the $M_s=1/2$ component,
while the spin-free results show almost no strong correlation present.

\begin{figure}[h]
\caption{
\label{spin4fecomplexanalysis}
Correlation analysis for the state ($S=3/2$) of the [Fe(SCH$_3$)$_4$]$^-$ complex, based on mutual information (left) and mutual correlation (right).
The bar graphs shows orbital entropies (left) or orbital sum of mutual information (right), while the color-coded size of the  mutual information (left) and mutual correlation (right) matrix elements is displayed below.
The weighted graphs combining the quantities are plotted as well.
The top row contains results for  $M_s=3/2$ , second row corresponds to  $M_s=1/2$, and the bottom gives the spin-free results, which are identical for all $M_s$.
In the correlation graph, the orbitals in their ascending energy order are placed in the clockwise direction starting from the gray indicated ``noon''.
Since mutual correlation systematically attains smaller values than mutual information, the mutual correlations have been rescaled (see text for details).
Configuration-averaged ROHF orbitals have been employed.
}
\begin{center}
\begin{tabular}{ccc}
\includegraphics[width=8cm]{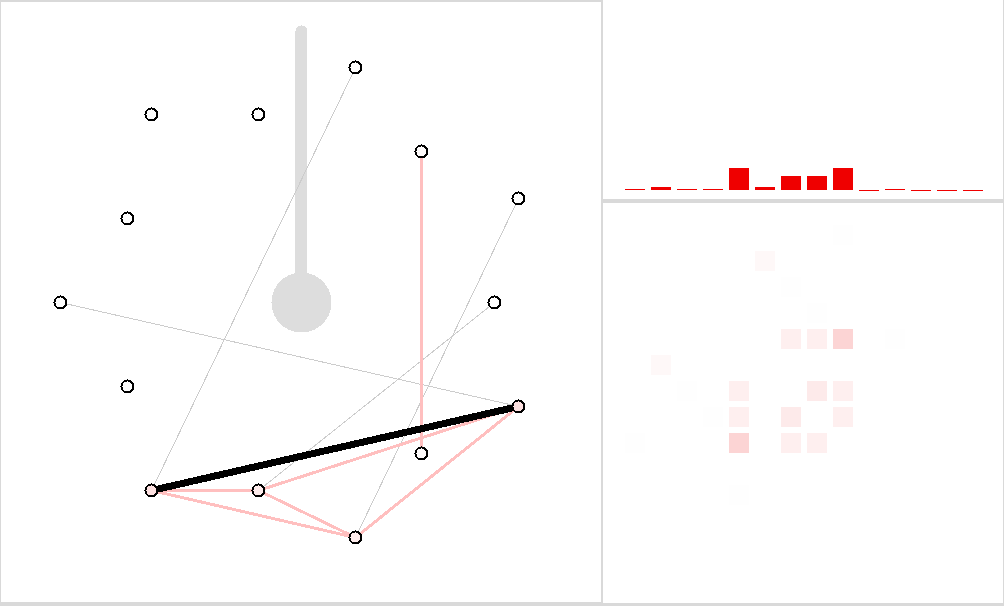} & $\;\;\;$ &\includegraphics[width=8cm]{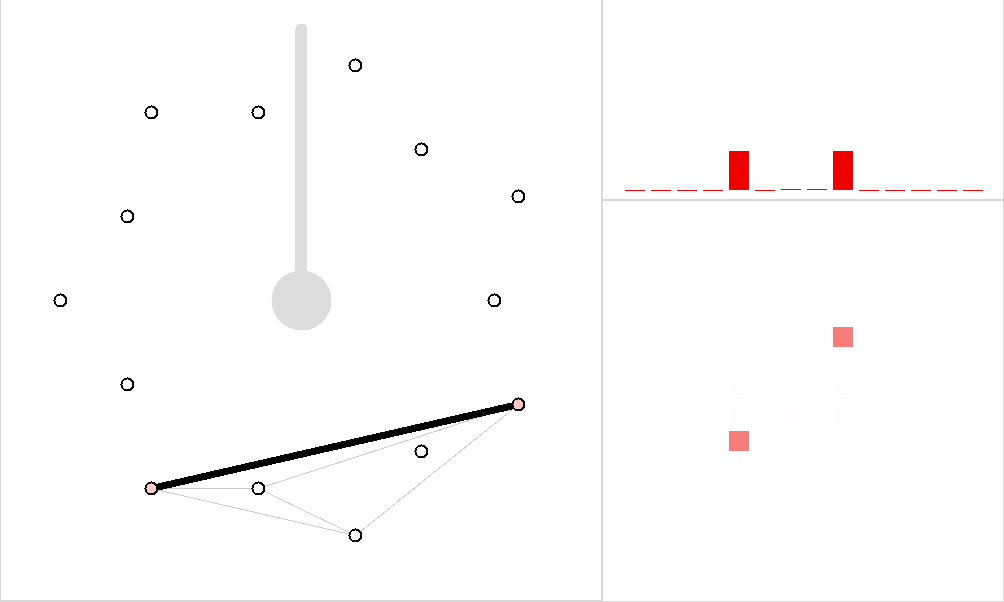}\\
\includegraphics[width=8cm]{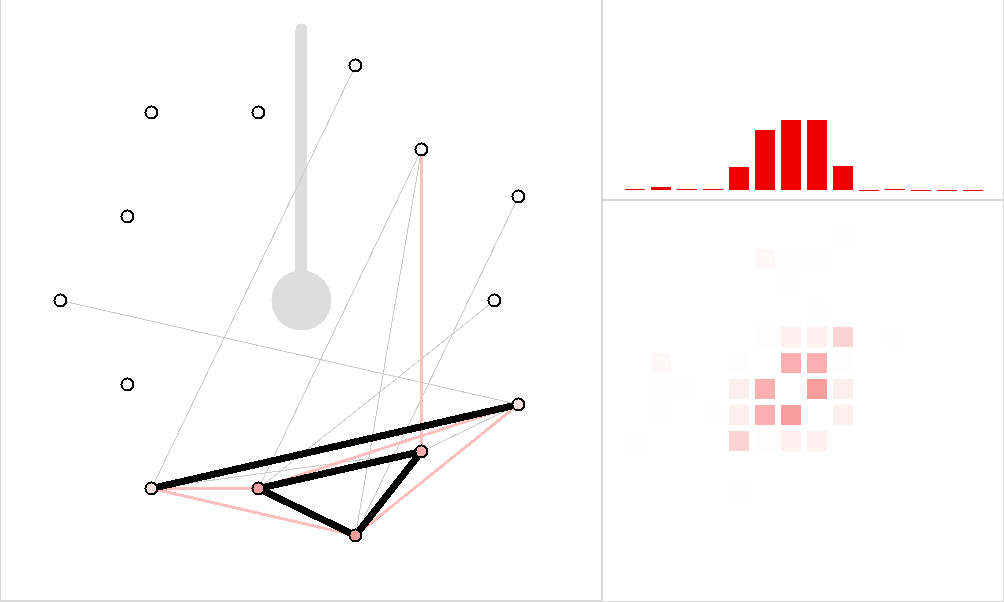} & $\;\;\;$ &\includegraphics[width=8cm]{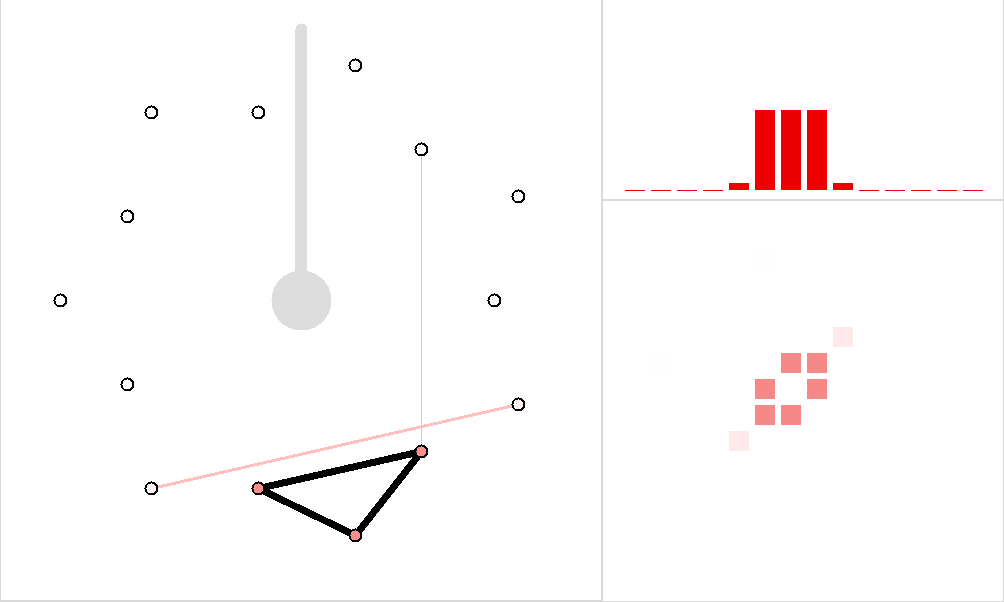}\\
\includegraphics[width=8cm]{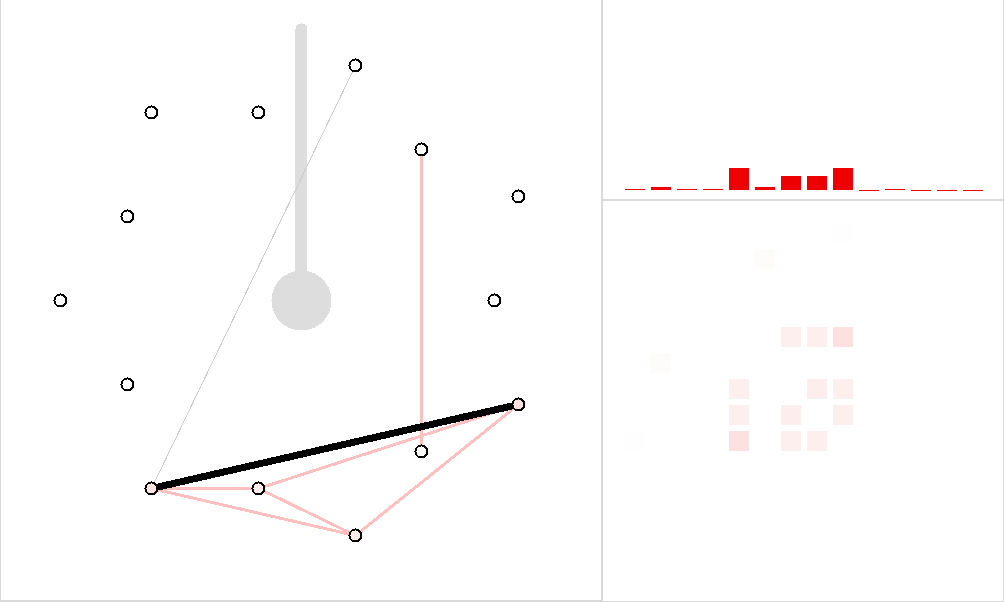} & $\;\;\;$ &\includegraphics[width=8cm]{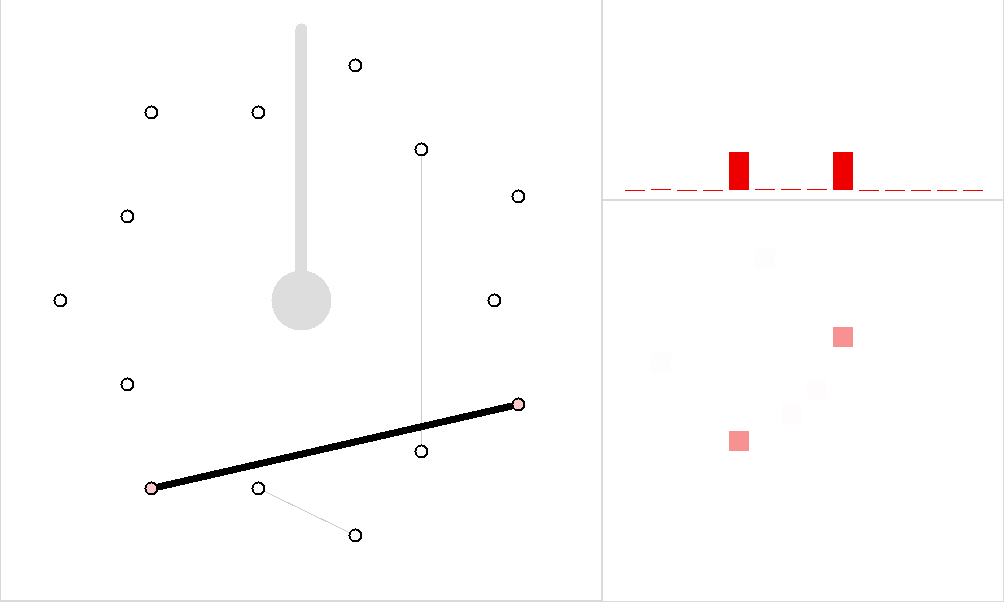}\\
\end{tabular}
\end{center}
\end{figure}

The first excited state  is a quartet, with its  $M_s=3/2$  component dominated by the determinant $2\alpha\alpha\alpha0$ with a non-negligible contribution of  $0\alpha\alpha\alpha2$, while the  $M_s=1/2$ component has equal dominant contribution of determinants $2\beta\alpha\alpha0$, $2\alpha\beta\alpha0$, and $2\alpha\alpha\beta0$ with a non-negligible contribution of analogous $0\ldots2$ determinants.
The correlation analysis for this state given in Fig.~\ref{spin4fecomplexanalysis} shows again a qualitative agreement between the mutual information and mutual correlation, although there are slight
deviations, in contrast to the ground state, where both columns were almost identical.

\begin{figure}[h]
\caption{
\label{spin2fecomplexanalysis}
Correlation analysis for the state ($S=1/2$) of the [Fe(SCH$_3$)$_4$]$^-$ complex, based on mutual information (left) and mutual correlation (right).
The bar graphs shows orbital entropies (left) or orbital sum of mutual information (right), while the color-coded size of the  mutual information (left) and mutual correlation (right) matrix elements is displayed below.
The weighted graphs combining the quantities are plotted as well.
The top row contains results for  $M_s=1/2$ , and the bottom gives the spin-free results.
In the correlation graph, the orbitals in their ascending energy order are placed in the clockwise direction starting from the gray indicated ``noon''.
Since mutual correlation systematically attains smaller values than mutual information, the mutual correlations have been rescaled (see text for details).
Configuration-averaged ROHF orbitals have been employed.
}
\begin{center}
\begin{tabular}{ccc}
\includegraphics[width=8cm]{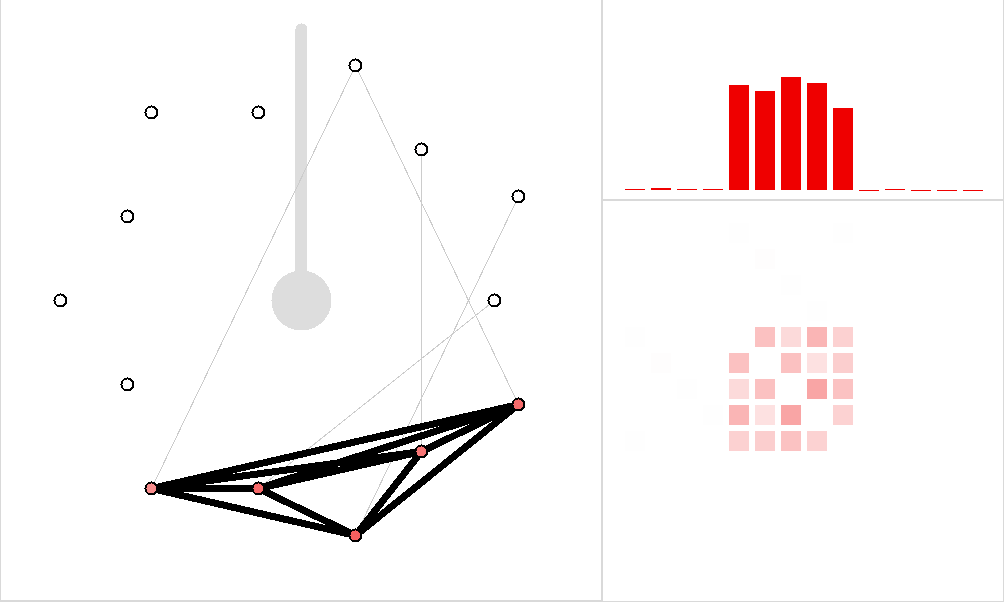} & $\;\;\;$ &\includegraphics[width=8cm]{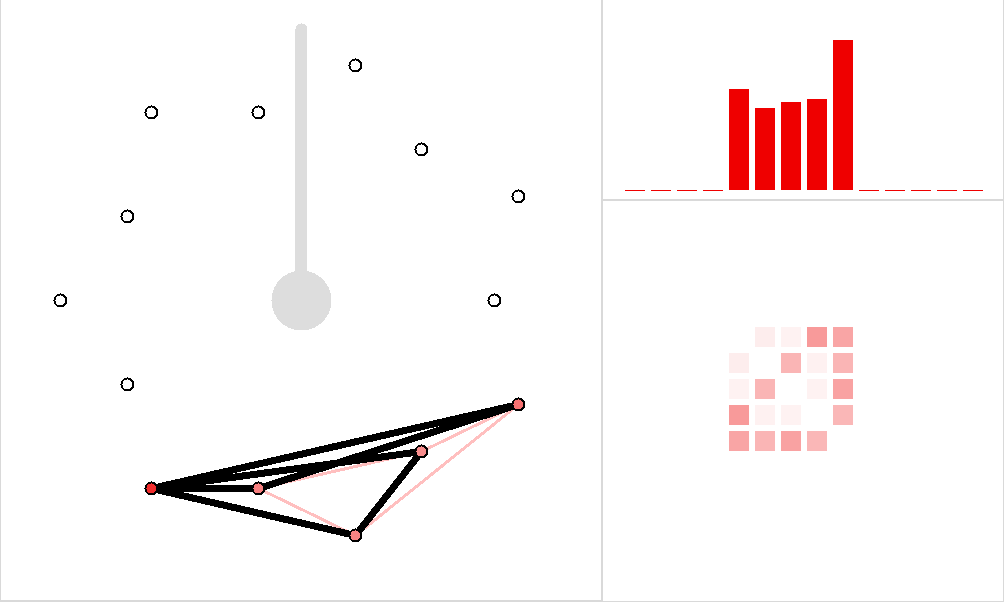}\\
\includegraphics[width=8cm]{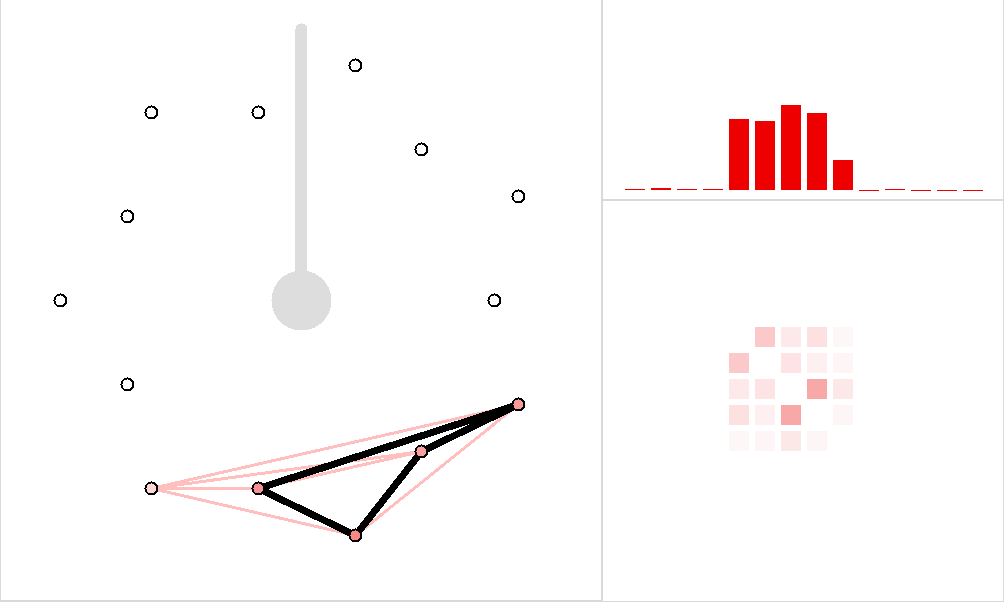} & $\;\;\;$ &\includegraphics[width=8cm]{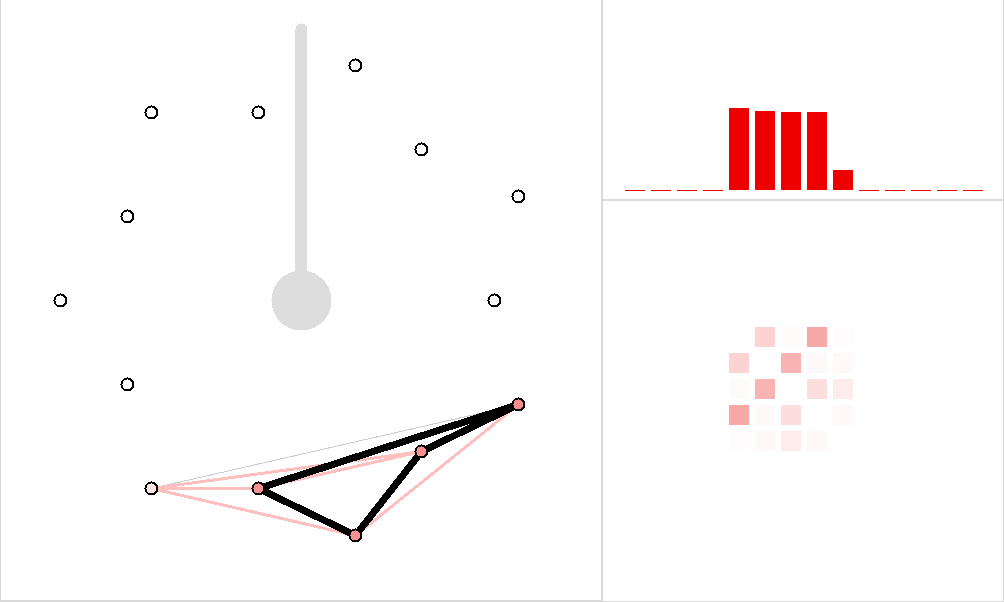}\\
\end{tabular}
\end{center}
\end{figure}

Figure~\ref{spin2fecomplexanalysis} then shows the corresponding correlation analysis for the second excited state, which is a doublet with 
its wave function dominated by several triples of equally contributing Slater determinants, where one of the 5 iron 3d MOs is doubly occupied, one empty, and the remaining three have two $\alpha$ and one $\beta$ electrons, in all three possible orderings. 
The static correlation due to spin couplings is thus inherently mixed with the strong non-dynamic correlation in this case, yielding a complex pattern, 
which is somewhat simplified in the spin-free case.
Again, one can observe an excellent qualitative agreement between the mutual information and mutual correlation analysis.

\begin{figure}[h]
\caption{
\label{fecomplexscatter1}
Scatter plots correlating orbital entropy, Von Neumann orbital from from 1RDM, and orbital sum of mutual correlation.
The quantities on the y axis were renormalized to match the norm of the orbital entropy vector (over all orbitals).
Colors code the state: $S=5/2$ red, $S=3/2$ orange, $S=1/2$ gray.
Configuration-averaged ROHF orbitals have been employed.
}
\begin{center}
\begin{tabular}{ccccc}
\includegraphics[width=5cm]{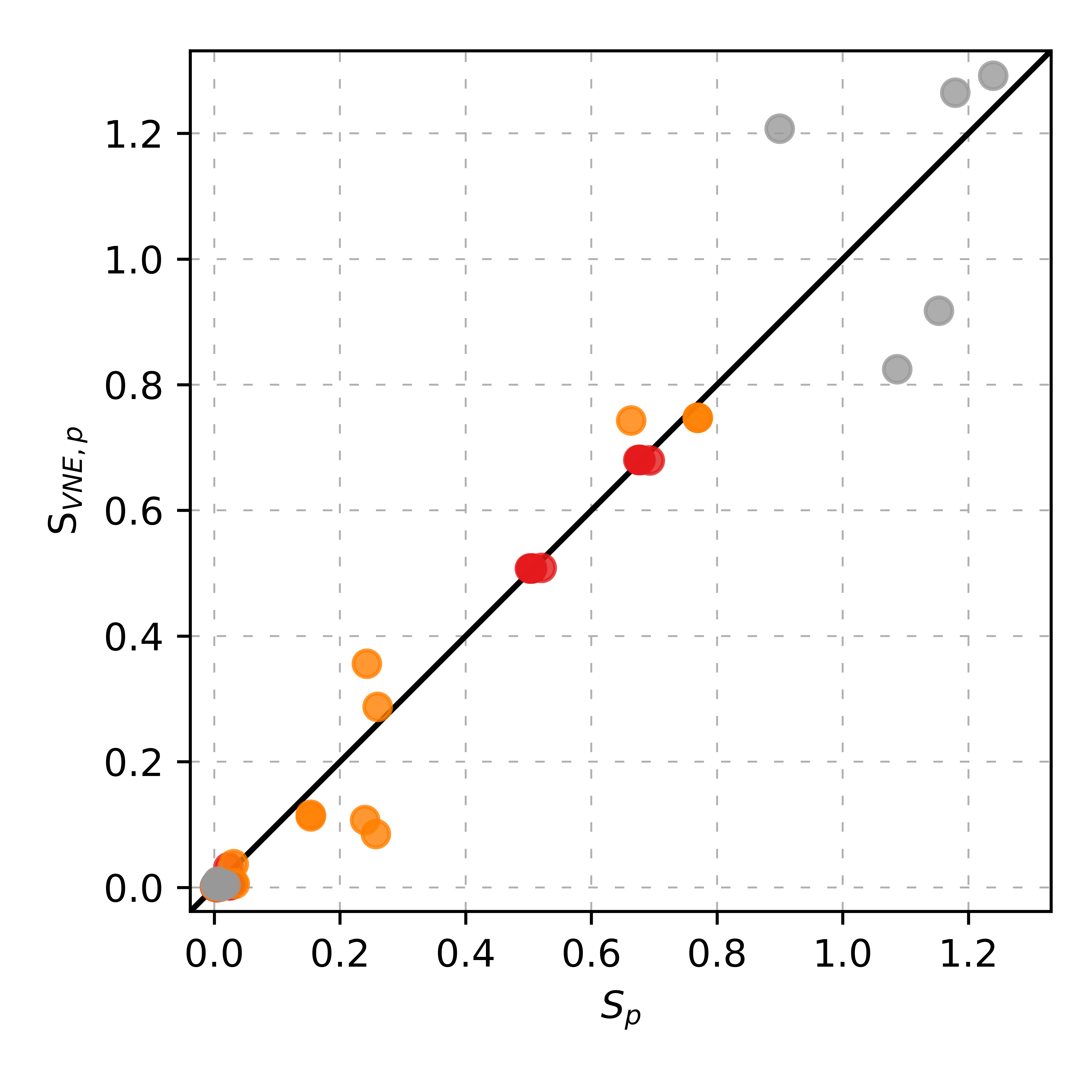} & $\;\;\;$ &\includegraphics[width=5cm]{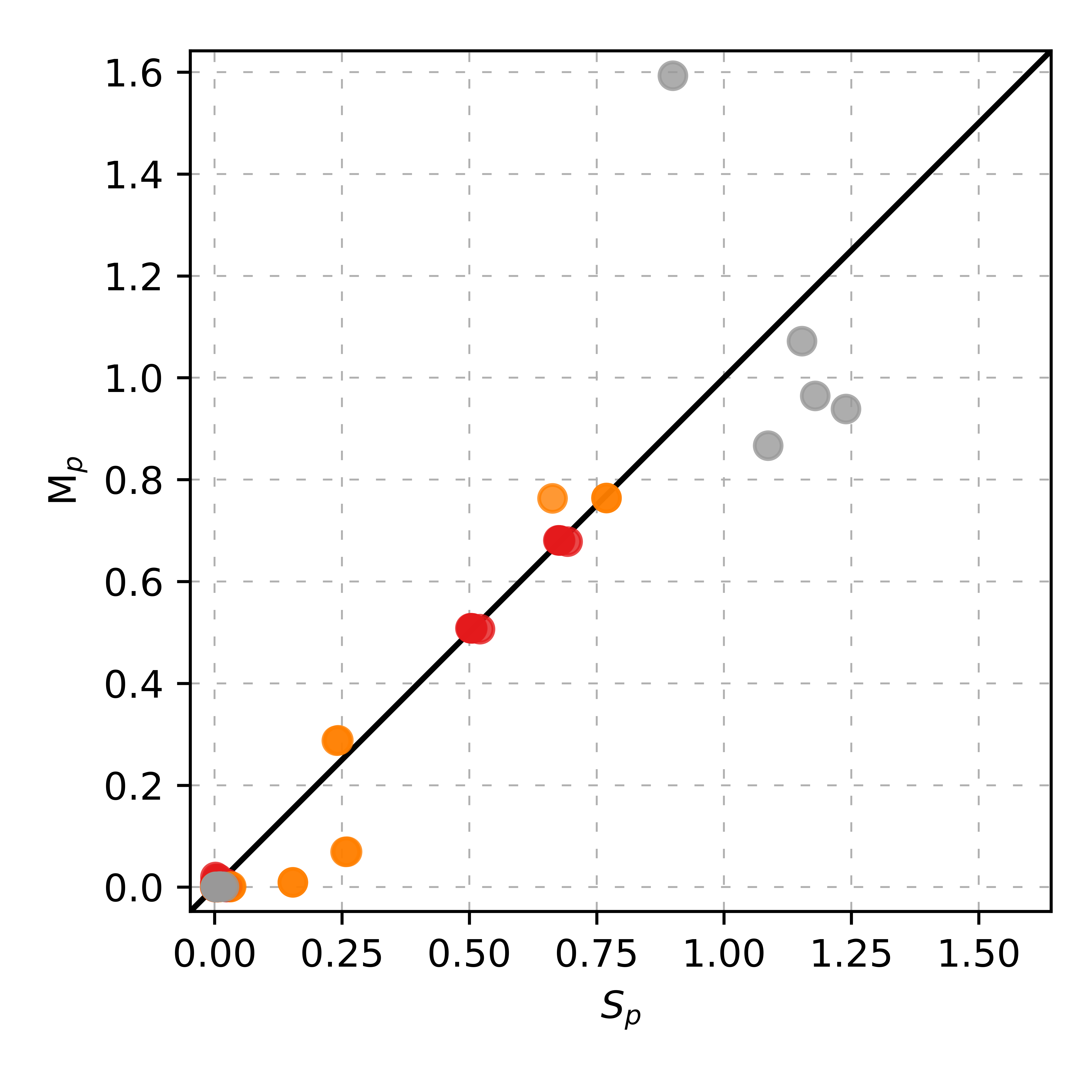} & $\;\;\;$ &\includegraphics[width=5cm]{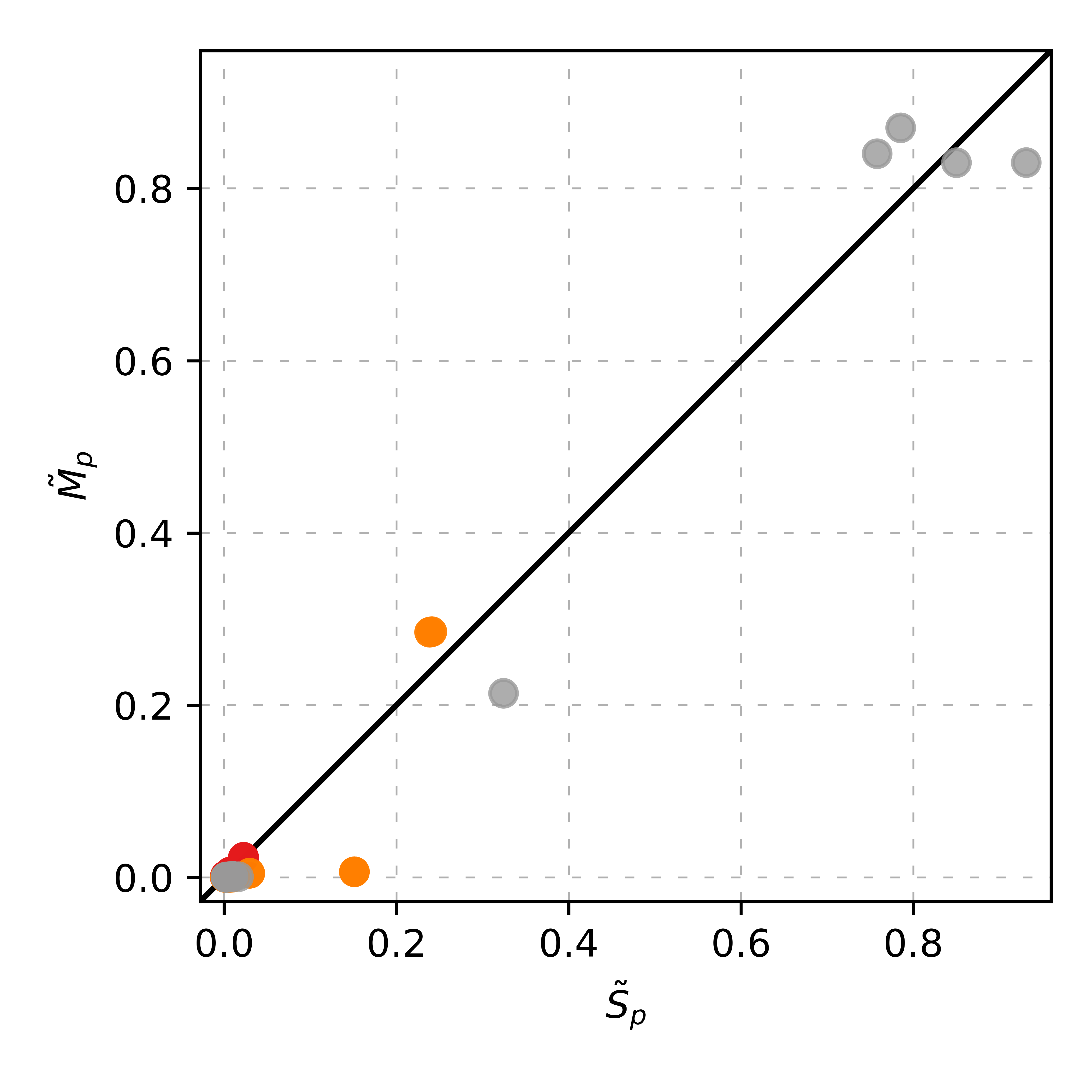}\\
\end{tabular}
\end{center}
\end{figure}

To facilitate the comparison between the different orbital correlation measures, in Fig.~\ref{fecomplexscatter1} we present scatter plots correlating 
the orbital entropy with orbital contributions to the the 1RDM-based Von Neumann  entropy (left), orbital entropy with the partial sums of mutual correlation (middle),
and the spin-free version thereof (right). Individual dots correspond to values for each orbital, color-coded by the $S$ quantum number of the state.
The correlation between the quantities is clearly excellent for the ground state and the first excited one, while being a bit looser for the second excited state.
However, the spin-free quantities correlate quite well for this state too.

\begin{figure}[h]
\caption{
\label{fecomplexscatter2}
Scatter plots correlating mutual information and  mutual correlation.
The quantities on the y axis were renormalized to match the norm of the mutual information vector (over all pairs orbitals).
Colors code the state: $S=5/2$ red, $S=3/2$ orange, $S=1/2$ gray.
Configuration-averaged ROHF orbitals have been employed.
}
\begin{center}
\begin{tabular}{ccc}
\includegraphics[width=5cm]{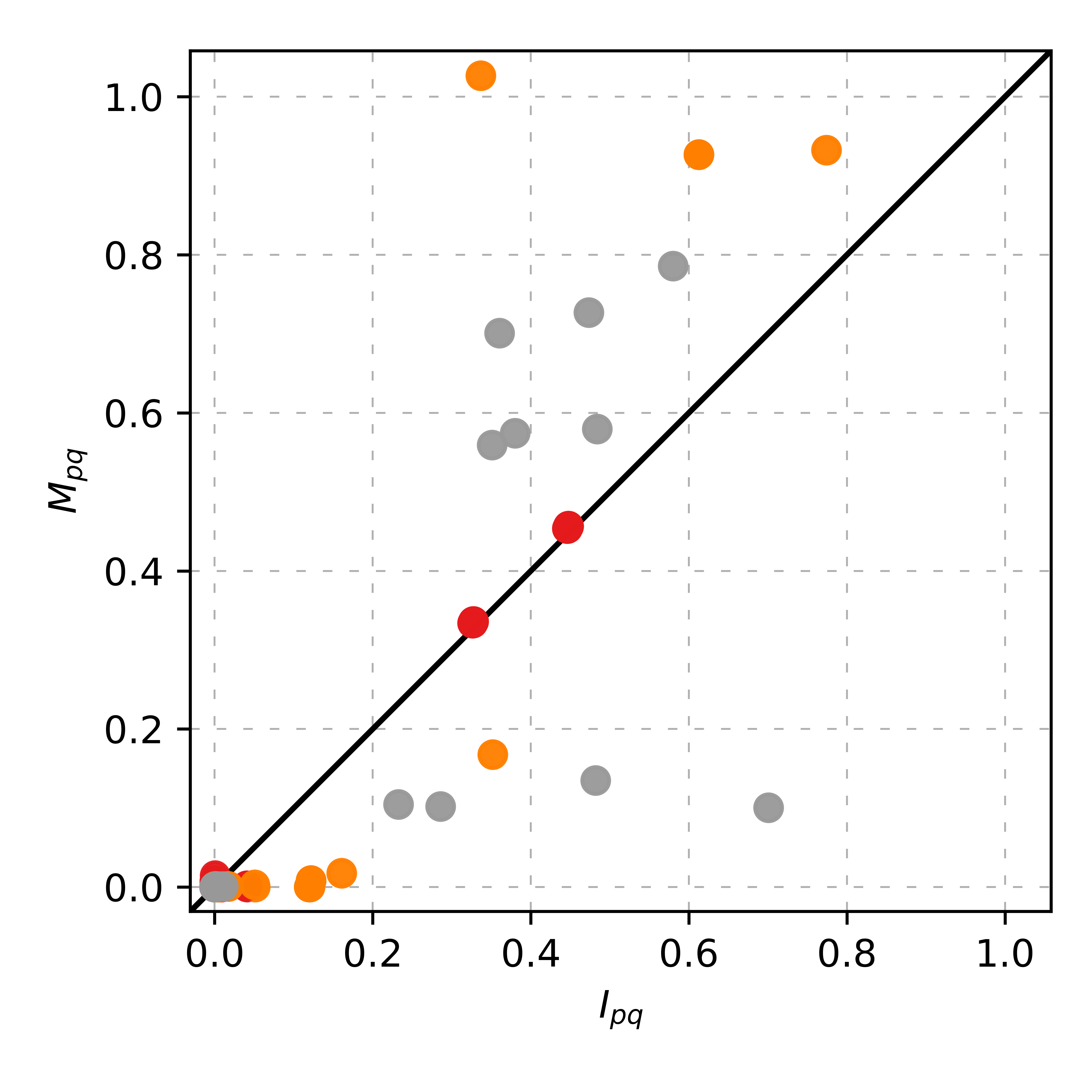} & $\;\;\;$ &\includegraphics[width=5cm]{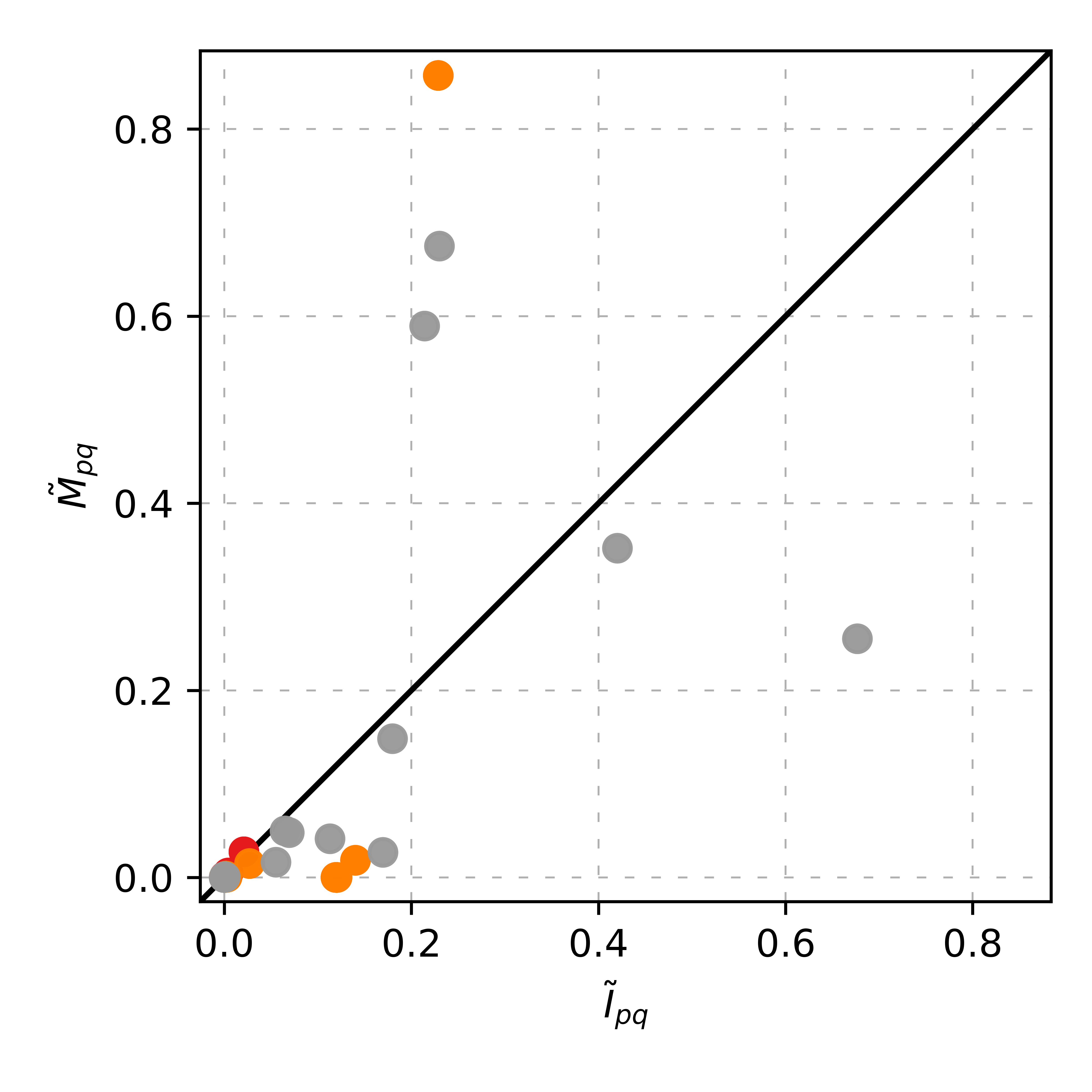}\\
\end{tabular}
\end{center}
\end{figure}

Similarly, Fig.~\ref{fecomplexscatter2} shows the scatter plots between mutual information and mutual correlation (spin-free version on the right).
 Individual dots correspond to values for each pair of orbitals, again color-coded by state.
For the ground state, the correlation is nearly perfect (and in the spin-free version all values are near zero), while the excited states show some spread, making the overall agreement more qualitative. 
For the pair quantities the correlation is not as good as for the orbital-based measures in the preceding figure.
Clearly the three- and four-body effects included in the mutual information, but not present in the mutual correlation, can have some impact in
molecules with a complicated electronic structure, like this iron complex
and one cannot expect a perfect agreement between the two correlation measures. 

\subsection{A complex with two iron atoms}

The  [Fe$_2$S$_2$(SCH$_3$)$_4$]$^{2-}$ complex is a  model of the biologically important iron-sulfur clusters, which act as
catalytic centers \cite{fe2s2complexgeom}. 
This species exhibits strong electronic correlation due to the coupling of the 3d-electrons
of the two iron atoms, while the dynamic correlation is very important to get the correct energetic order of the states of
different spin multiplicities (i.e. singlet ground state) \cite{fe2s2veis2024}.
In \cite{myspinfree2025} we have shown that DMRG(22,21) already gives a qualitatively correct energy ordering of the low-lying states
compared to a post-DMRG treatment, so we employed this active space for the correlation analysis.

Table~\ref{fe2s2complextable} gives an overall comparison of the  spin-including total quantum informations, total Von Neumann entropy from 1RDM, and total mutual information and mutual correlation for the lowest states with $S=0\ldots 5$ of this complex and their individual $M_s$ components.
In some cases we did not achieve convergence of the higher roots 
and the accuracy of the $M_s$-invariance of the spin-free quantities is in some cases limited due to the fixed bond dimension employed.
Nevertheless, one can clearly see that all the spin-free measures $\tilde{S}_{\rm tot}$, and  $\tilde{M}_{\rm tot}$
qualitative agree on the trend of decreasing strong correlation with the increasing spin quantum number $S$.
The spin-free mutual information $\tilde{I}_{\rm tot}$, however, stays very close around 4.4 for all states except $S=5$,
while the spin-free mutual correlation $\tilde{M}_{\rm tot}$ decreases monotonously in significant steps.
The $M_s$-dependent  quantities with a few exceptions agree on the trend of increasing static correlation with $M_s$ for each multiplet
while their variation with $M_s$ is very strong and hides the decreasing trend with growing $S$, particularly for the mutual information.

\begin{table}
\caption{
\label{fe2s2complextable}
Spin-free and spin-including total quantum informations $\tilde{S}_{\rm tot}$, $S_{\rm tot}$, total Von Neumann entropy from 1RDM  $S_{\rm VNE}$, mutual informations  $\tilde{I}_{\rm tot}$, $I_{\rm tot}$ and mutual correlations  $\tilde{M}_{\rm tot}$, $M_{\rm tot}$ for the ground ($S=0$) and low lying excited states of the  [Fe$_2$S$_2$(SCH$_3$)$_4$]$^{2-}$  complex and their $M_s$ components.
}
\begin{tabular}{ccrrrrrrr}
\hline
$S$ & $M_s$ &  $\tilde{S}_{\rm tot}$ & $S_{\rm tot}$ &  $S_{\rm VNE}$ &  $\tilde{I}_{\rm tot}$ & $I_{\rm tot}$ &  $\tilde{M}_{\rm tot}$ & $M_{\rm tot}$ \\
\hline
0 & 0 & 11.028 & 13.642 &6.799 & 4.456 & 8.470 &2.428 & 1.391\\
1 & 1 & 10.925 & 13.346 &6.616 & 4.342 & 8.276 &2.111 & 1.240 \\
1 & 0 & 10.920 & 13.854 &6.822 & 4.340 & 8.224 &2.110 & 1.436 \\
2 & 2 & 10.490 & 12.210 &6.033 & 4.374 & 7.854 &1.535 & 0.953 \\
2 & 1 & 10.488 & 13.612 &6.662 & 4.374 & 7.560 &1.535 & 0.960 \\
2 & 0 & 10.473 & 14.021 &6.865 & 4.362 & 7.948 &1.545 & 1.043 \\
3 & 3 &  9.303 & 10.134 &4.993 & 4.398 & 6.910 &0.857 & 0.559 \\
3 & 2 &  9.302 & 12.367 &6.101 & 4.400 & 8.400 &0.857 & 0.758 \\
3 & 1 &  9.300 & 13.402 &6.721 & 4.398 & 10.158 &0.857&  1.211 \\
3 & 0 & \multicolumn{7}{c}{not conv.}\\
4 & 4 &  6.717 &  6.791 &3.343 & 4.404 & 5.504 &0.270  &0.183\\
4 & 3 &  6.717 &  9.761 &5.074 & 4.404 & 10.578 &0.270 & 0.708 \\
4 & 2 &  6.716 & 11.239 &6.162 & 4.404 & 13.840 &0.270 & 1.768 \\
4 & 1 &  6.618 & 11.966 &6.774 & 4.262 & 15.914 &0.270 & 2.678 \\
4 & 0 &  \multicolumn{7}{c}{not conv.}\\
5 & 5 &  0.684 &  0.687 &0.346 & 0.262 & 0.642 &0.015  &0.000 \\
5 & 4 &  0.684 &  3.936 &3.395 & 0.262 & 7.370 &0.015  &0.897 \\
5 & 3 &  0.684 &  5.689 &5.103 & 0.260 & 12.030 &0.015 & 2.834 \\
5 & 2 &  0.684 &  6.793 &6.184 & 0.260 & 15.472 &0.015 & 4.881 \\
5 & 1 &  \multicolumn{7}{c}{not conv.}\\
5 & 0 &  \multicolumn{7}{c}{not conv.}\\
\hline
\end{tabular}

{\small $\tilde{S}_{\rm tot}$,$S_{\rm tot}$,$\tilde{I}_{\rm tot}$, $I_{\rm tot}$ from Ref.~\cite{myspinfree2025}}
\end{table}

\begin{figure}[h]
\caption{
\label{fe2s2complexS0}
Correlation analysis for the $S=0$ (ground) state of the [Fe$_2$S$_2$(SCH$_3$)$_4$]$^{2-}$ complex, based on mutual information (left) and mutual correlation (right).
The bar graphs shows orbital entropies (left) or orbital sum of mutual information (right), while the color-coded size of the  mutual information (left) and mutual correlation (right) matrix elements is displayed below.
The weighted graphs combining the quantities are plotted as well.
The top row contains results for  $M_s=0$ , and the bottom gives the spin-free results.
In the correlation graph, the orbitals in their ascending energy order are placed in the clockwise direction starting from the gray indicated ``noon''.
Since mutual correlation systematically attains smaller values than mutual information, the mutual correlations have been rescaled (see text for details).
}
\begin{center}
\begin{tabular}{ccc}
\includegraphics[width=8cm]{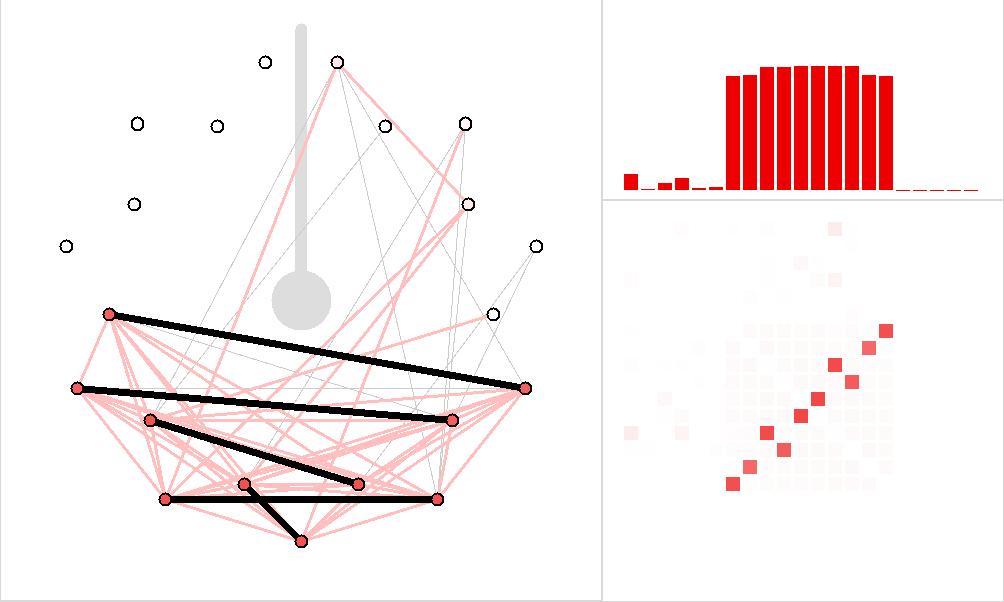} & $\;\;\;$ &\includegraphics[width=8cm]{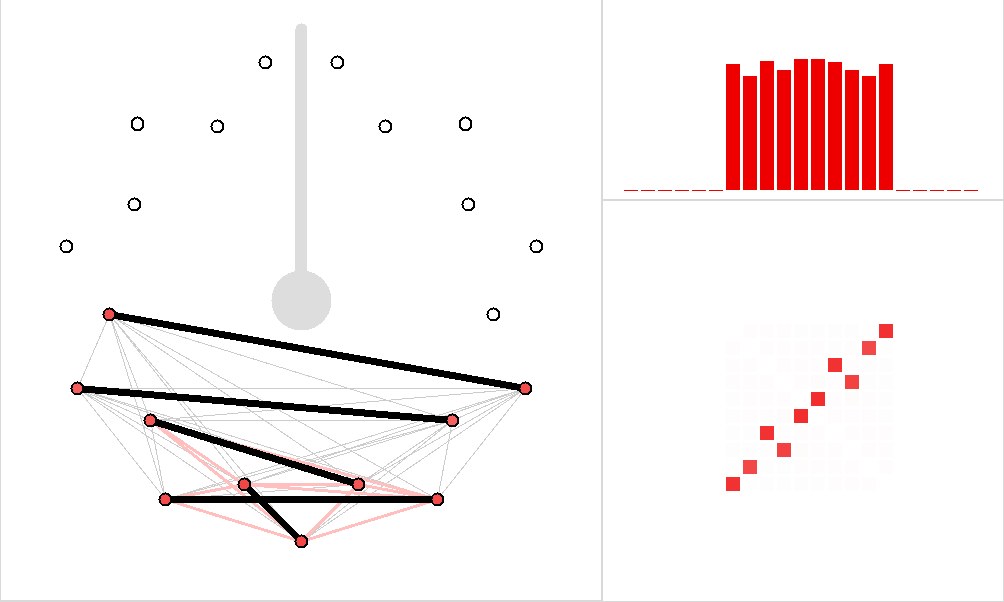}\\
\includegraphics[width=8cm]{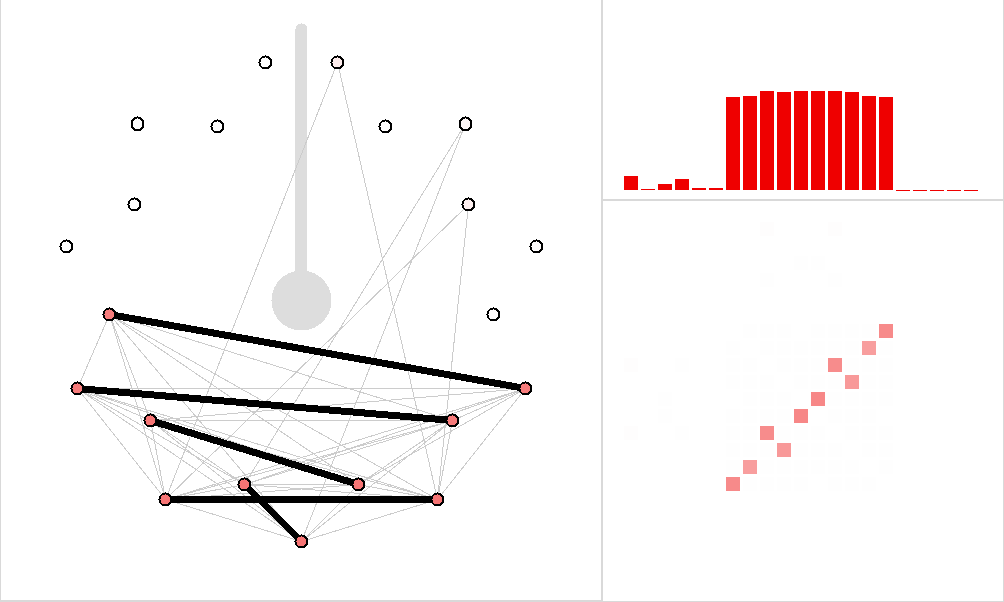} & $\;\;\;$ &\includegraphics[width=8cm]{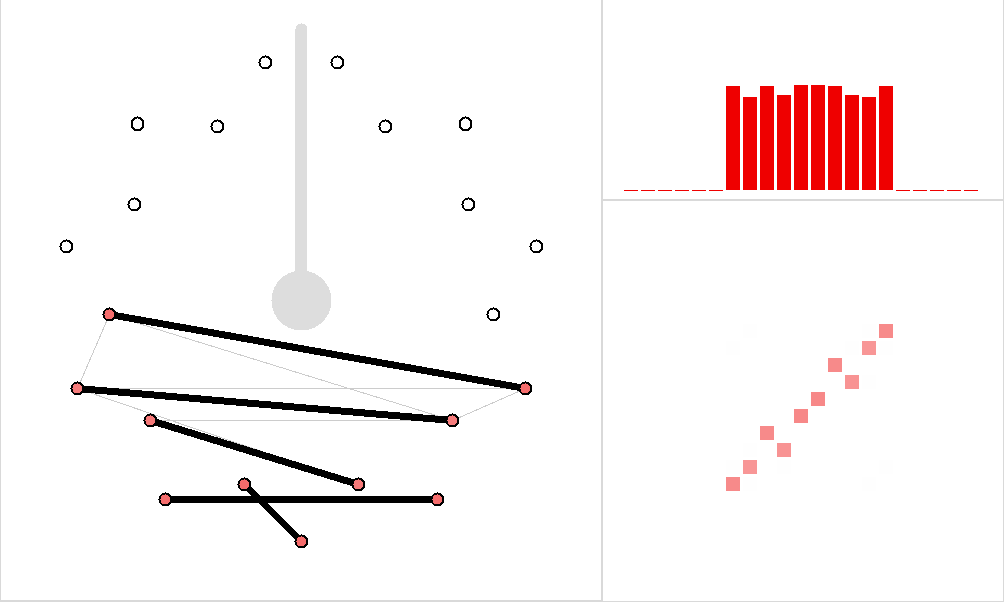}\\
\end{tabular}
\end{center}
\end{figure}

\begin{figure}[h]
\caption{
\label{fe2s2complexS1}
Correlation analysis for the $S=1$ state of the [Fe$_2$S$_2$(SCH$_3$)$_4$]$^{2-}$ complex, based on mutual information (left) and mutual correlation (right).
The bar graphs shows orbital entropies (left) or orbital sum of mutual information (right), while the color-coded size of the  mutual information (left) and mutual correlation (right) matrix elements is displayed below.
The weighted graphs combining the quantities are plotted as well.
The top row contains results for  $M_s=1$ , the second one $M_s=0$, and the bottom gives the spin-free results.
In the correlation graph, the orbitals in their ascending energy order are placed in the clockwise direction starting from the gray indicated ``noon''.
Since mutual correlation systematically attains smaller values than mutual information, the mutual correlations have been rescaled (see text for details).
}
\begin{center}
\begin{tabular}{ccc}
\includegraphics[width=8cm]{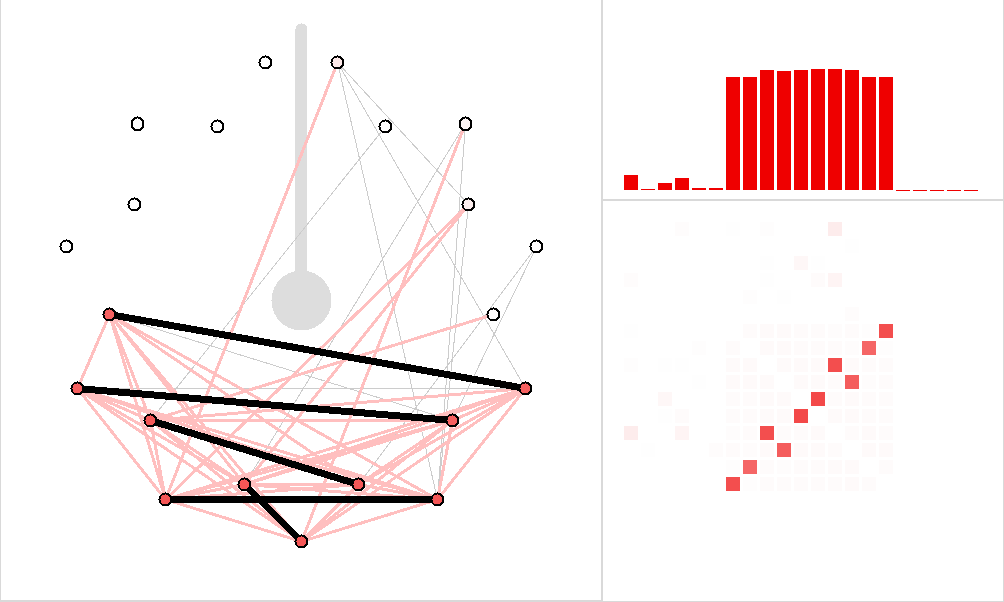} & $\;\;\;$ &\includegraphics[width=8cm]{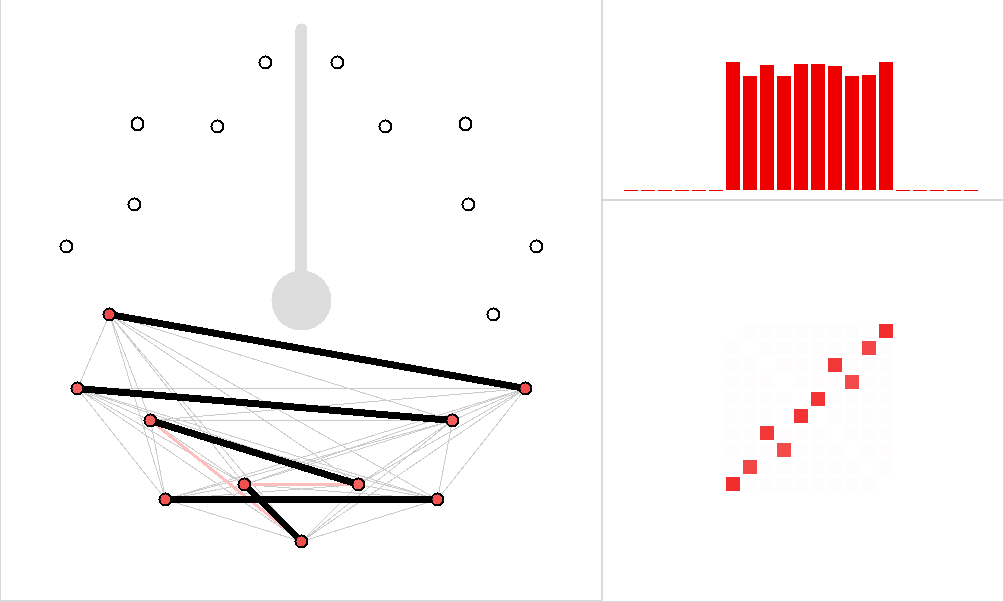}\\
\includegraphics[width=8cm]{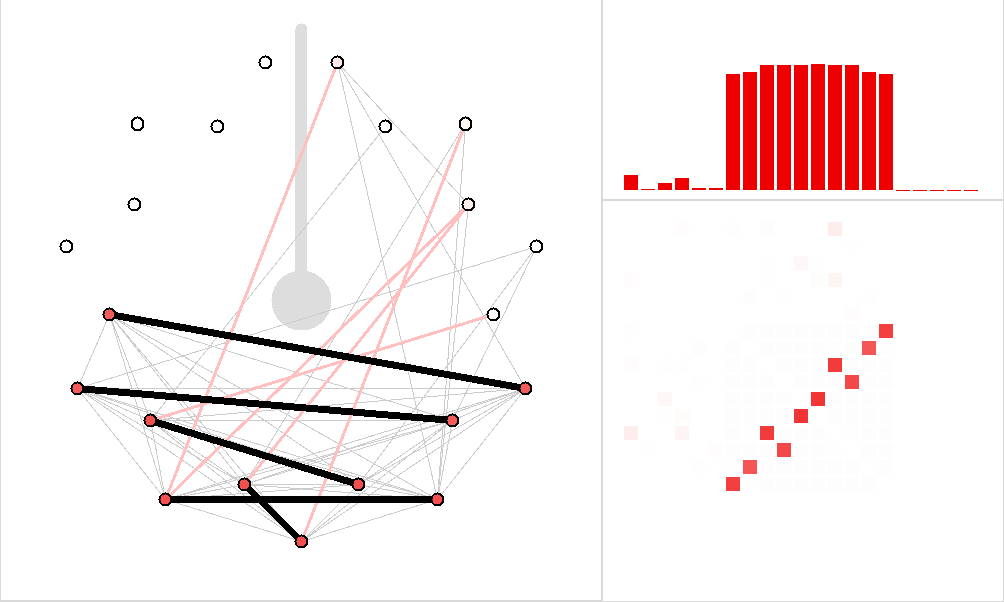} & $\;\;\;$ &\includegraphics[width=8cm]{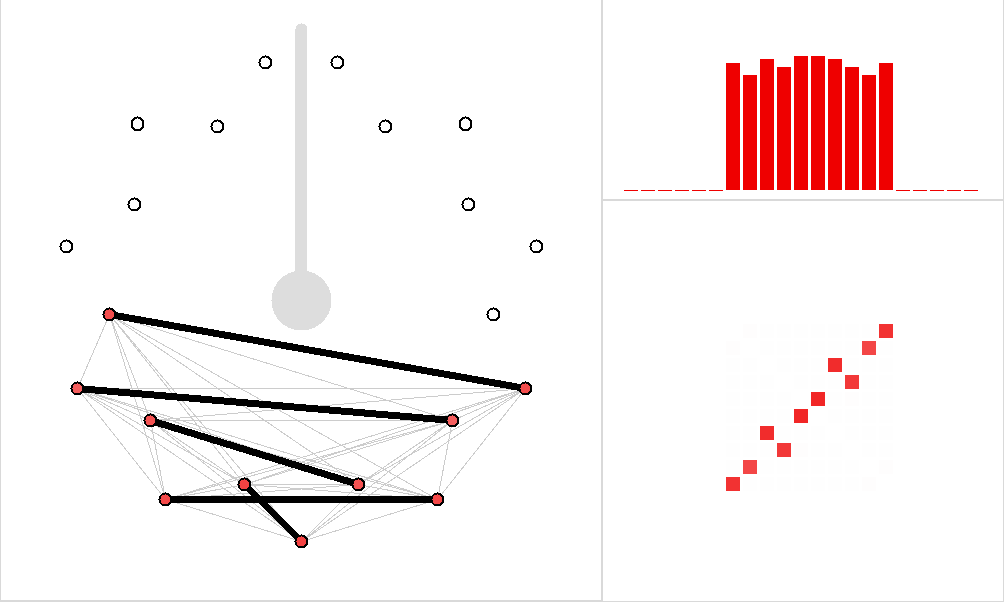}\\
\includegraphics[width=8cm]{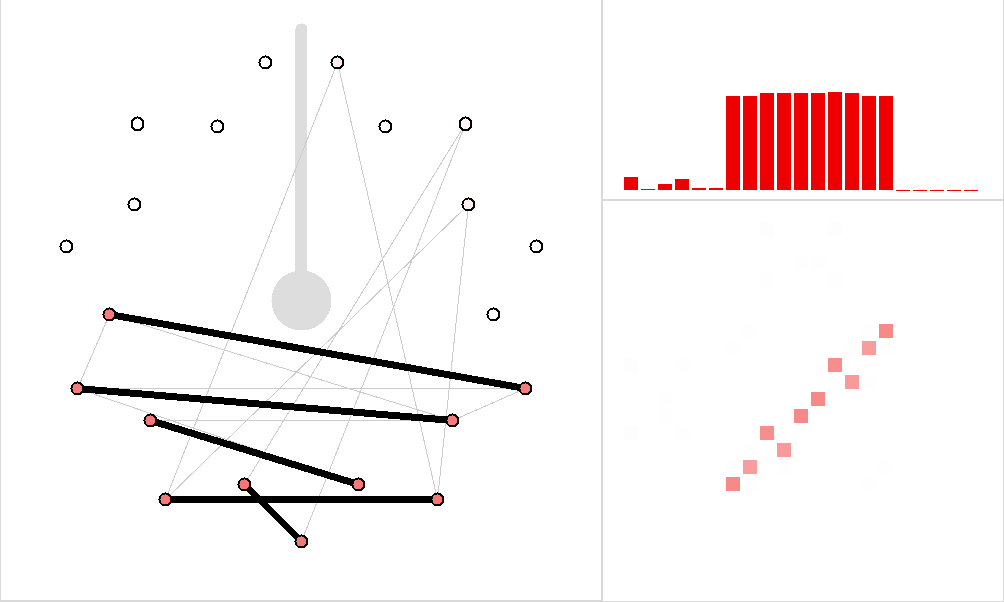} & $\;\;\;$ &\includegraphics[width=8cm]{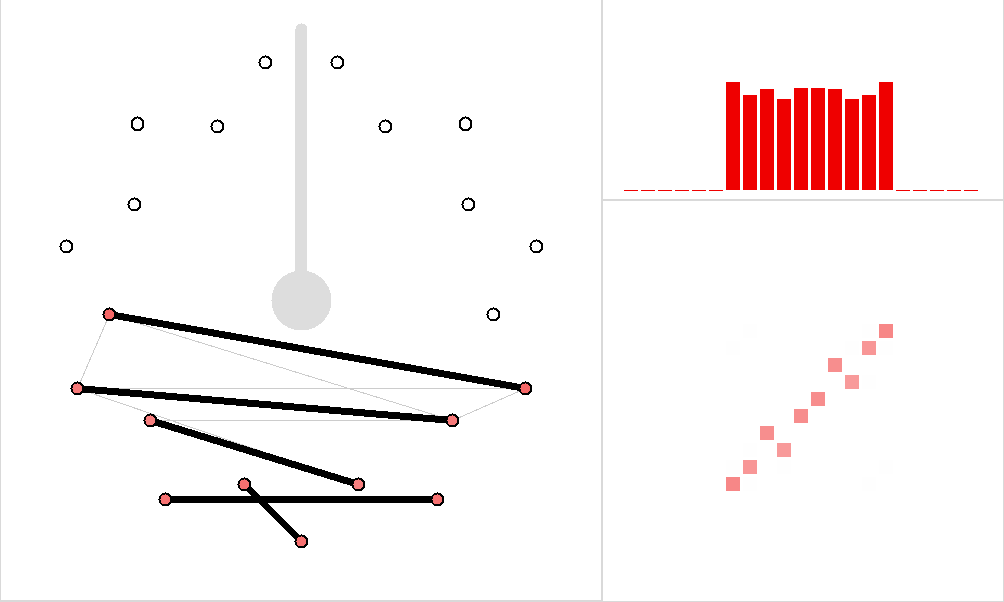}\\
\end{tabular}
\end{center}
\end{figure}

\begin{figure}[h]
\caption{
\label{fe2s2complexS2}
Correlation analysis for the $S=2$ state of the [Fe$_2$S$_2$(SCH$_3$)$_4$]$^{2-}$ complex, based on mutual information (left) and mutual correlation (right).
The bar graphs shows orbital entropies (left) or orbital sum of mutual information (right), while the color-coded size of the  mutual information (left) and mutual correlation (right) matrix elements is displayed below.
The weighted graphs combining the quantities are plotted as well.
The top row contains results for  $M_s=2$ , the second one $M_s=1$, third one $M_s=0$, and the bottom gives the spin-free results.
In the correlation graph, the orbitals in their ascending energy order are placed in the clockwise direction starting from the gray indicated ``noon''.
Since mutual correlation systematically attains smaller values than mutual information, the mutual correlations have been rescaled (see text for details).
}
\begin{center}
\begin{tabular}{ccc}
\includegraphics[width=8cm]{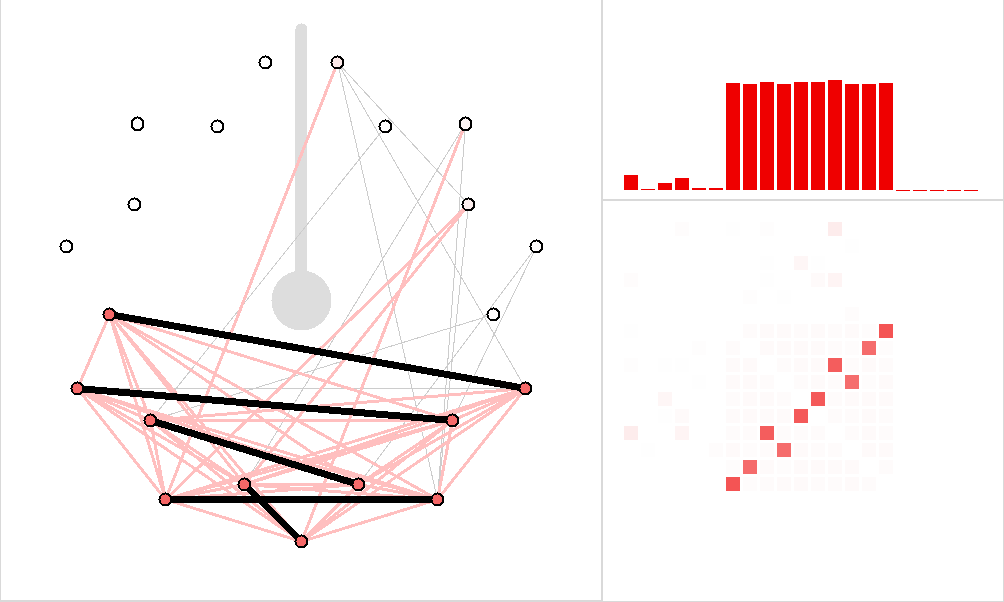} & $\;\;\;$ &\includegraphics[width=8cm]{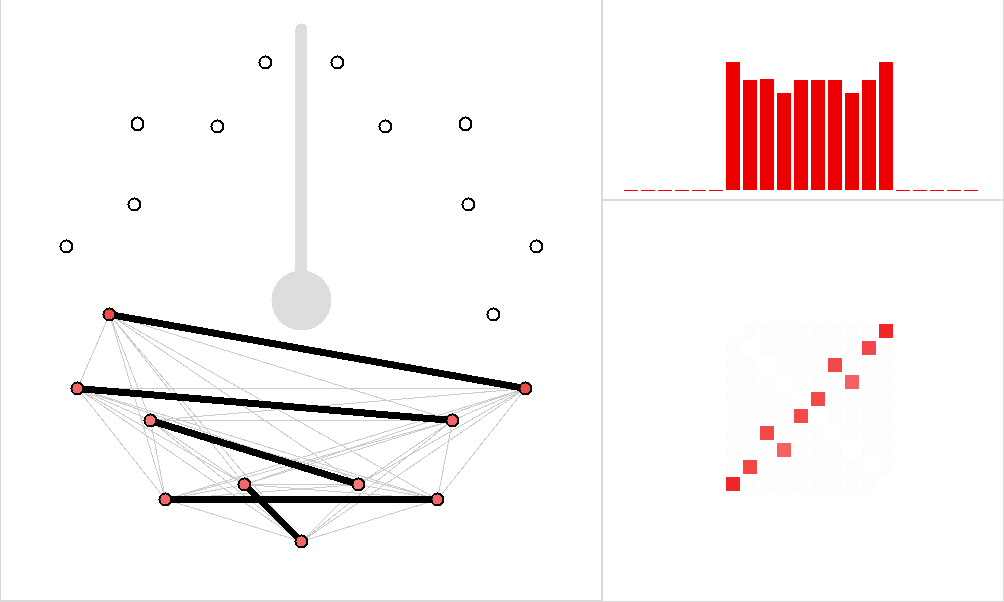}\\
\includegraphics[width=8cm]{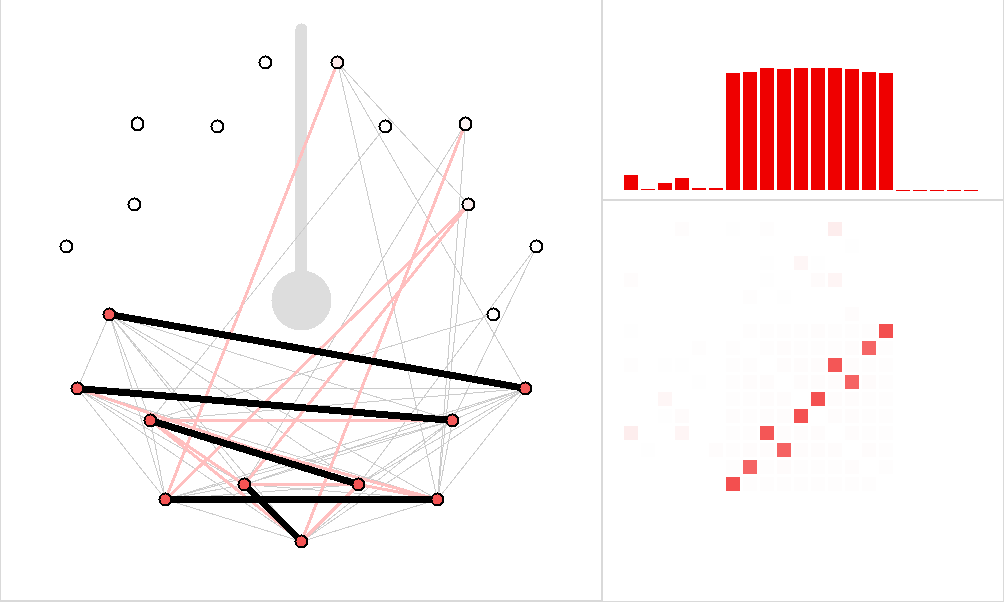} & $\;\;\;$ &\includegraphics[width=8cm]{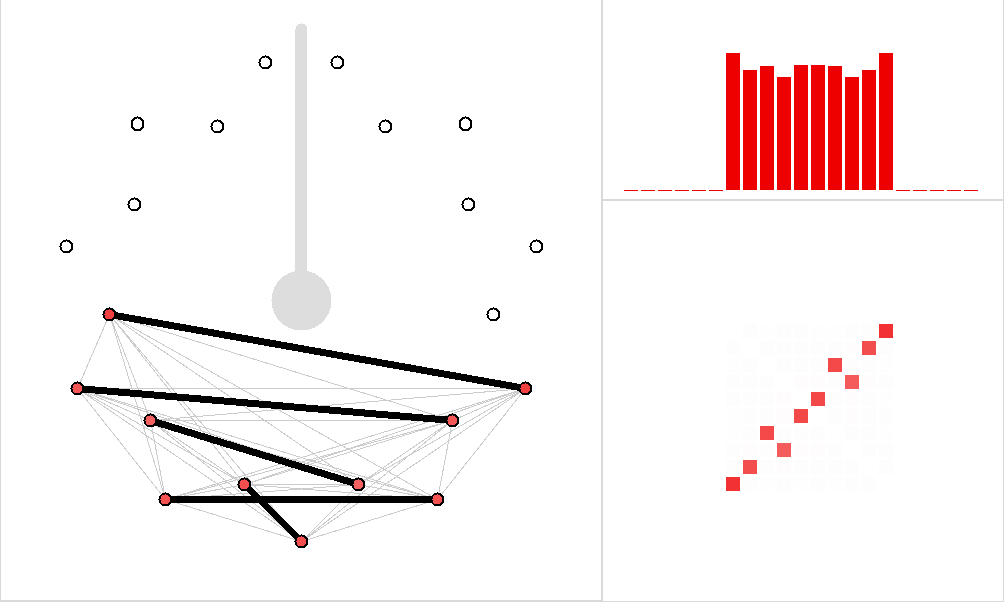}\\
\includegraphics[width=8cm]{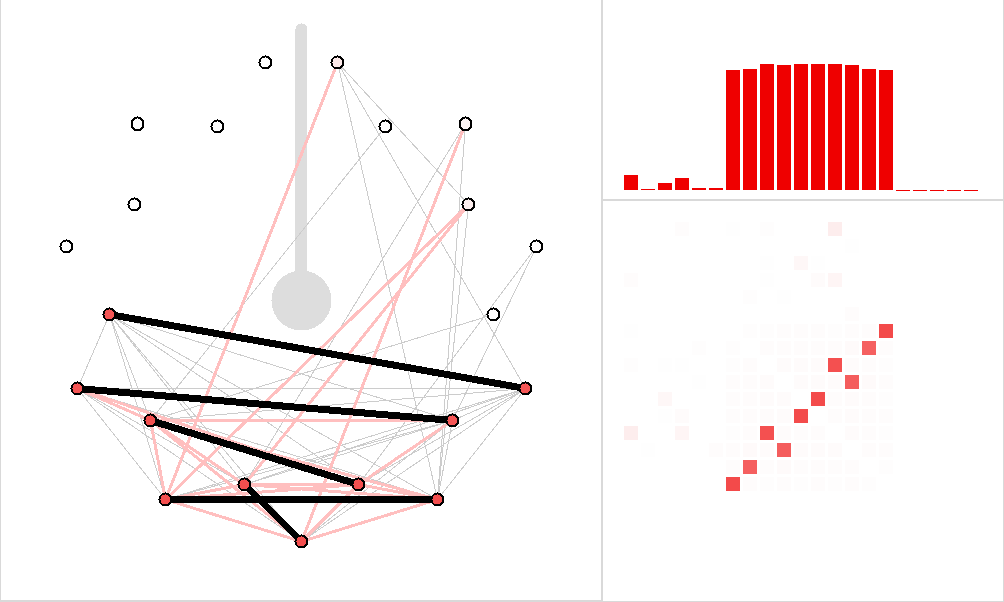} & $\;\;\;$ &\includegraphics[width=8cm]{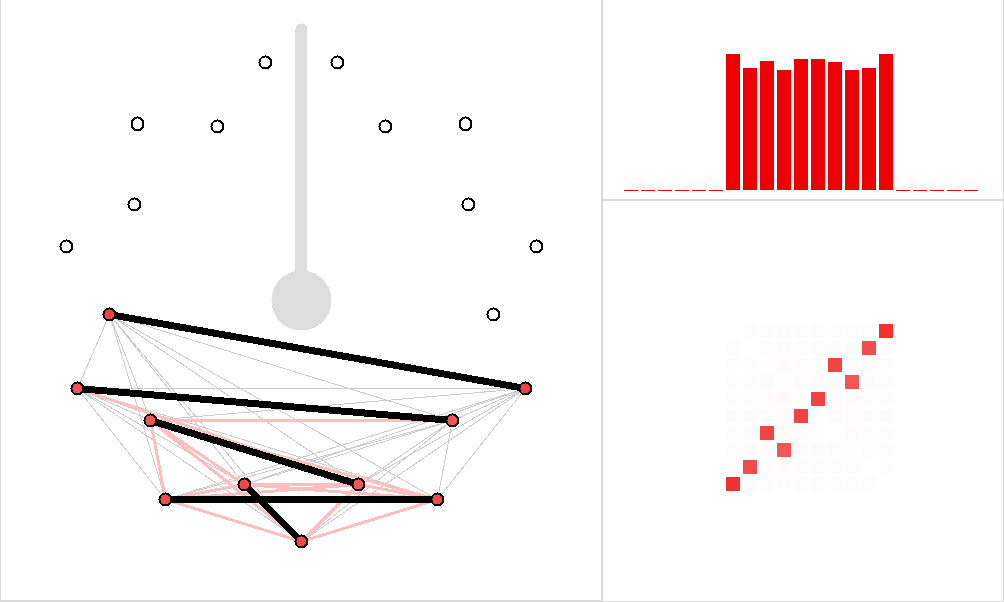}\\
\includegraphics[width=8cm]{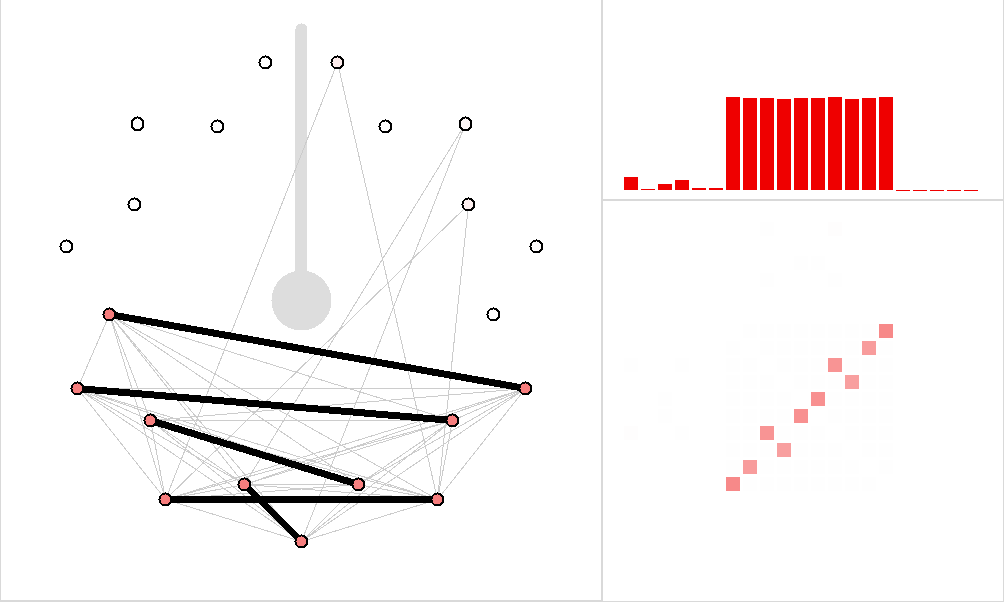} & $\;\;\;$ &\includegraphics[width=8cm]{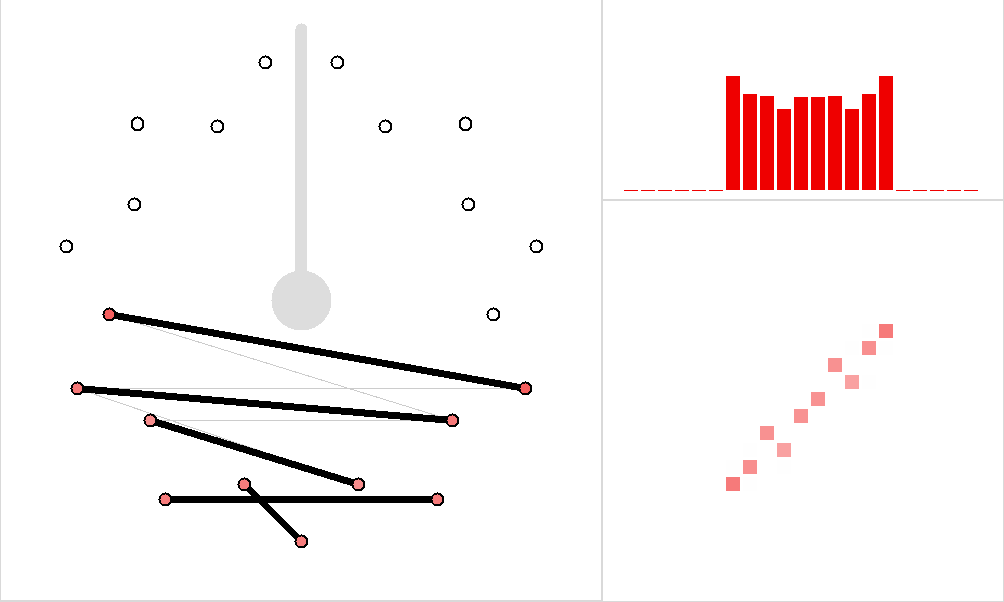}\\
\end{tabular}
\end{center}
\end{figure}

\begin{figure}[h]
\caption{
\label{fe2s2complexS3}
Correlation analysis for the $S=3$ state of the [Fe$_2$S$_2$(SCH$_3$)$_4$]$^{2-}$ complex, based on mutual information (left) and mutual correlation (right).
The bar graphs shows orbital entropies (left) or orbital sum of mutual information (right), while the color-coded size of the  mutual information (left) and mutual correlation (right) matrix elements is displayed below.
The weighted graphs combining the quantities are plotted as well.
The top row contains results for  $M_s=3$ , the second one $M_s=2$, third one $M_s=1$, and the bottom gives the spin-free results.
In the correlation graph, the orbitals in their ascending energy order are placed in the clockwise direction starting from the gray indicated ``noon''.
Since mutual correlation systematically attains smaller values than mutual information, the mutual correlations have been rescaled (see text for details).
}
\begin{center}
\begin{tabular}{ccc}
\includegraphics[width=8cm]{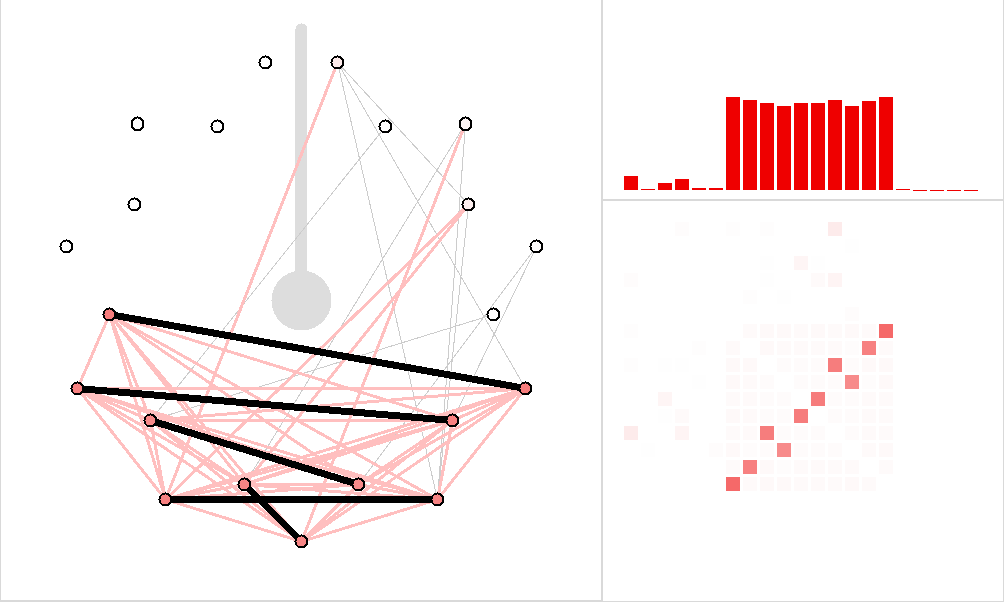} & $\;\;\;$ &\includegraphics[width=8cm]{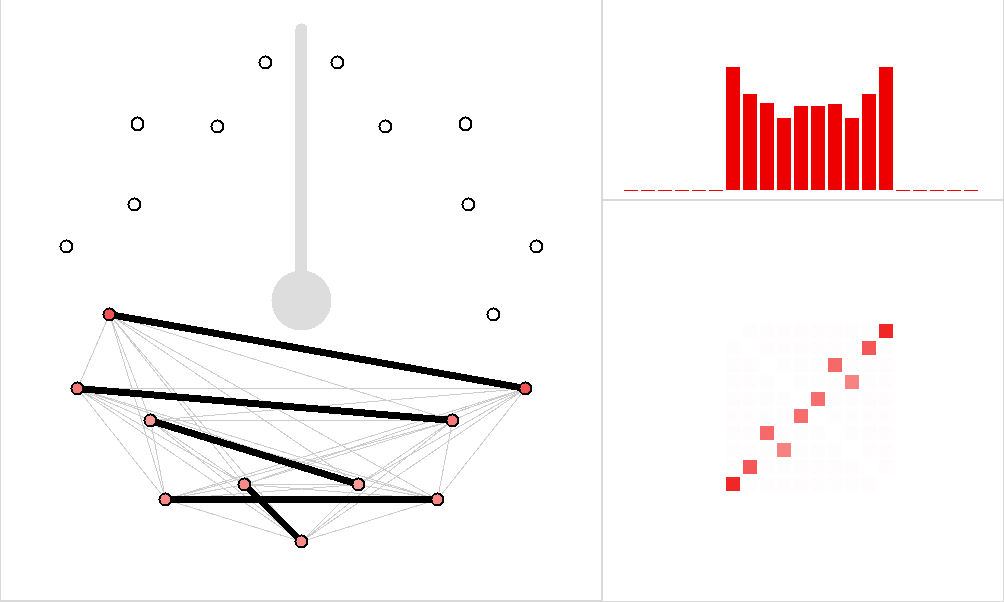}\\
\includegraphics[width=8cm]{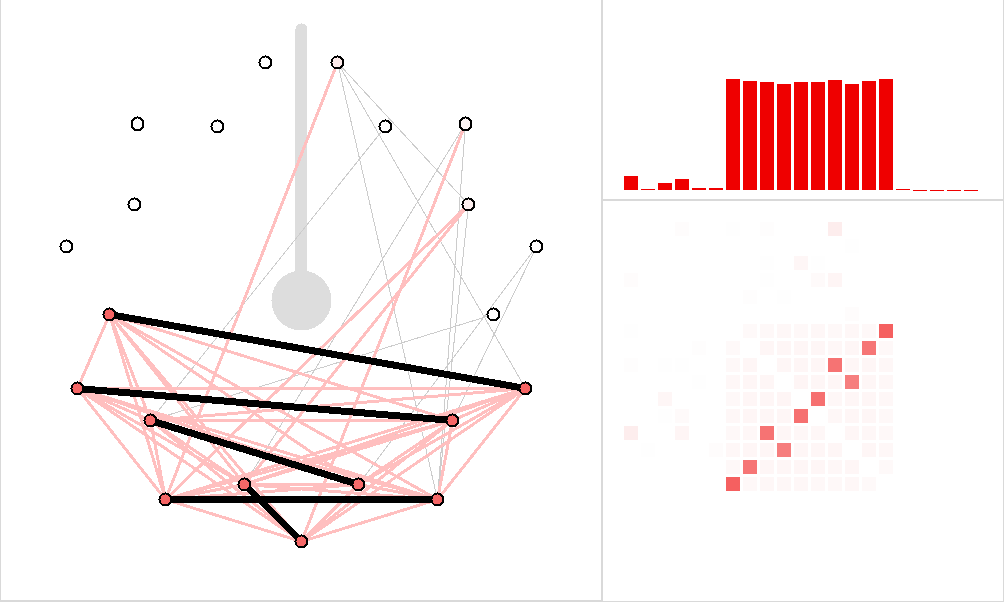} & $\;\;\;$ &\includegraphics[width=8cm]{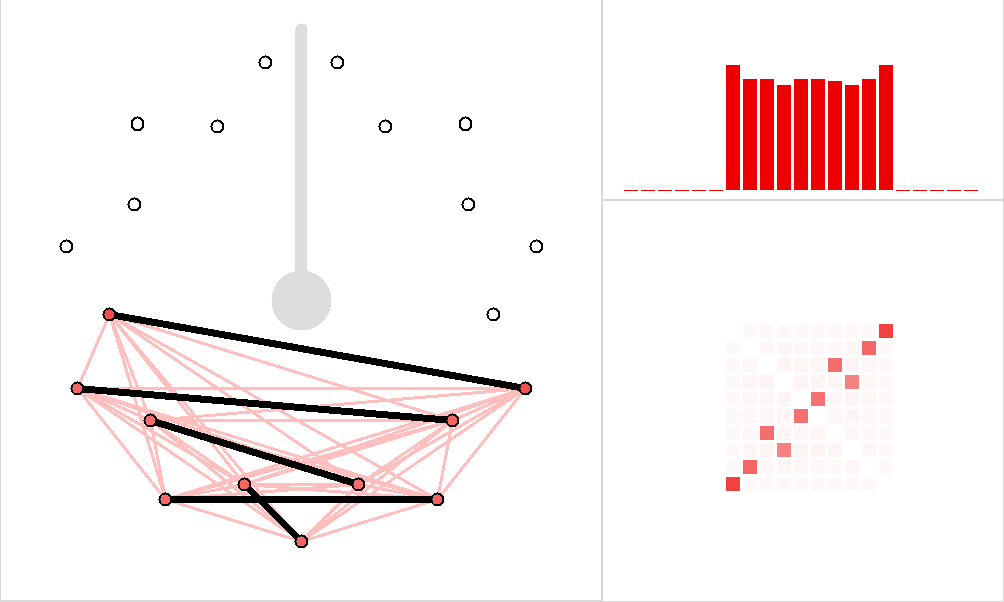}\\
\includegraphics[width=8cm]{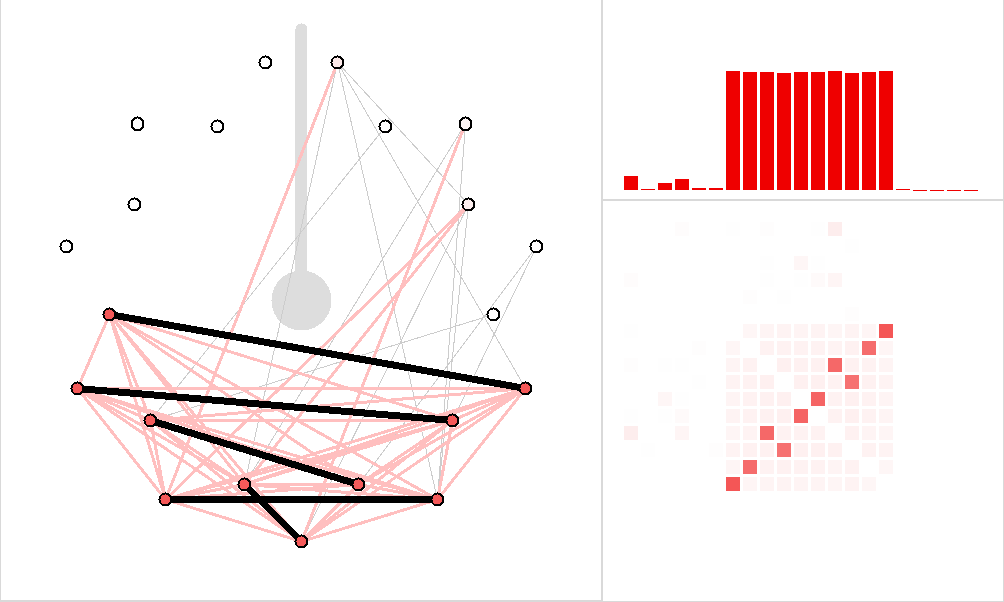} & $\;\;\;$ &\includegraphics[width=8cm]{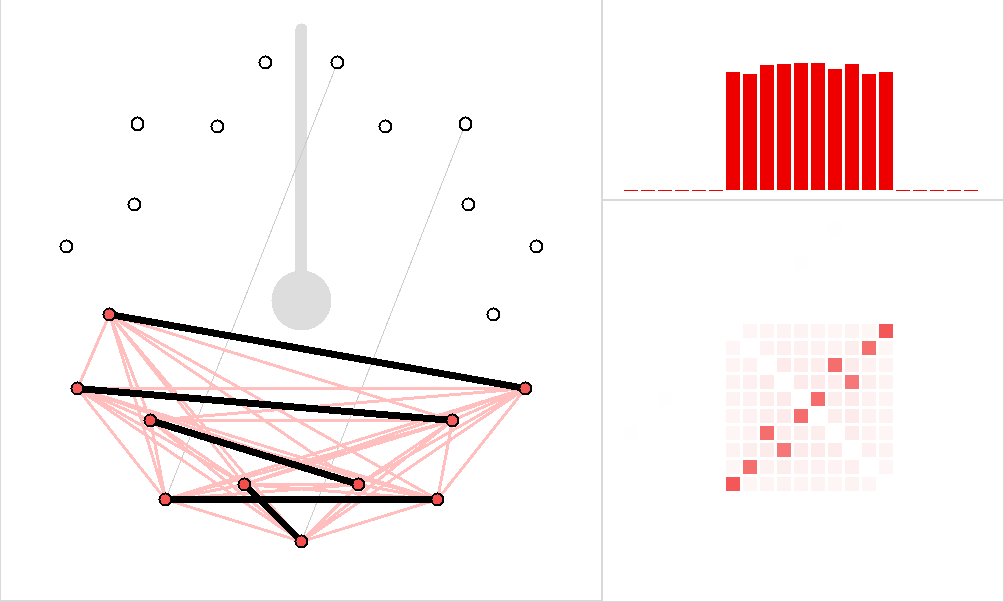}\\
\includegraphics[width=8cm]{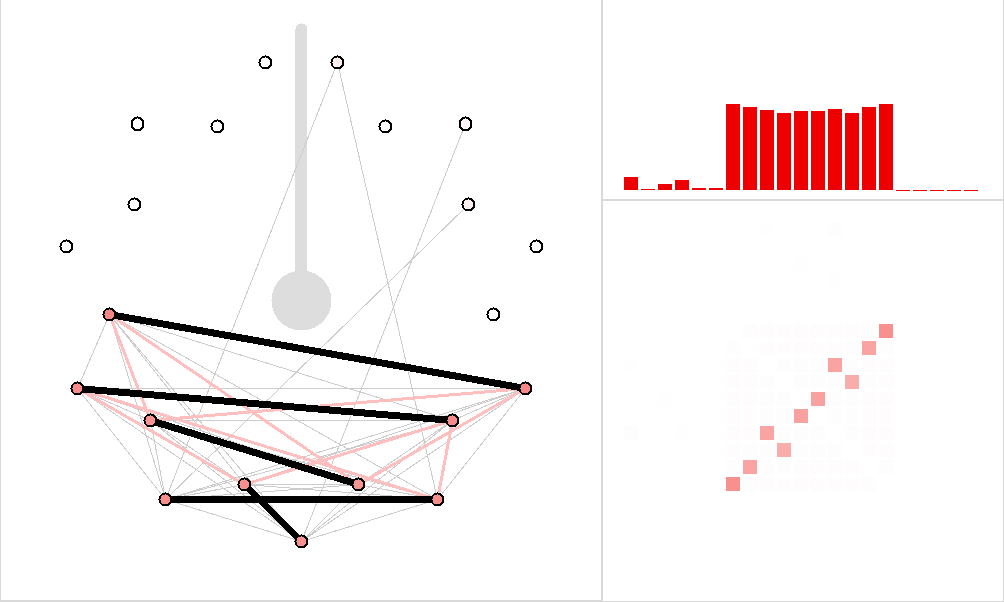} & $\;\;\;$ &\includegraphics[width=8cm]{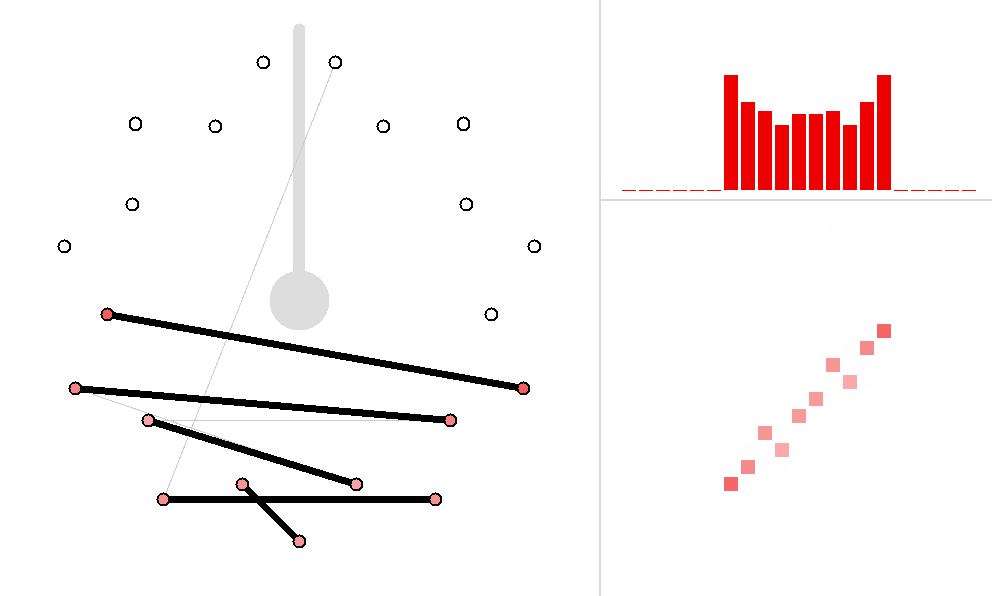}\\
\end{tabular}
\end{center}
\end{figure}

\begin{figure}[h]
\caption{
\label{fe2s2complexS4}
Correlation analysis for the $S=4$ state of the [Fe$_2$S$_2$(SCH$_3$)$_4$]$^{2-}$ complex, based on mutual information (left) and mutual correlation (right).
The bar graphs shows orbital entropies (left) or orbital sum of mutual information (right), while the color-coded size of the  mutual information (left) and mutual correlation (right) matrix elements is displayed below.
The weighted graphs combining the quantities are plotted as well.
The top row contains results for  $M_s=4$ , the second one $M_s=3$, third one $M_s=2$, fourth one $M_s=1$, and the bottom gives the spin-free results.
In the correlation graph, the orbitals in their ascending energy order are placed in the clockwise direction starting from the gray indicated ``noon''.
Since mutual correlation systematically attains smaller values than mutual information, the mutual correlations have been rescaled (see text for details).
}
\begin{center}
\begin{tabular}{ccc}
\includegraphics[width=7cm]{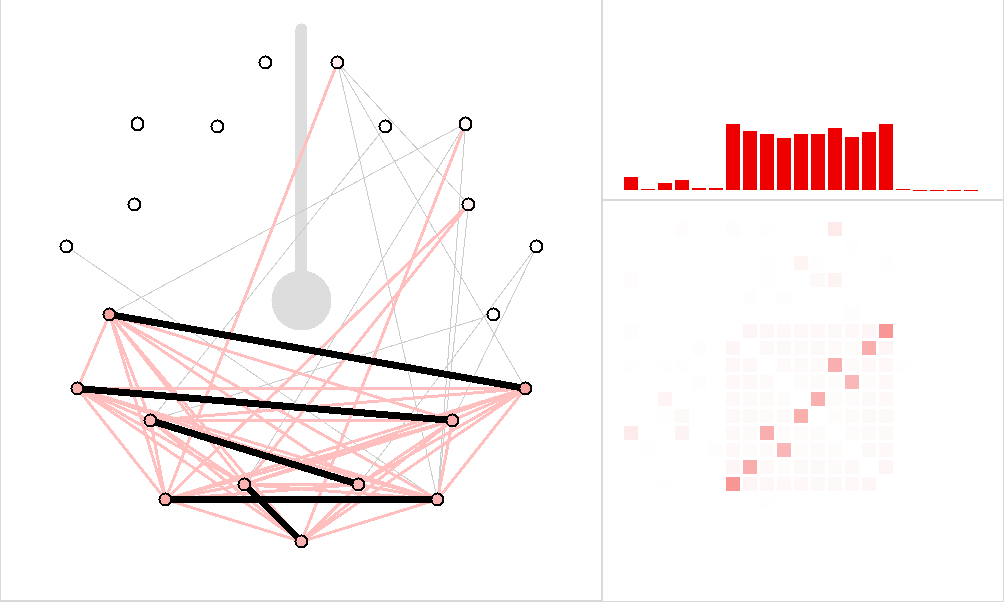} & $\;\;\;$ &\includegraphics[width=7cm]{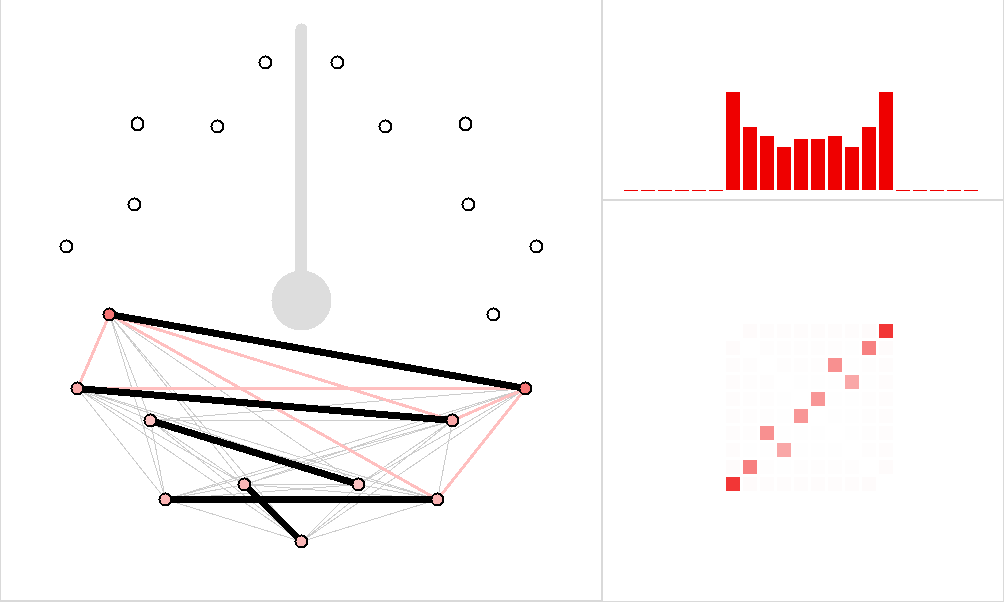}\\
\includegraphics[width=7cm]{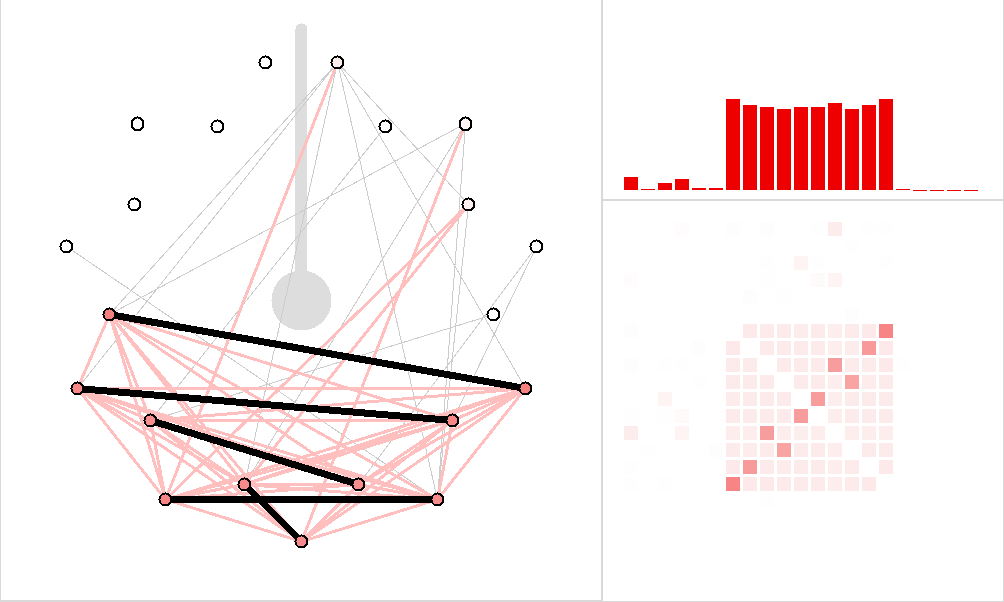} & $\;\;\;$ &\includegraphics[width=7cm]{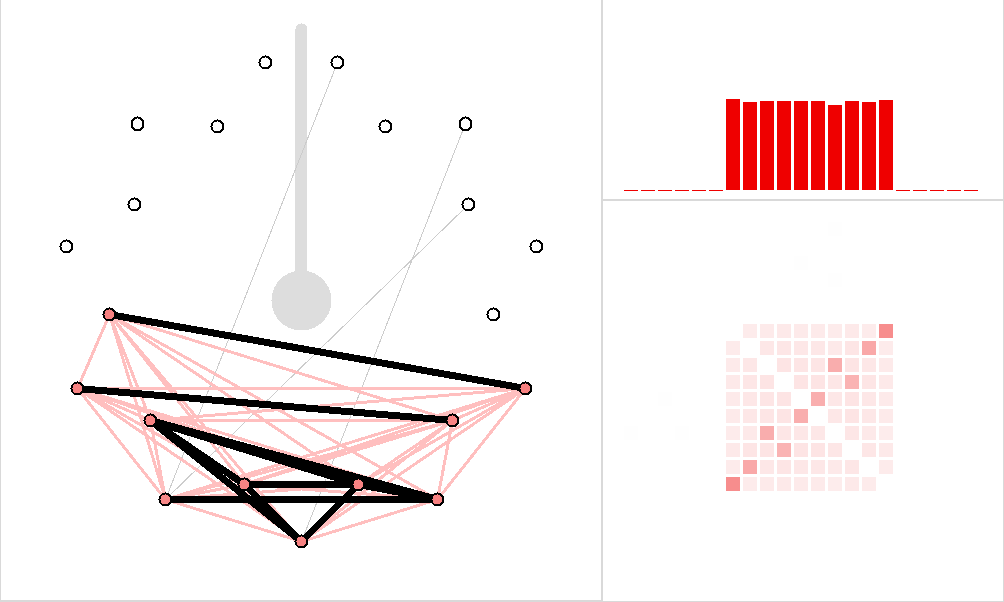}\\
\includegraphics[width=7cm]{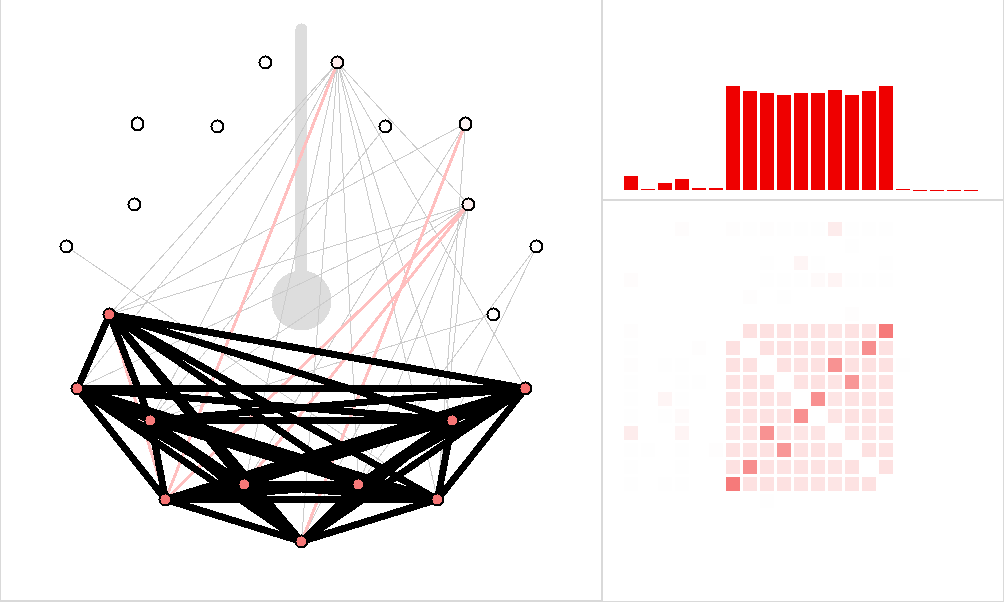} & $\;\;\;$ &\includegraphics[width=7cm]{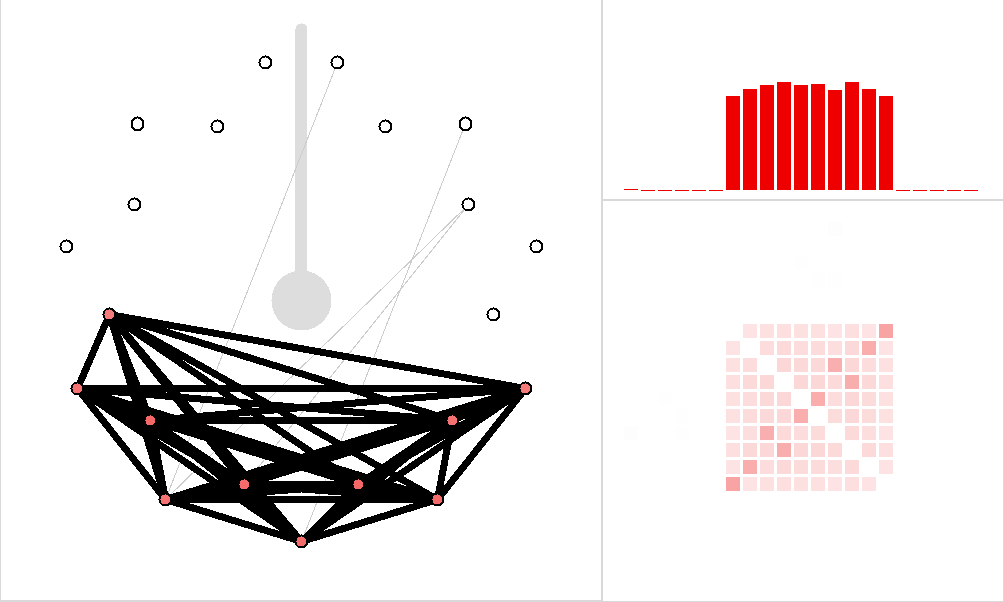}\\
\includegraphics[width=7cm]{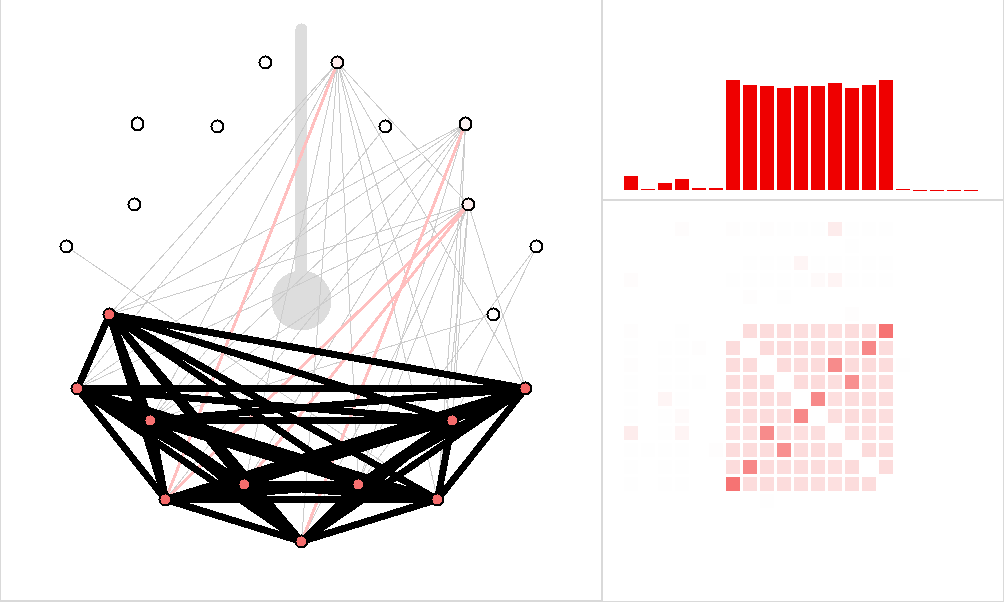} & $\;\;\;$ &\includegraphics[width=7cm]{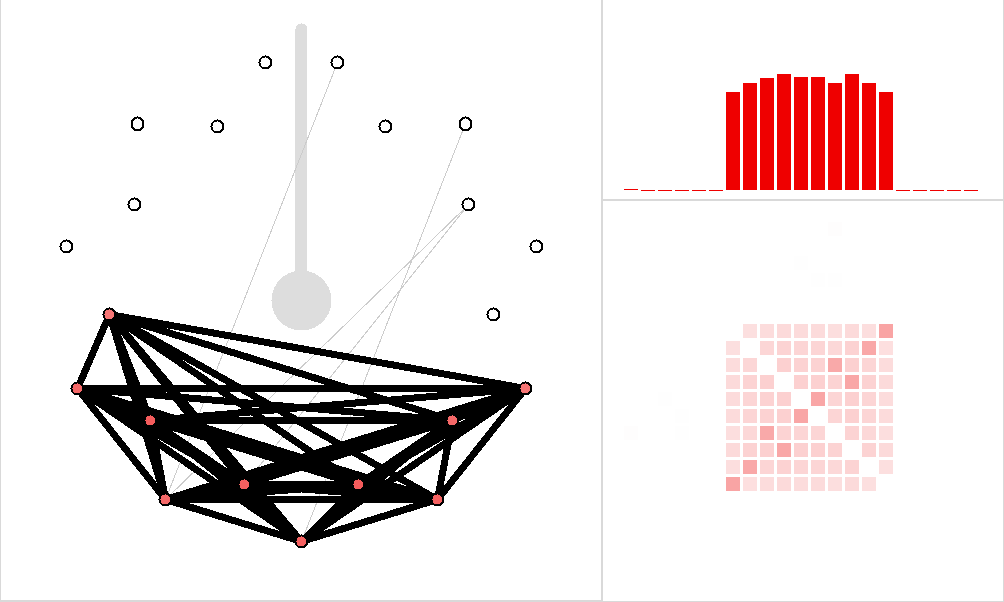}\\
\includegraphics[width=7cm]{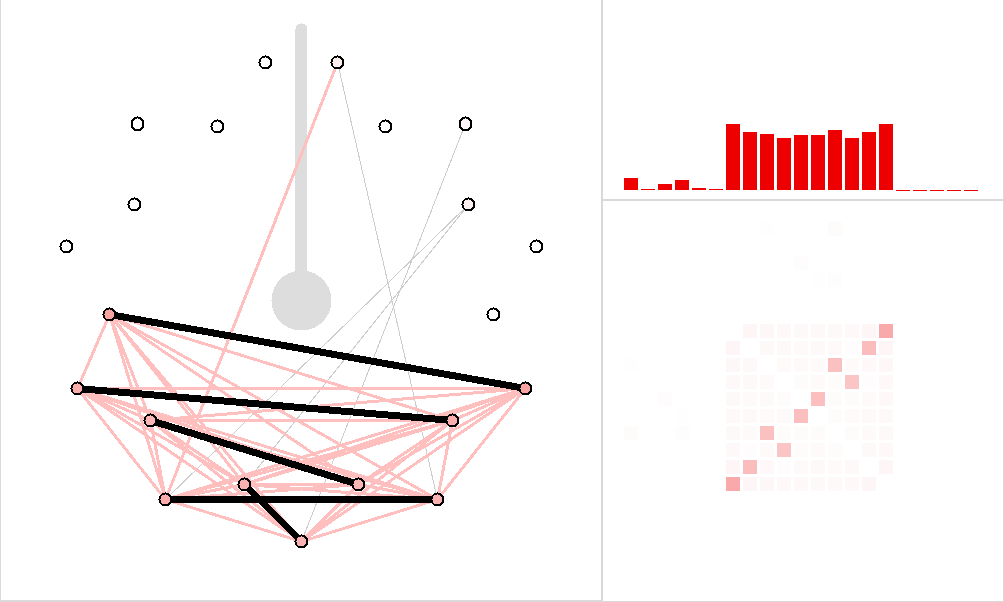} & $\;\;\;$ &\includegraphics[width=7cm]{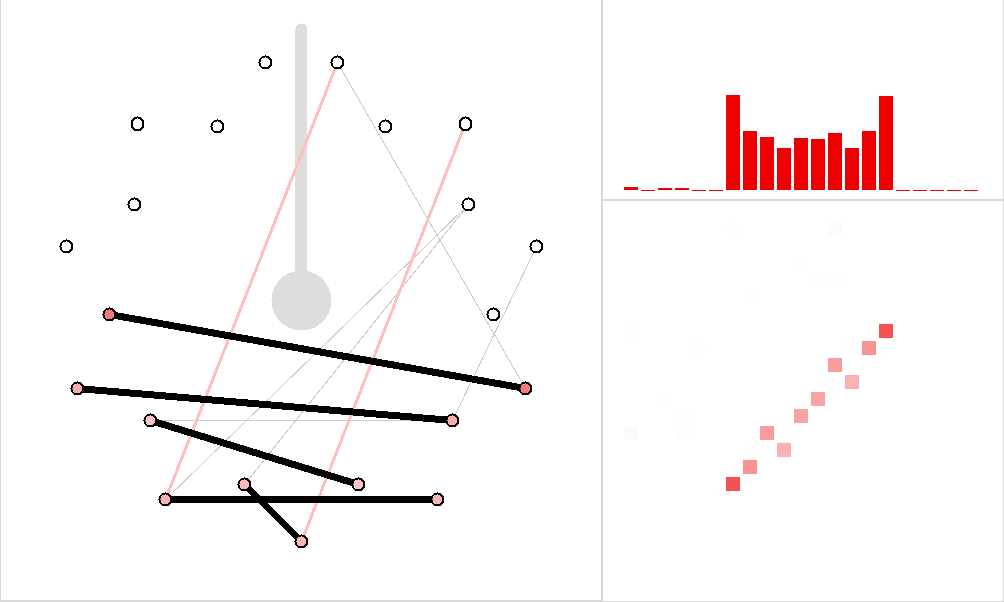}\\
\end{tabular}
\end{center}
\end{figure}

\begin{figure}[h]
\caption{
\label{fe2s2complexS5}
Correlation analysis for the $S=5$ state of the [Fe$_2$S$_2$(SCH$_3$)$_4$]$^{2-}$ complex, based on mutual information (left) and mutual correlation (right).
The bar graphs shows orbital entropies (left) or orbital sum of mutual information (right), while the color-coded size of the  mutual information (left) and mutual correlation (right) matrix elements is displayed below.
The weighted graphs combining the quantities are plotted as well.
The top row contains results for  $M_s=5$ , the second one $M_s=4$, third one $M_s=3$, fourth one $M_s=2$, and the bottom gives the spin-free results.
In the correlation graph, the orbitals in their ascending energy order are placed in the clockwise direction starting from the gray indicated ``noon''.
Since mutual correlation systematically attains smaller values than mutual information, the mutual correlations have been rescaled (see text for details).
}
\begin{center}
\begin{tabular}{ccc}
\includegraphics[width=7cm]{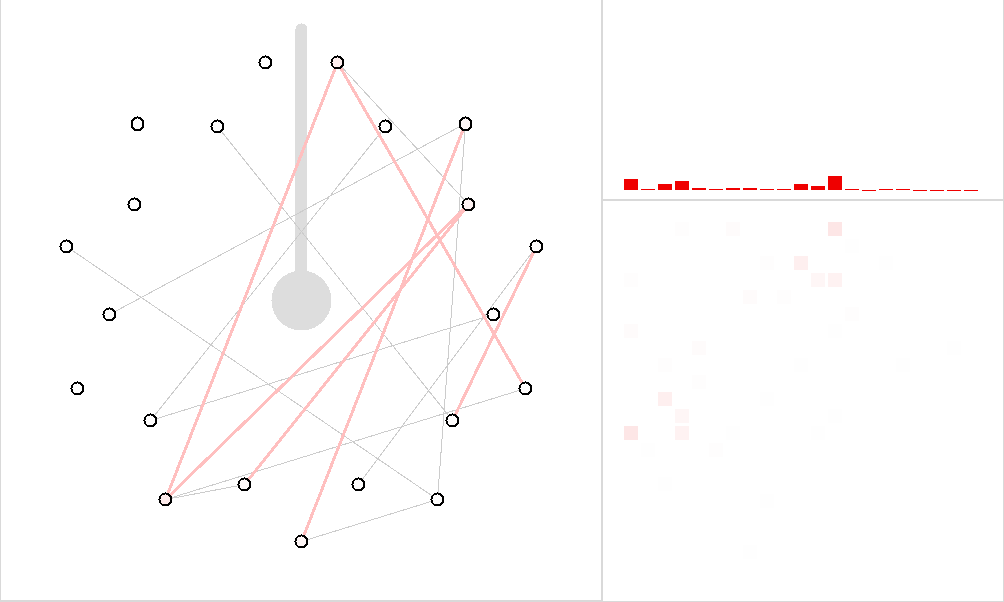} & $\;\;\;$ &\includegraphics[width=7cm]{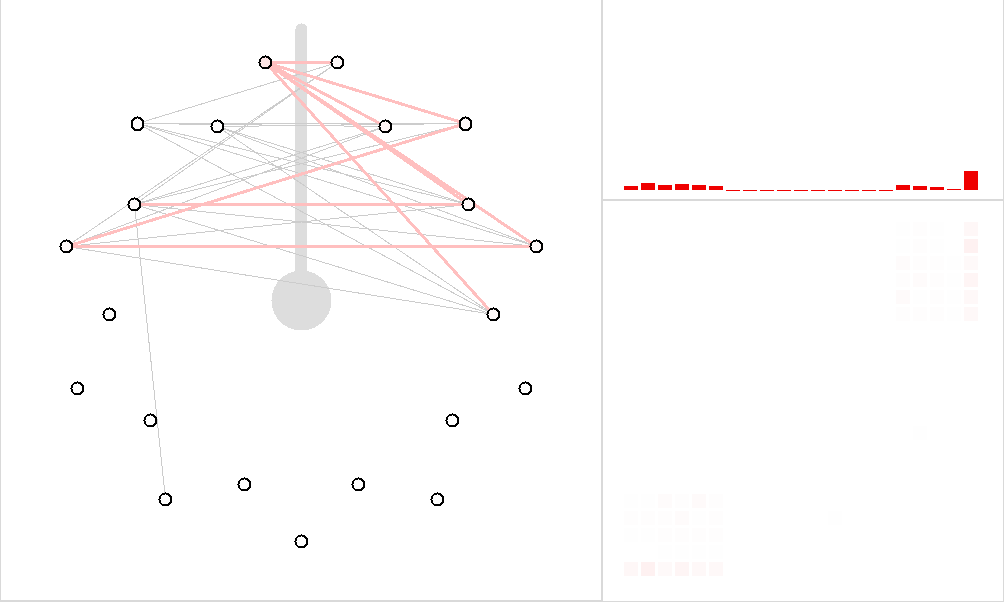}\\
\includegraphics[width=7cm]{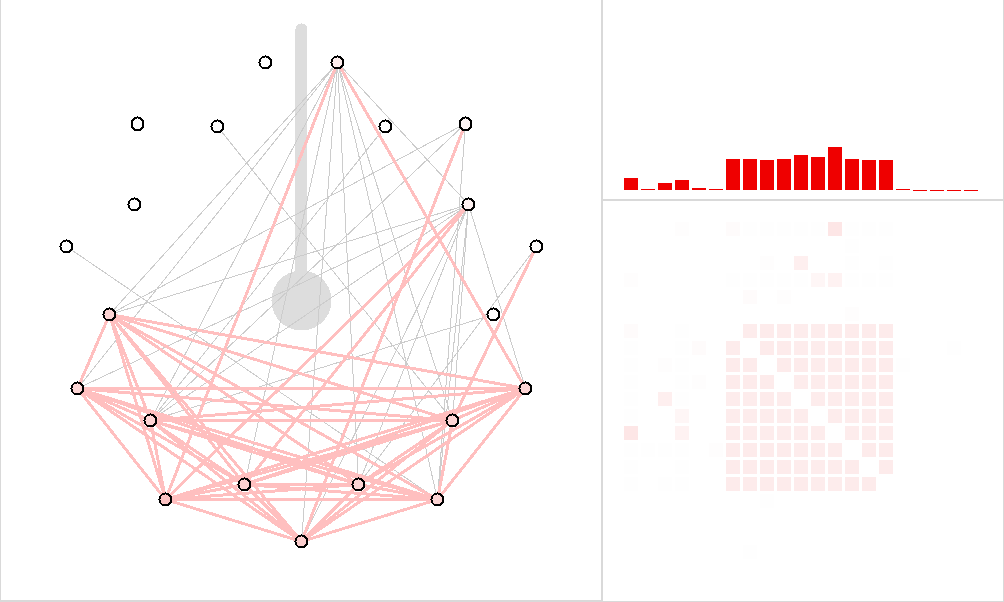} & $\;\;\;$ &\includegraphics[width=7cm]{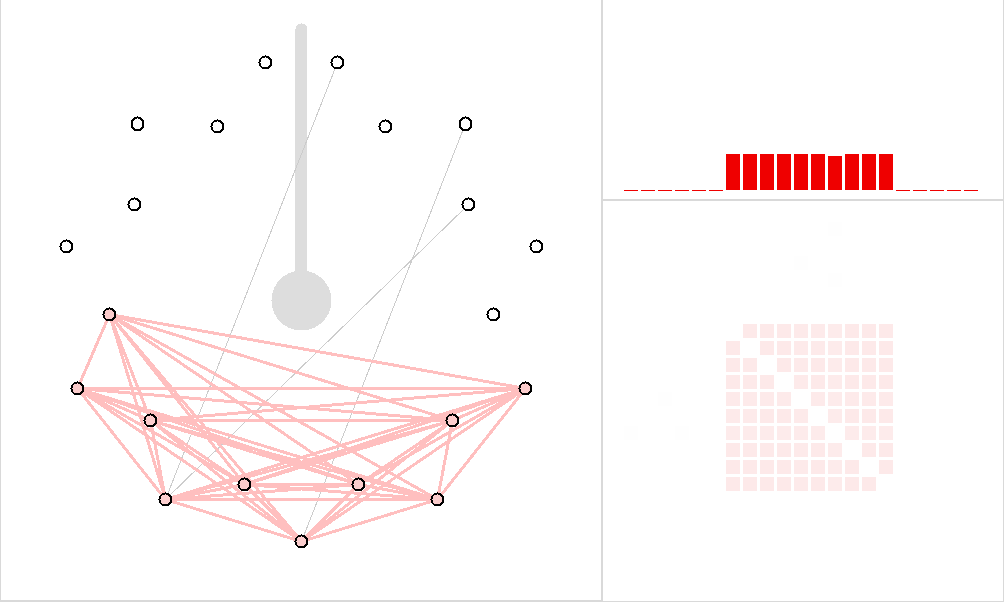}\\
\includegraphics[width=7cm]{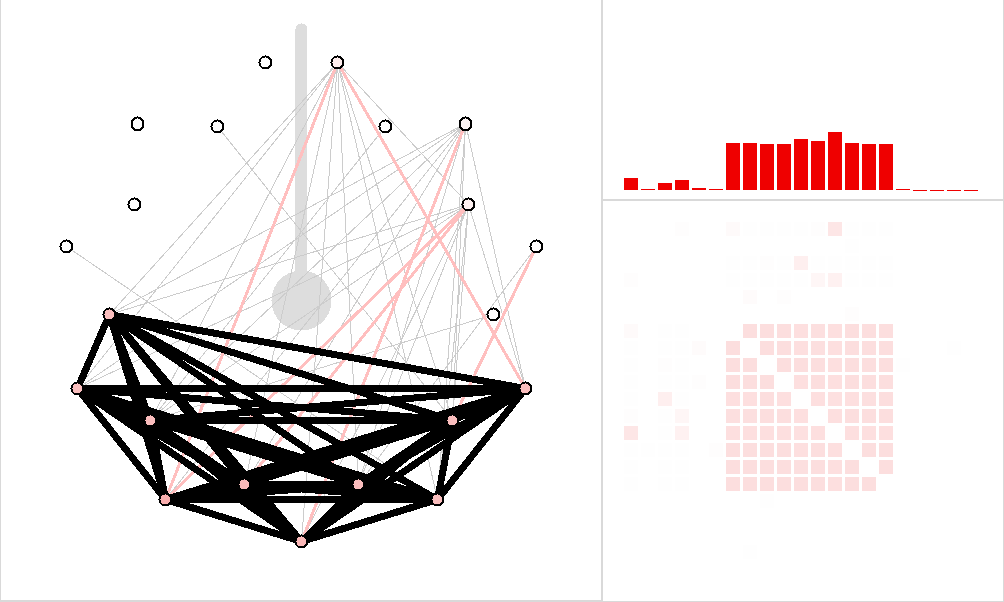} & $\;\;\;$ &\includegraphics[width=7cm]{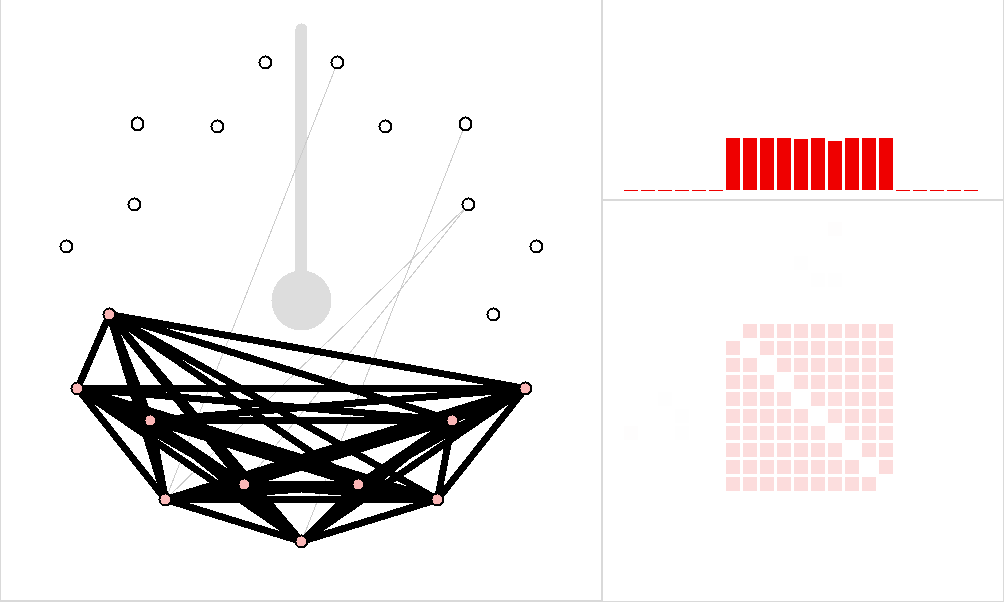}\\
\includegraphics[width=7cm]{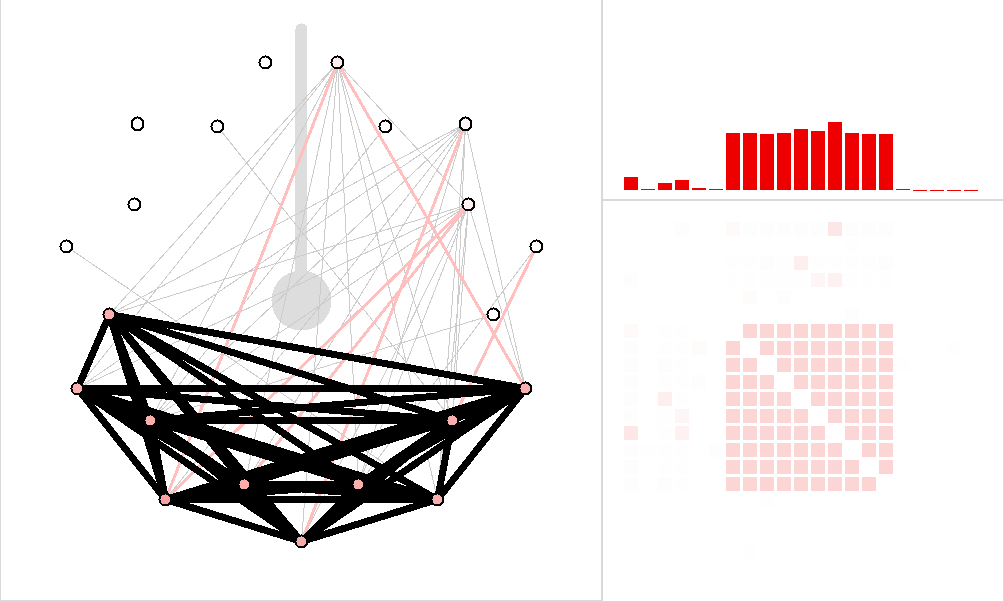} & $\;\;\;$ &\includegraphics[width=7cm]{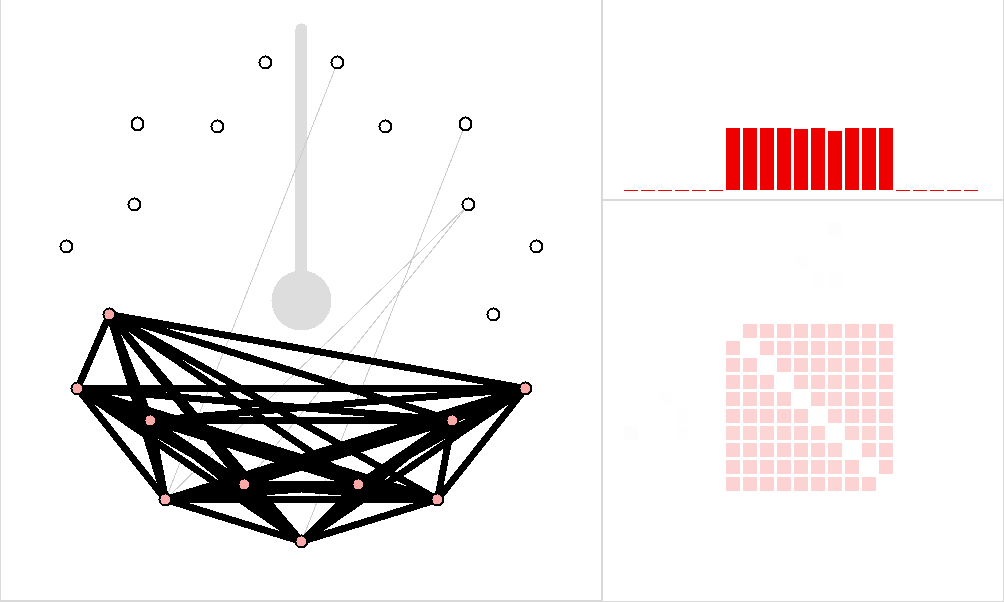}\\
\includegraphics[width=7cm]{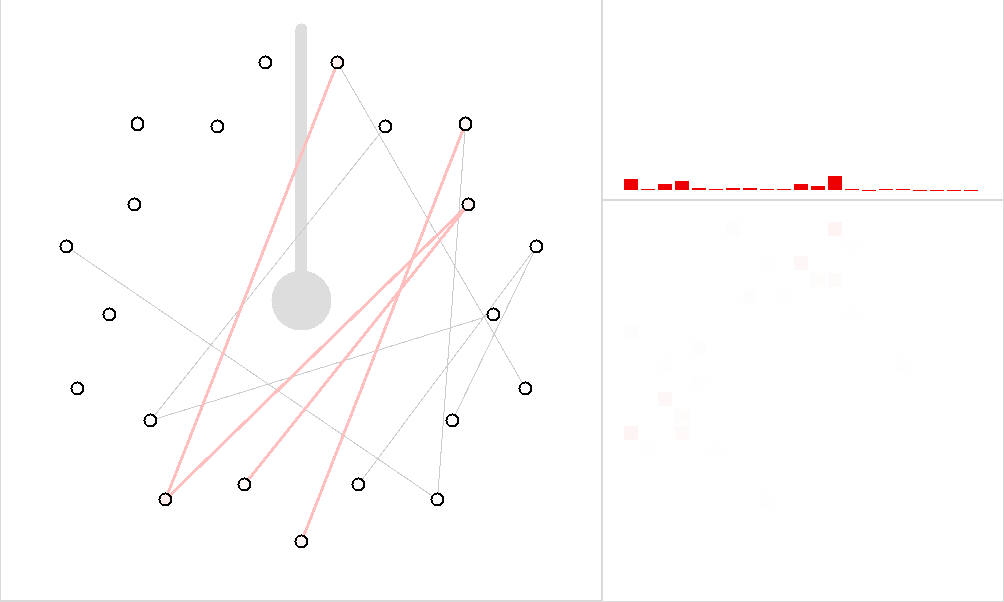} & $\;\;\;$ &\includegraphics[width=7cm]{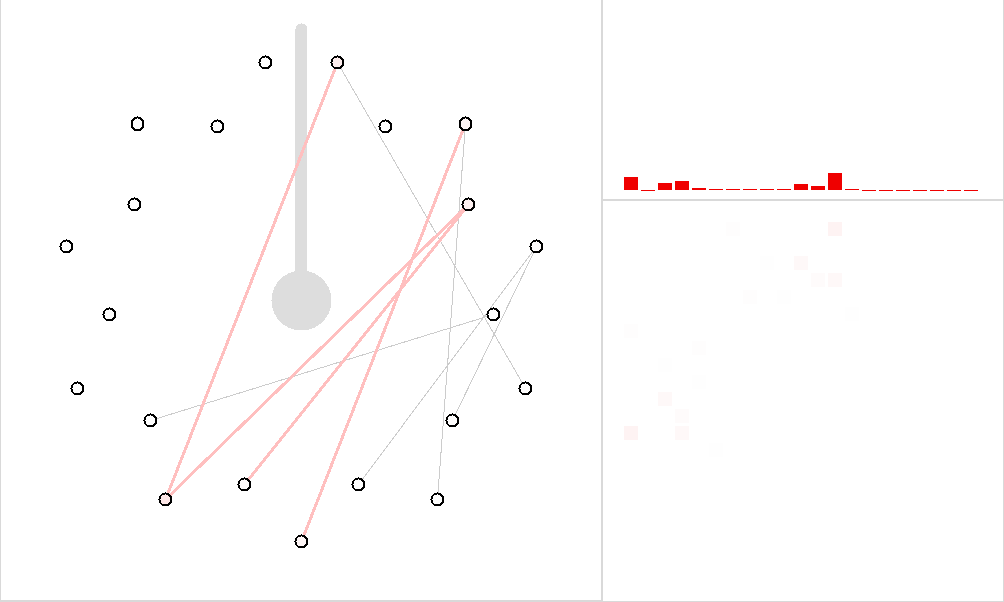}\\
\end{tabular}
\end{center}
\end{figure}

Fig.~\ref{fe2s2complexS0} show the correlation analysis by mutual information (left) and mutual correlation (right) for the ground singlet state.
An excellent qualitative agreement can be observed, although mutual information diagram shows more orbital pair correlations of medium strength the mutual correlation. The simplification of the pattern when transitioning to the spin-free quantities (bottom row) occurs similarly in both cases.
We performed the analogous analysis for the remaining five states $S=1$ (Fig.~\ref{fe2s2complexS1}),  $S=2$ (Fig.~\ref{fe2s2complexS2}), $S=3$ (Fig.~\ref{fe2s2complexS3}),$S=4$ (Fig.~\ref{fe2s2complexS4}), and $S=5$ (Fig.~\ref{fe2s2complexS5}).
For $S=1,2,3$, the situation is very similar to the ground state, while it is possible to see a slight increase of the orbital pair correlations with decreasing $M_s$.
For $S=4,5$, this increase becomes quite dramatic for $M_s=2,1$, while the spin-free analysis, which separates strong correlation from the spin couplings,
simplifies the picture dramatically, for $S=5$ confirming qualitatively single configurational nature of the wave function, as expected.
In all cases, the mutual information and mutual correlation yield qualitatively the same picture and are thus interchangeable for this purpose.

\begin{figure}[h]
\caption{
\label{fe2s2complexscatter1}
Scatter plots correlating orbital entropy, Von Neumann orbital from from 1RDM, and orbital sum of mutual correlation.
The quantities on the y axis were renormalized to match the norm of the orbital entropy vector (over all orbitals).
Colors code the state: $S=0$ red, $S=1$ blue, $S=2$ violet, $S=3$ yellow, $S=4$ pink, $S=5$ gray.
}
\begin{center}
\begin{tabular}{ccccc}
\includegraphics[width=5cm]{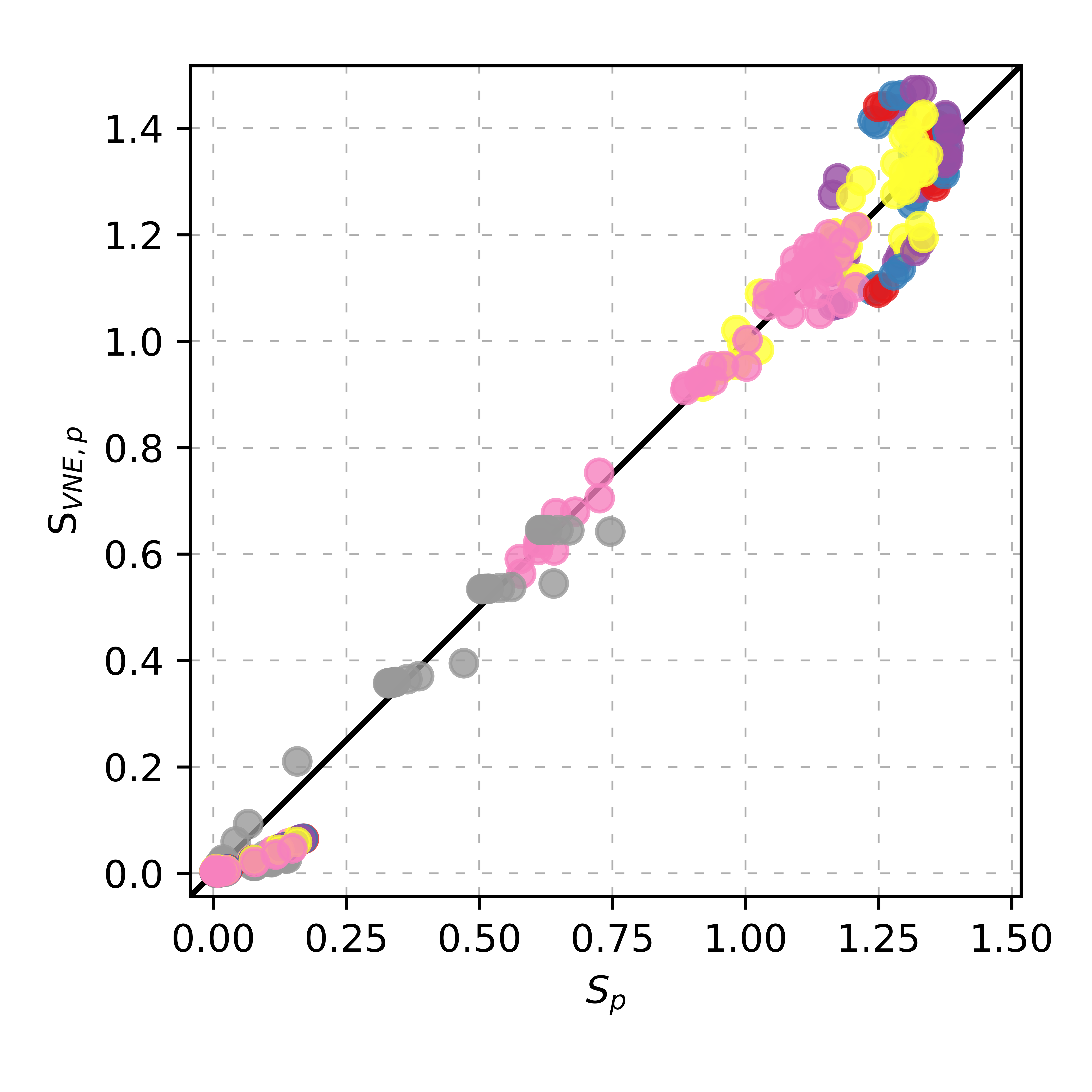} & $\;\;\;$ &\includegraphics[width=5cm]{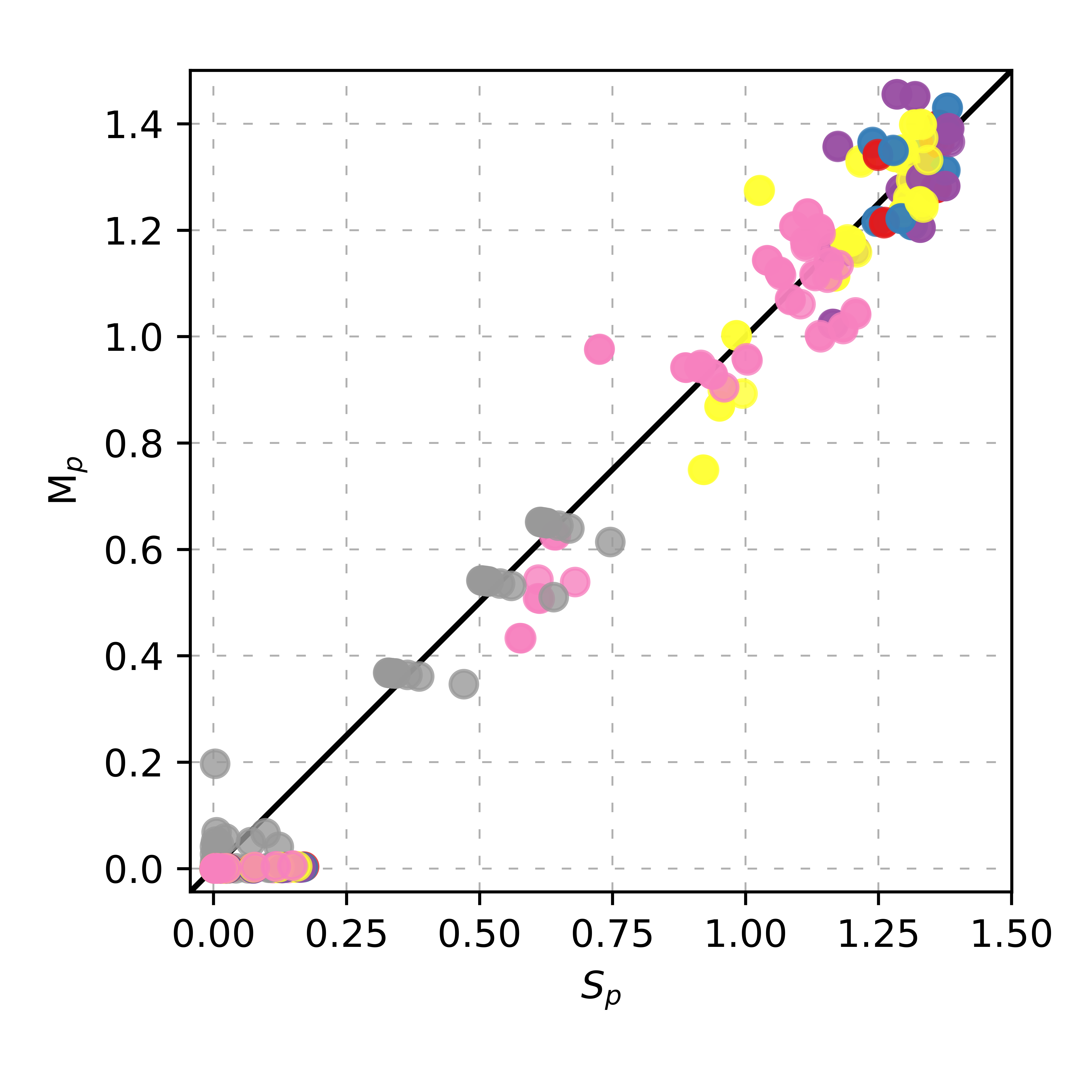} & $\;\;\;$ &\includegraphics[width=5cm]{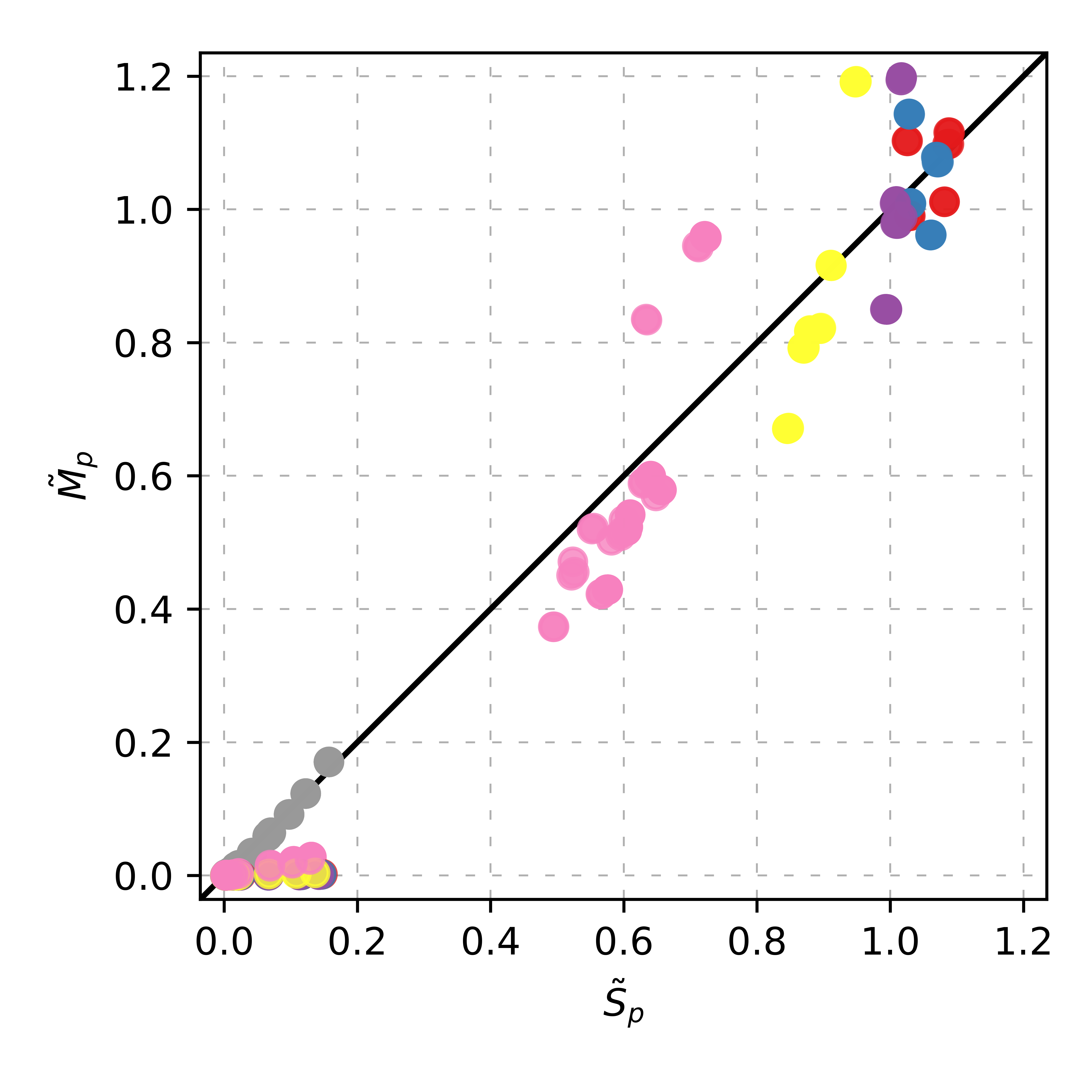}\\
\end{tabular}
\end{center}
\end{figure}

\begin{figure}[h]
\caption{
\label{fe2s2complexscatter2}
Scatter plots correlating mutual information and  mutual correlation.
The quantities on the y axis were renormalized to match the norm of the mutual information vector (over all pairs orbitals).
Colors code the state: $S=0$ red, $S=1$ blue, $S=2$ violet, $S=3$ yellow, $S=4$ pink, $S=5$ gray.
}
\begin{center}
\begin{tabular}{ccc}
\includegraphics[width=5cm]{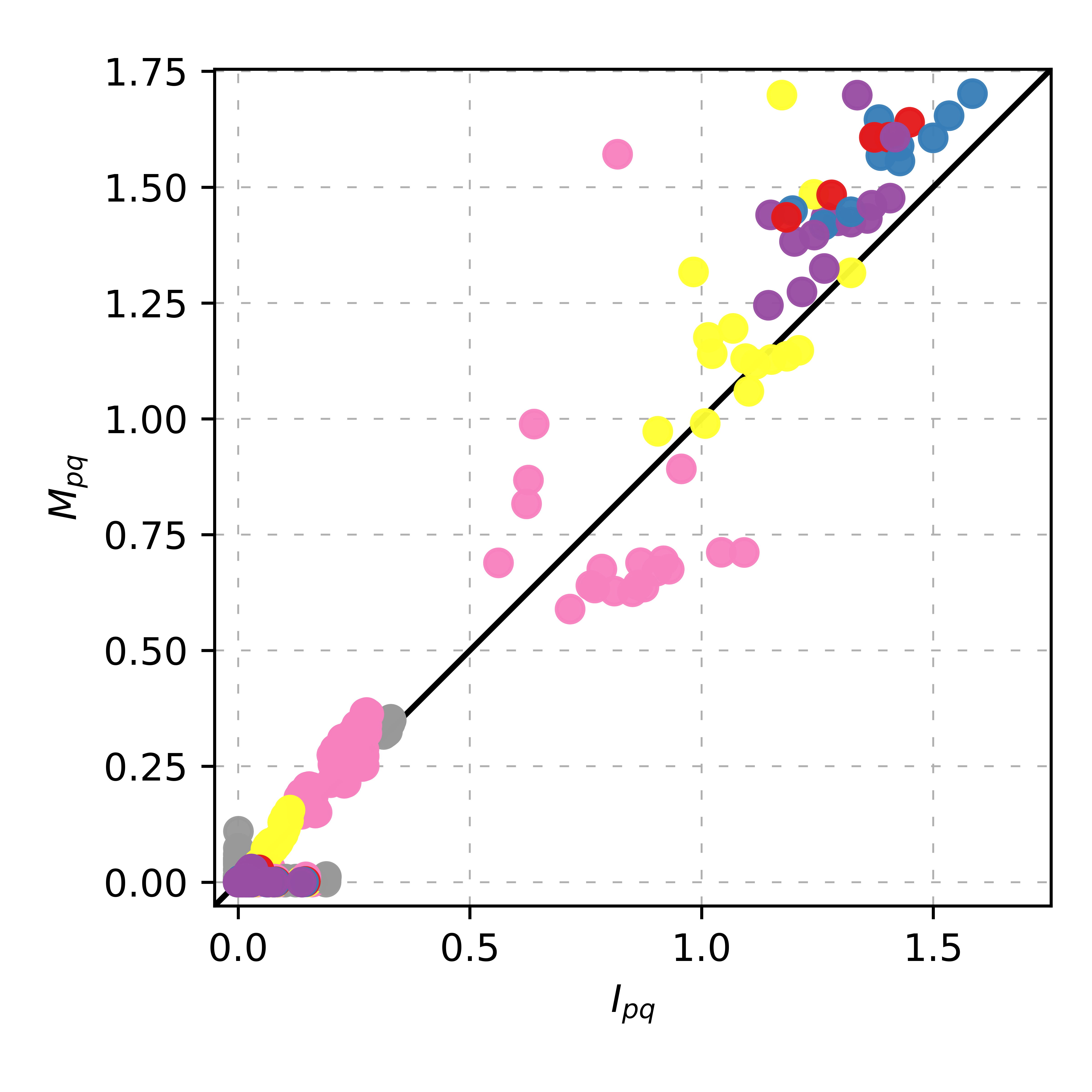} & $\;\;\;$ &\includegraphics[width=5cm]{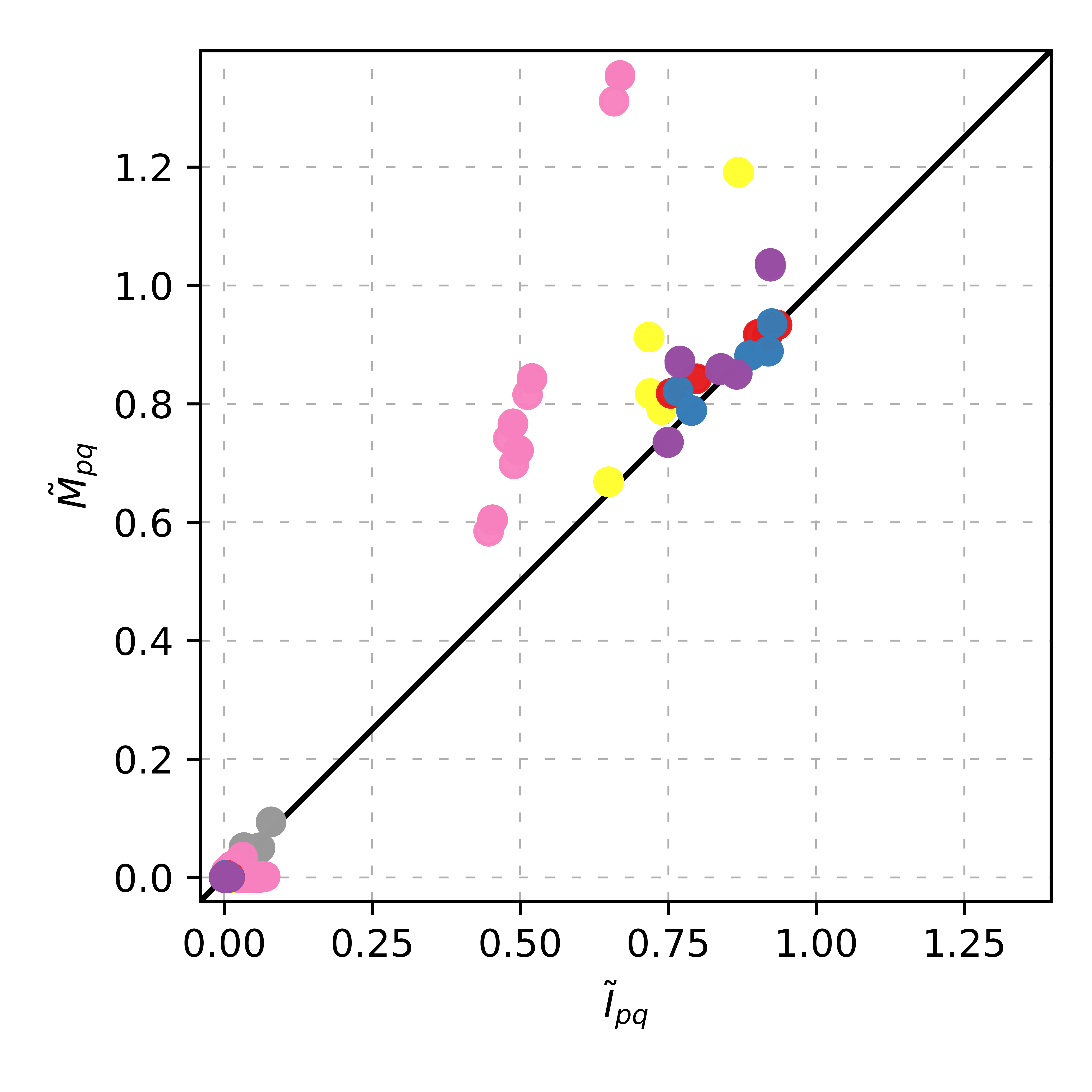}\\
\end{tabular}
\end{center}
\end{figure}

Finally, in Fig.~\ref{fe2s2complexscatter1} we present scatter plots correlating
the orbital entropy with orbital contributions to the the 1RDM-based Von Neumann  entropy (left), orbital entropy with the partial sums of mutual correlation (middle),
and the spin-free version thereof (right), 
where individual dots correspond to values for each orbital, color-coded by the $S$ quantum number of the state.
Similarly to the single-iron complex,
the correlation between all the quantities is very good, but this time it is best for the $S=5$ highest state (which has no strong correlation),
not for the ground state.  
In other words, the correlation improves with the increasing $S$ value for both complexes.
The scatter plot for the correlation of mutual information and mutual correlation is shown in Fig.~\ref{fe2s2complexscatter2}.
Similarly to the single-iron complex, the correlation for the orbital-pair quantities is not as perfect as for the single-orbitals quantities.
The largest deviations can be observed for the $S=4$ state, in agreement with the slight discrepancy between the left and right panels in the
second row of Fig.~\ref{fe2s2complexS4}.
The ground state,  first excited ($S=1$), and the $S=5$ state exhibit excellent correlation, almost perfect one in the spin-free case.

\section{Conclusions}

In analogy with spin-free orbital entropy and mutual information, which we introduced previously \cite{myspinfree2025},
we have defined a spin-free analogue of the mutual correlation, which is invariant with respect to the spin projection quantum number
for a give spin multiplet. 
Although for the spin-free mutual correlation it cannot be proven to be unsharply smaller than the original mutual correlation (which is the case for mutual information),
the correlation analysis becomes simplified in the spin-free picture for the mutual correlation as well.
The spin-free analysis is thus able to filter out the `static' correlation due to spin couplings from the `genuine' strong correlation.
We studied two realistic examples of complexes with iron-sulfur bonds which exhibit strong correlation and by comparison of the spin-free
and spin-including correlation measures we were able to identify to which extent the correlation is due to spin couplings.
We have compared the analysis based on mutual information and on mutual correlation and have seen that both yield qualitatively the same
results, making both quantities well interchangeable for this purpose.
This has been also confirmed by scatter plots which show a high degree of correlation between the two quantities.
Since the computation of the spin-free  correlation analysis, both for mutual information and for mutual correlation, does not require any significant additional computational cost compared to the spin-dependent ones,
we think that it can be routinely used as a useful supplement to the analysis using the ``traditional'' spin-dependent orbital entropy and mutual information/correlation.

\section*{Acknowledgments}

The work has been supported by the Advanced Multiscale Materials for Key Enabling Technologies project of the Ministry of Education, Youth, and Sports of the Czech Republic. Project No. CZ.02.01.01/00/22\_008/0004558, Co-funded by the European Union.
The computational time was supported by the Ministry of Education, Youth and Sports of the Czech Republic through the e-INFRA CZ project (ID:90254).

\bibliographystyle{achemso}
\bibliography{entropy,ors,cc,perspective}

\providecommand{\latin}[1]{#1}
\makeatletter
\providecommand{\doi}
  {\begingroup\let\do\@makeother\dospecials
  \catcode`\{=1 \catcode`\}=2 \doi@aux}
\providecommand{\doi@aux}[1]{\endgroup\texttt{#1}}
\makeatother
\providecommand*\mcitethebibliography{\thebibliography}
\csname @ifundefined\endcsname{endmcitethebibliography}
  {\let\endmcitethebibliography\endthebibliography}{}
\begin{mcitethebibliography}{69}
\providecommand*\natexlab[1]{#1}
\providecommand*\mciteSetBstSublistMode[1]{}
\providecommand*\mciteSetBstMaxWidthForm[2]{}
\providecommand*\mciteBstWouldAddEndPuncttrue
  {\def\EndOfBibitem{\unskip.}}
\providecommand*\mciteBstWouldAddEndPunctfalse
  {\let\EndOfBibitem\relax}
\providecommand*\mciteSetBstMidEndSepPunct[3]{}
\providecommand*\mciteSetBstSublistLabelBeginEnd[3]{}
\providecommand*\EndOfBibitem{}
\mciteSetBstSublistMode{f}
\mciteSetBstMaxWidthForm{subitem}{(\alph{mcitesubitemcount})}
\mciteSetBstSublistLabelBeginEnd
  {\mcitemaxwidthsubitemform\space}
  {\relax}
  {\relax}

\bibitem[Sinanoglu and tai Tuan(1963)Sinanoglu, and tai Tuan]{sinanoglu1963}
Sinanoglu,~O.; tai Tuan,~F. Many-Electron Theory of Atoms and Molecules. m.
  Effect of Correlation on Orbitals. \emph{J. Chem. Phys.} \textbf{1963},
  \emph{38}, 1740\relax
\mciteBstWouldAddEndPuncttrue
\mciteSetBstMidEndSepPunct{\mcitedefaultmidpunct}
{\mcitedefaultendpunct}{\mcitedefaultseppunct}\relax
\EndOfBibitem
\bibitem[Ziesche(1995)]{Ziesche-1995}
Ziesche,~P. Correlation strength and information entropy. \emph{Int. J. Quant.
  Chem.} \textbf{1995}, \emph{56}, 363--369\relax
\mciteBstWouldAddEndPuncttrue
\mciteSetBstMidEndSepPunct{\mcitedefaultmidpunct}
{\mcitedefaultendpunct}{\mcitedefaultseppunct}\relax
\EndOfBibitem
\bibitem[Nagy and Parr(1996)Nagy, and Parr]{Nagy-1996}
Nagy,~A.; Parr,~R.~G. Information entropy as a measure of the quality of an
  approximate electronic wave function. \emph{Int. J. Quant. Chem.}
  \textbf{1996}, \emph{58}, 323--327\relax
\mciteBstWouldAddEndPuncttrue
\mciteSetBstMidEndSepPunct{\mcitedefaultmidpunct}
{\mcitedefaultendpunct}{\mcitedefaultseppunct}\relax
\EndOfBibitem
\bibitem[Nalewajski(2000)]{Nalewajski-2000}
Nalewajski,~R.~F. Entropic Measures of Bond Multiplicity from the Information
  Theory. \emph{J. Phys. Chem. A} \textbf{2000}, \emph{104}, 11940--11951\relax
\mciteBstWouldAddEndPuncttrue
\mciteSetBstMidEndSepPunct{\mcitedefaultmidpunct}
{\mcitedefaultendpunct}{\mcitedefaultseppunct}\relax
\EndOfBibitem
\bibitem[Szalay \latin{et~al.}(2021)Szalay, Zimbor\'{a}s, M\'{a}t\'{e}, Barcza,
  Schilling, and \"{O}rs Legeza]{legeza2021fermionic}
Szalay,~S.; Zimbor\'{a}s,~Z.; M\'{a}t\'{e},~M.; Barcza,~G.; Schilling,~C.;
  \"{O}rs Legeza Fermionic systems for quantum information people. \emph{J.
  Phys. A: Math. Theor.} \textbf{2021}, \emph{54}, 393001\relax
\mciteBstWouldAddEndPuncttrue
\mciteSetBstMidEndSepPunct{\mcitedefaultmidpunct}
{\mcitedefaultendpunct}{\mcitedefaultseppunct}\relax
\EndOfBibitem
\bibitem[Ding and Schilling(2020)Ding, and Schilling]{schilling2020H2}
Ding,~L.; Schilling,~C. Correlation Paradox of the Dissociation Limit: A
  Quantum Information Perspective. \emph{J. Chem. Theory Comput.}
  \textbf{2020}, \emph{16}, 4159\relax
\mciteBstWouldAddEndPuncttrue
\mciteSetBstMidEndSepPunct{\mcitedefaultmidpunct}
{\mcitedefaultendpunct}{\mcitedefaultseppunct}\relax
\EndOfBibitem
\bibitem[Ding \latin{et~al.}(2021)Ding, Mardazad, Das, Szalay, Schollw\"{o}ck,
  Zimbor\'{a}s, and Schilling]{schilling2021jctc}
Ding,~L.; Mardazad,~S.; Das,~S.; Szalay,~S.; Schollw\"{o}ck,~U.;
  Zimbor\'{a}s,~Z.; Schilling,~C. Concept of Orbital Entanglement and
  Correlation in Quantum Chemistry. \emph{J. Chem. Theory Comput.}
  \textbf{2021}, \emph{17}, 79--95\relax
\mciteBstWouldAddEndPuncttrue
\mciteSetBstMidEndSepPunct{\mcitedefaultmidpunct}
{\mcitedefaultendpunct}{\mcitedefaultseppunct}\relax
\EndOfBibitem
\bibitem[Ding \latin{et~al.}(2024)Ding, D\"{u}nnweber, and
  Schilling]{schilling2024qst}
Ding,~L.; D\"{u}nnweber,~G.; Schilling,~C. Physical entanglement between
  localized orbitals. \emph{Quantum Sci. Technol.} \textbf{2024}, \emph{9},
  015005\relax
\mciteBstWouldAddEndPuncttrue
\mciteSetBstMidEndSepPunct{\mcitedefaultmidpunct}
{\mcitedefaultendpunct}{\mcitedefaultseppunct}\relax
\EndOfBibitem
\bibitem[Legeza and S\'olyom(2003)Legeza, and S\'olyom]{Legeza-2003b}
Legeza,~O.; S\'olyom,~J. Optimizing the density-matrix renormalization group
  method using quantum information entropy. \emph{Phys. Rev. B} \textbf{2003},
  \emph{68}, 195116\relax
\mciteBstWouldAddEndPuncttrue
\mciteSetBstMidEndSepPunct{\mcitedefaultmidpunct}
{\mcitedefaultendpunct}{\mcitedefaultseppunct}\relax
\EndOfBibitem
\bibitem[Legeza and S\'olyom(2004)Legeza, and S\'olyom]{Legeza-2004a}
Legeza,~O.; S\'olyom,~J. Optimizing density-matrix renormalization group method
  using quantum information entropy. International Workshop on Recent Progress
  and Prospects in Density-Matrix Renormalization. 2004\relax
\mciteBstWouldAddEndPuncttrue
\mciteSetBstMidEndSepPunct{\mcitedefaultmidpunct}
{\mcitedefaultendpunct}{\mcitedefaultseppunct}\relax
\EndOfBibitem
\bibitem[Legeza and S\'olyom(2004)Legeza, and S\'olyom]{Legeza-2004b}
Legeza,~O.; S\'olyom,~J. Quantum data compression, quantum information
  generation, and the density-matrix renormalization-group method. \emph{Phys.
  Rev. B} \textbf{2004}, \emph{70}, 205118\relax
\mciteBstWouldAddEndPuncttrue
\mciteSetBstMidEndSepPunct{\mcitedefaultmidpunct}
{\mcitedefaultendpunct}{\mcitedefaultseppunct}\relax
\EndOfBibitem
\bibitem[Legeza and S\'olyom(2006)Legeza, and S\'olyom]{Legeza-2006a}
Legeza,~{\"O}.; S\'olyom,~J. Two-Site Entropy and Quantum Phase Transitions in
  Low-Dimensional Models. \emph{Phys. Rev. Lett.} \textbf{2006}, \emph{96},
  116401\relax
\mciteBstWouldAddEndPuncttrue
\mciteSetBstMidEndSepPunct{\mcitedefaultmidpunct}
{\mcitedefaultendpunct}{\mcitedefaultseppunct}\relax
\EndOfBibitem
\bibitem[Rissler \latin{et~al.}(2006)Rissler, Noack, and White]{Rissler-2006}
Rissler,~J.; Noack,~R.~M.; White,~S.~R. Measuring orbital interaction using
  quantum information theory. \emph{Chem. Phys.} \textbf{2006}, \emph{323}, 519
  -- 531\relax
\mciteBstWouldAddEndPuncttrue
\mciteSetBstMidEndSepPunct{\mcitedefaultmidpunct}
{\mcitedefaultendpunct}{\mcitedefaultseppunct}\relax
\EndOfBibitem
\bibitem[Barcza \latin{et~al.}(2011)Barcza, Legeza, Marti, and
  Reiher]{Barcza-2011}
Barcza,~G.; Legeza,~{\"O}.; Marti,~K.~H.; Reiher,~M. Quantum-information
  analysis of electronic states of different molecular structures. \emph{Phys.
  Rev. A} \textbf{2011}, \emph{83}, 012508\relax
\mciteBstWouldAddEndPuncttrue
\mciteSetBstMidEndSepPunct{\mcitedefaultmidpunct}
{\mcitedefaultendpunct}{\mcitedefaultseppunct}\relax
\EndOfBibitem
\bibitem[Boguslawski \latin{et~al.}(2012)Boguslawski, Tecmer, Legeza, and
  Reiher]{Boguslawski-2012b}
Boguslawski,~K.; Tecmer,~P.; Legeza,~O.; Reiher,~M. Entanglement Measures for
  Single- and Multireference Correlation Effects. \emph{J. Phys. Chem. Lett.}
  \textbf{2012}, \emph{3}, 3129--3135\relax
\mciteBstWouldAddEndPuncttrue
\mciteSetBstMidEndSepPunct{\mcitedefaultmidpunct}
{\mcitedefaultendpunct}{\mcitedefaultseppunct}\relax
\EndOfBibitem
\bibitem[Boguslawski \latin{et~al.}(2013)Boguslawski, Tecmer, Barcza, Legeza,
  and Reiher]{Boguslawski-2013}
Boguslawski,~K.; Tecmer,~P.; Barcza,~G.; Legeza,~O.; Reiher,~M. Orbital
  Entanglement in Bond-Formation Processes. \emph{J. Chem. Theory Comput.}
  \textbf{2013}, \emph{9}, 2959--2973\relax
\mciteBstWouldAddEndPuncttrue
\mciteSetBstMidEndSepPunct{\mcitedefaultmidpunct}
{\mcitedefaultendpunct}{\mcitedefaultseppunct}\relax
\EndOfBibitem
\bibitem[Barcza \latin{et~al.}(2015)Barcza, Noack, S\'olyom, and
  Legeza]{Barcza-2014}
Barcza,~G.; Noack,~R.~M.; S\'olyom,~J.; Legeza,~{\"O}. Entanglement patterns
  and generalized correlation functions in quantum many-body systems.
  \emph{Phys. Rev. B} \textbf{2015}, \emph{92}, 125140\relax
\mciteBstWouldAddEndPuncttrue
\mciteSetBstMidEndSepPunct{\mcitedefaultmidpunct}
{\mcitedefaultendpunct}{\mcitedefaultseppunct}\relax
\EndOfBibitem
\bibitem[Boguslawski and Tecmer(2015)Boguslawski, and Tecmer]{Boguslawski-2014}
Boguslawski,~K.; Tecmer,~P. Orbital entanglement in quantum chemistry.
  \emph{Int. J. Quant. Chem.} \textbf{2015}, \emph{115}, 1289--1295\relax
\mciteBstWouldAddEndPuncttrue
\mciteSetBstMidEndSepPunct{\mcitedefaultmidpunct}
{\mcitedefaultendpunct}{\mcitedefaultseppunct}\relax
\EndOfBibitem
\bibitem[Boguslawski \latin{et~al.}(2016)Boguslawski, Tecmer, and \"{O}rs
  Legeza]{boguslawski2016prb}
Boguslawski,~K.; Tecmer,~P.; \"{O}rs Legeza Analysis of two-orbital
  correlations in wave functions restricted to electron-pair states.
  \emph{Phys. Rev. B} \textbf{2016}, \emph{94}, 155126\relax
\mciteBstWouldAddEndPuncttrue
\mciteSetBstMidEndSepPunct{\mcitedefaultmidpunct}
{\mcitedefaultendpunct}{\mcitedefaultseppunct}\relax
\EndOfBibitem
\bibitem[Nowak \latin{et~al.}(2021)Nowak, \"{O}rs Legeza, and
  Boguslawski]{boguslawski2021jcp}
Nowak,~A.; \"{O}rs Legeza; Boguslawski,~K. Orbital entanglement and correlation
  from pCCD-tailored coupled cluster wave functions. \emph{J. Chem. Phys.}
  \textbf{2021}, \emph{154}, 084111\relax
\mciteBstWouldAddEndPuncttrue
\mciteSetBstMidEndSepPunct{\mcitedefaultmidpunct}
{\mcitedefaultendpunct}{\mcitedefaultseppunct}\relax
\EndOfBibitem
\bibitem[Leszczyk \latin{et~al.}(2022)Leszczyk, Dome, Tecmer, Kedziera, and
  Boguslawski]{boguslawski2022pccp}
Leszczyk,~A.; Dome,~T.; Tecmer,~P.; Kedziera,~D.; Boguslawski,~K. Resolving the
  $\pi$-assisted U-N $\sigma_f$-bond formation using quantum information
  theory. \emph{Phys. Chem. Chem. Phys.} \textbf{2022}, \emph{24},
  21296--21307\relax
\mciteBstWouldAddEndPuncttrue
\mciteSetBstMidEndSepPunct{\mcitedefaultmidpunct}
{\mcitedefaultendpunct}{\mcitedefaultseppunct}\relax
\EndOfBibitem
\bibitem[Verstraete \latin{et~al.}(2023)Verstraete, Nishino, amd Mari
  Carmen~Ba\'{n}uls, Chan, and Stoudenmire]{dmrg2023}
Verstraete,~F.; Nishino,~T.; amd Mari Carmen~Ba\'{n}uls,~U.~S.; Chan,~G. K.-L.;
  Stoudenmire,~M.~E. Density matrix renormalization group, 30 years on.
  \emph{Nat. rev. phys.} \textbf{2023}, \emph{5}, 273--276\relax
\mciteBstWouldAddEndPuncttrue
\mciteSetBstMidEndSepPunct{\mcitedefaultmidpunct}
{\mcitedefaultendpunct}{\mcitedefaultseppunct}\relax
\EndOfBibitem
\bibitem[Marti \latin{et~al.}(2008)Marti, Ond\'ik, Moritz, and
  Reiher]{Marti-2008a}
Marti,~K.~H.; Ond\'ik,~I.~M.; Moritz,~G.; Reiher,~M. Density matrix
  renormalization group calculations on relative energies of transition metal
  complexes and clusters. \emph{J. Chem. Phys.} \textbf{2008}, \emph{128},
  014104\relax
\mciteBstWouldAddEndPuncttrue
\mciteSetBstMidEndSepPunct{\mcitedefaultmidpunct}
{\mcitedefaultendpunct}{\mcitedefaultseppunct}\relax
\EndOfBibitem
\bibitem[Chan and Sharma(2011)Chan, and Sharma]{Chan-2011}
Chan,~G. K.-L.; Sharma,~S. The Density Matrix Renormalization Group in Quantum
  Chemistry. \emph{Ann. Rev. Phys. Chem.} \textbf{2011}, \emph{62}, 465--481,
  PMID: 21219144\relax
\mciteBstWouldAddEndPuncttrue
\mciteSetBstMidEndSepPunct{\mcitedefaultmidpunct}
{\mcitedefaultendpunct}{\mcitedefaultseppunct}\relax
\EndOfBibitem
\bibitem[Chan(2012)]{Chan-2012}
Chan,~G. K.-L. Low entanglement wavefunctions. \emph{Wiley Int. Revs.: Comp.
  Mol. Sci.} \textbf{2012}, \emph{2}, 907--920\relax
\mciteBstWouldAddEndPuncttrue
\mciteSetBstMidEndSepPunct{\mcitedefaultmidpunct}
{\mcitedefaultendpunct}{\mcitedefaultseppunct}\relax
\EndOfBibitem
\bibitem[Harris \latin{et~al.}(2014)Harris, Kurashige, Yanai, and
  Morokuma]{Harris-2014}
Harris,~T.~V.; Kurashige,~Y.; Yanai,~T.; Morokuma,~K. Ab initio density matrix
  renormalization group study of magnetic coupling in dinuclear iron and
  chromium complexes. \emph{J. Chem. Phys.} \textbf{2014}, \emph{140},
  054303\relax
\mciteBstWouldAddEndPuncttrue
\mciteSetBstMidEndSepPunct{\mcitedefaultmidpunct}
{\mcitedefaultendpunct}{\mcitedefaultseppunct}\relax
\EndOfBibitem
\bibitem[Sharma \latin{et~al.}(2014)Sharma, Sivalingam, Neese, and
  Chan]{fe2s2chan2014}
Sharma,~S.; Sivalingam,~K.; Neese,~F.; Chan,~G. K.-L. Low-energy spectrum of
  iron-sulfur clusters directly from many-particle quantum mechanics.
  \emph{Nat. Chem.} \textbf{2014}, \emph{6}, 927--933\relax
\mciteBstWouldAddEndPuncttrue
\mciteSetBstMidEndSepPunct{\mcitedefaultmidpunct}
{\mcitedefaultendpunct}{\mcitedefaultseppunct}\relax
\EndOfBibitem
\bibitem[Duperrouzel \latin{et~al.}(2015)Duperrouzel, Tecmer, Boguslawski,
  Barcza, \"{O}rs Legeza, and Ayers]{boguslawski2015cpl}
Duperrouzel,~C.; Tecmer,~P.; Boguslawski,~K.; Barcza,~G.; \"{O}rs Legeza;
  Ayers,~P.~W. A quantum informational approach for dissecting chemical
  reactions. \emph{Chem. Phys. Lett.} \textbf{2015}, \emph{621}, 160--164\relax
\mciteBstWouldAddEndPuncttrue
\mciteSetBstMidEndSepPunct{\mcitedefaultmidpunct}
{\mcitedefaultendpunct}{\mcitedefaultseppunct}\relax
\EndOfBibitem
\bibitem[Freitag \latin{et~al.}(2015)Freitag, Knecht, Keller, Delcey,
  Aquilante, Pedersen, Lindh, Reiher, and Gonz\'{a}lez]{gonzalez2015pccp}
Freitag,~L.; Knecht,~S.; Keller,~S.~F.; Delcey,~M.~G.; Aquilante,~F.;
  Pedersen,~T.~B.; Lindh,~R.; Reiher,~M.; Gonz\'{a}lez,~L. Orbital entanglement
  and CASSCF analysis of the Ru-NO bond in a Ruthenium nitrosyl complex.
  \emph{Phys. Chem. Chem. Phys.} \textbf{2015}, \emph{17}, 14383--14392\relax
\mciteBstWouldAddEndPuncttrue
\mciteSetBstMidEndSepPunct{\mcitedefaultmidpunct}
{\mcitedefaultendpunct}{\mcitedefaultseppunct}\relax
\EndOfBibitem
\bibitem[Tzeli \latin{et~al.}(2024)Tzeli, Golub, Brabec, Matou\v{s}ek, Pernal,
  Veis, Raugei, and Xantheas]{fe2s2veis2024}
Tzeli,~D.; Golub,~P.; Brabec,~J.; Matou\v{s}ek,~M.; Pernal,~K.; Veis,~L.;
  Raugei,~S.; Xantheas,~S.~S. Importance of Electron Correlation on the
  Geometry and Electronic Structure of [2Fe-2S] Systems: A Benchmark Study of
  the [Fe$_2$S$_2$(SCH$_3$)$_4$]$^{2-,3-,4-}$, [Fe$_2$S$_2$(SCys)$_4$]$^{2-}$,
  [Fe$_2$S$_2$(S-$p$-tol)$_4$]$^{2-}$, and [Fe$_2$S$_2$(S-$o$-xyl)$_4$]$^{2-}$
  Complexes. \emph{J. Chem. Theory Comput.} \textbf{2024}, \emph{20},
  10406--10423\relax
\mciteBstWouldAddEndPuncttrue
\mciteSetBstMidEndSepPunct{\mcitedefaultmidpunct}
{\mcitedefaultendpunct}{\mcitedefaultseppunct}\relax
\EndOfBibitem
\bibitem[Brower \latin{et~al.}(2026)Brower, Rodriguez~Bernabeu, Hammond,
  Gunnels, Xantheas, Ganahl, Menczer, and Legeza]{legeza2026femoco}
Brower,~C.; Rodriguez~Bernabeu,~S.; Hammond,~J.; Gunnels,~J.; Xantheas,~S.~S.;
  Ganahl,~M.; Menczer,~A.; Legeza,~O. Mixed-Precision Ab Initio Tensor Network
  State Methods Adapted for NVIDIA Blackwell Technology via Emulated FP64
  Arithmetic. \emph{J. Chem. Theory Comput.} \textbf{2026}, \relax
\mciteBstWouldAddEndPunctfalse
\mciteSetBstMidEndSepPunct{\mcitedefaultmidpunct}
{}{\mcitedefaultseppunct}\relax
\EndOfBibitem
\bibitem[Moritz \latin{et~al.}(2005)Moritz, Hess, and Reiher]{Moritz-2005a}
Moritz,~G.; Hess,~B.~A.; Reiher,~M. Convergence behavior of the density-matrix
  renormalization group algorithm for optimized orbital orderings. \emph{The
  Journal of Chemical Physics} \textbf{2005}, \emph{122}, 024107\relax
\mciteBstWouldAddEndPuncttrue
\mciteSetBstMidEndSepPunct{\mcitedefaultmidpunct}
{\mcitedefaultendpunct}{\mcitedefaultseppunct}\relax
\EndOfBibitem
\bibitem[Stein and Reiher(2016)Stein, and Reiher]{stein-reiher-2016}
Stein,~C.~J.; Reiher,~M. Automated Selection of Active Orbital Spaces. \emph{J.
  Chem. Theory Comput.} \textbf{2016}, \emph{12}, 1760--1771\relax
\mciteBstWouldAddEndPuncttrue
\mciteSetBstMidEndSepPunct{\mcitedefaultmidpunct}
{\mcitedefaultendpunct}{\mcitedefaultseppunct}\relax
\EndOfBibitem
\bibitem[Ding \latin{et~al.}(2023)Ding, Knecht, and
  Schilling]{schilling2023jpcl}
Ding,~L.; Knecht,~S.; Schilling,~C. Quantum Information-Assisted Complete
  Active Space Optimization (QICAS). \emph{J. Phys. Chem. Lett.} \textbf{2023},
  \emph{14}, 11022--11029\relax
\mciteBstWouldAddEndPuncttrue
\mciteSetBstMidEndSepPunct{\mcitedefaultmidpunct}
{\mcitedefaultendpunct}{\mcitedefaultseppunct}\relax
\EndOfBibitem
\bibitem[Ding \latin{et~al.}(2023)Ding, Knecht, Zimbor\'{a}s, and
  Schilling]{schilling2023qst}
Ding,~L.; Knecht,~S.; Zimbor\'{a}s,~Z.; Schilling,~C. Quantum correlations in
  molecules: from quantum resourcing to chemical bonding. \emph{Quantum Sci.
  Technol.} \textbf{2023}, \emph{8}, 015015\relax
\mciteBstWouldAddEndPuncttrue
\mciteSetBstMidEndSepPunct{\mcitedefaultmidpunct}
{\mcitedefaultendpunct}{\mcitedefaultseppunct}\relax
\EndOfBibitem
\bibitem[Aliverti-Piuri \latin{et~al.}(2024)Aliverti-Piuri, Chatterjee, Ding,
  Liao, Liebert, and Schilling]{schilling2024faraday}
Aliverti-Piuri,~D.; Chatterjee,~K.; Ding,~L.; Liao,~K.; Liebert,~J.;
  Schilling,~C. What can quantum information theory offer to quantum chemistry?
  \emph{Faraday Discuss.} \textbf{2024}, \emph{254}, 76\relax
\mciteBstWouldAddEndPuncttrue
\mciteSetBstMidEndSepPunct{\mcitedefaultmidpunct}
{\mcitedefaultendpunct}{\mcitedefaultseppunct}\relax
\EndOfBibitem
\bibitem[Tenti \latin{et~al.}(2024)Tenti, Peeters, Giner, and
  Angeli]{angeli2024}
Tenti,~L.; Peeters,~S.; Giner,~E.; Angeli,~C. Entanglement and Mutual
  Information in Molecules: Comparing Localized and Delocalized Orbitals.
  \emph{J. Chem. Theory Comput.} \textbf{2024}, \emph{20}, 10861--10874\relax
\mciteBstWouldAddEndPuncttrue
\mciteSetBstMidEndSepPunct{\mcitedefaultmidpunct}
{\mcitedefaultendpunct}{\mcitedefaultseppunct}\relax
\EndOfBibitem
\bibitem[Brandejs \latin{et~al.}(2019)Brandejs, Veis, Szalay, Barcza, Pittner,
  and Legeza]{Brandejs2019}
Brandejs,~J.; Veis,~L.; Szalay,~S.; Barcza,~G.; Pittner,~J.; Legeza,~O. Quantum
  information-based analysis of electron-deficient bonds. \emph{J. Chem. Phys.}
  \textbf{2019}, \emph{150}, 204117\relax
\mciteBstWouldAddEndPuncttrue
\mciteSetBstMidEndSepPunct{\mcitedefaultmidpunct}
{\mcitedefaultendpunct}{\mcitedefaultseppunct}\relax
\EndOfBibitem
\bibitem[L\"owdin(1955)]{lowdin1}
L\"owdin,~P.-O. Quantum Theory of Many-Particle Systems. I. Physical
  Interpretations by Means of Density Matrices, Natural Spin-Orbitals, and
  Convergence Problems in the Method of Configurational Interaction.
  \emph{Phys. Rev.} \textbf{1955}, \emph{97}, 1474--1489\relax
\mciteBstWouldAddEndPuncttrue
\mciteSetBstMidEndSepPunct{\mcitedefaultmidpunct}
{\mcitedefaultendpunct}{\mcitedefaultseppunct}\relax
\EndOfBibitem
\bibitem[Krumnow \latin{et~al.}(2016)Krumnow, Veis, Legeza, and
  Eisert]{legeza2016prl}
Krumnow,~C.; Veis,~L.; Legeza,~O.; Eisert,~J. Fermionic Orbital Optimization in
  Tensor Network States. \emph{Phys. Rev. Lett.} \textbf{2016}, \emph{117},
  210402\relax
\mciteBstWouldAddEndPuncttrue
\mciteSetBstMidEndSepPunct{\mcitedefaultmidpunct}
{\mcitedefaultendpunct}{\mcitedefaultseppunct}\relax
\EndOfBibitem
\bibitem[M\'{a}t\'{e} \latin{et~al.}(2023)M\'{a}t\'{e}, Petrov, Szalay, and
  \"{O}rs Legeza]{legeza2023jmc}
M\'{a}t\'{e},~M.; Petrov,~K.; Szalay,~S.; \"{O}rs Legeza Compressing
  multireference character of wave functions via fermionic mode optimization.
  \emph{J. Math. Chem.} \textbf{2023}, \emph{61}, 362--375\relax
\mciteBstWouldAddEndPuncttrue
\mciteSetBstMidEndSepPunct{\mcitedefaultmidpunct}
{\mcitedefaultendpunct}{\mcitedefaultseppunct}\relax
\EndOfBibitem
\bibitem[{Li Manni} \latin{et~al.}(2023){Li Manni}, Kats, and
  Liebermann]{limanni2023jctc}
{Li Manni},~G.; Kats,~D.; Liebermann,~N. Resolution of Electronic States in
  Heisenberg Cluster Models within the Unitary Group Approach. \emph{J. Chem.
  Theory Comput.} \textbf{2023}, \emph{19}, 1218--1230\relax
\mciteBstWouldAddEndPuncttrue
\mciteSetBstMidEndSepPunct{\mcitedefaultmidpunct}
{\mcitedefaultendpunct}{\mcitedefaultseppunct}\relax
\EndOfBibitem
\bibitem[Dobrautz \latin{et~al.}(2022)Dobrautz, Katukuri, Bogdanov, Kats, {Li
  Manni}, and Alavi]{dobrautz2022prb}
Dobrautz,~W.; Katukuri,~V.~M.; Bogdanov,~N.~A.; Kats,~D.; {Li Manni},~G.;
  Alavi,~A. Combined unitary and symmetric group approach applied to
  low-dimensional Heisenberg spin systems. \emph{Phys. Rev. B} \textbf{2022},
  \emph{105}, 195123\relax
\mciteBstWouldAddEndPuncttrue
\mciteSetBstMidEndSepPunct{\mcitedefaultmidpunct}
{\mcitedefaultendpunct}{\mcitedefaultseppunct}\relax
\EndOfBibitem
\bibitem[{Li Manni}(2021)]{limanni2021pccp}
{Li Manni},~G. Modeling magnetic interactions in high-valent trinuclear
  [Mn$_3$(IV)O$_4$]$^{4+}$ complexes through highly compressed
  multi-configurational wave functions. \emph{Phys. Chem. Chem. Phys.}
  \textbf{2021}, \emph{23}, 19766\relax
\mciteBstWouldAddEndPuncttrue
\mciteSetBstMidEndSepPunct{\mcitedefaultmidpunct}
{\mcitedefaultendpunct}{\mcitedefaultseppunct}\relax
\EndOfBibitem
\bibitem[{Li Manni} \latin{et~al.}(2020){Li Manni}, Dobrautz, and
  Alavi]{limanni2020jctc}
{Li Manni},~G.; Dobrautz,~W.; Alavi,~A. Compression of Spin-Adapted
  Multiconfgurational Wave Functions in Exchange-Coupled Polynuclear Spin
  Systems. \emph{J. Chem. Theory Comput.} \textbf{2020}, \emph{16},
  2202--2215\relax
\mciteBstWouldAddEndPuncttrue
\mciteSetBstMidEndSepPunct{\mcitedefaultmidpunct}
{\mcitedefaultendpunct}{\mcitedefaultseppunct}\relax
\EndOfBibitem
\bibitem[{Li Manni} \latin{et~al.}(2021){Li Manni}, Dobrautz, Bogdanov, Guther,
  and Alavi]{limanni2021jpca}
{Li Manni},~G.; Dobrautz,~W.; Bogdanov,~N.~A.; Guther,~K.; Alavi,~A. Resolution
  of Low-Energy States in Spin-Exchange Transition-Metal Clusters: Case Study
  of Singlet States in [Fe(III)$_4$S$_4$] Cubanes. \emph{J. Phys. Chem. A}
  \textbf{2021}, \emph{125}, 4727--4740\relax
\mciteBstWouldAddEndPuncttrue
\mciteSetBstMidEndSepPunct{\mcitedefaultmidpunct}
{\mcitedefaultendpunct}{\mcitedefaultseppunct}\relax
\EndOfBibitem
\bibitem[Dobrautz \latin{et~al.}(2021)Dobrautz, Weser, Bogdanov, Alavi, and {Li
  Manni}]{dobrautz2021jctc}
Dobrautz,~W.; Weser,~O.; Bogdanov,~N.~A.; Alavi,~A.; {Li Manni},~G. Spin-Pure
  Stochastic-CASSCF via GUGA-FCIQMC Applied to Iron-Sulfur Clusters. \emph{J.
  Chem. Theory Comput.} \textbf{2021}, \emph{17}, 5684--5703\relax
\mciteBstWouldAddEndPuncttrue
\mciteSetBstMidEndSepPunct{\mcitedefaultmidpunct}
{\mcitedefaultendpunct}{\mcitedefaultseppunct}\relax
\EndOfBibitem
\bibitem[Liao \latin{et~al.}(2024)Liao, Ding, and Schilling]{schilling2024jpcl}
Liao,~K.; Ding,~L.; Schilling,~C. Quantum Information Orbitals (QIO): Unveiling
  Intrinsic Many-Body Complexity by Compressing Single-Body Triviality.
  \emph{J. Phys. Chem. Lett.} \textbf{2024}, \emph{15}, 6782--6790\relax
\mciteBstWouldAddEndPuncttrue
\mciteSetBstMidEndSepPunct{\mcitedefaultmidpunct}
{\mcitedefaultendpunct}{\mcitedefaultseppunct}\relax
\EndOfBibitem
\bibitem[Li(2025)]{entanglement-minimized2025}
Li,~Z. Entanglement-minimized orbitals enable faster quantum simulation of
  molecules. \emph{arXiv} \textbf{2025}, 2506.13386\relax
\mciteBstWouldAddEndPuncttrue
\mciteSetBstMidEndSepPunct{\mcitedefaultmidpunct}
{\mcitedefaultendpunct}{\mcitedefaultseppunct}\relax
\EndOfBibitem
\bibitem[McWeeny(1967)]{mcweeny1967rdm}
McWeeny,~R. The nature of electron correlation in molecules. \emph{Int. J.
  Quant. Chem.} \textbf{1967}, \emph{1}, 351--359\relax
\mciteBstWouldAddEndPuncttrue
\mciteSetBstMidEndSepPunct{\mcitedefaultmidpunct}
{\mcitedefaultendpunct}{\mcitedefaultseppunct}\relax
\EndOfBibitem
\bibitem[Kutzelnigg and Mukherjee(1999)Kutzelnigg, and
  Mukherjee]{kutzelnigg1999cumulant}
Kutzelnigg,~W.; Mukherjee,~D. Cumulant expansion of the reduced density
  matrices. \emph{J. Chem. Phys.} \textbf{1999}, \emph{110}, 2800--2809\relax
\mciteBstWouldAddEndPuncttrue
\mciteSetBstMidEndSepPunct{\mcitedefaultmidpunct}
{\mcitedefaultendpunct}{\mcitedefaultseppunct}\relax
\EndOfBibitem
\bibitem[Mazziotti(2012)]{mazziotti2012review}
Mazziotti,~D.~A. Two-Electron Reduced Density Matrix as the Basic Variable in
  Many-Electron Quantum Chemistry and Physics. \emph{Chem. Rev.} \textbf{2012},
  \emph{112}, 244--262\relax
\mciteBstWouldAddEndPuncttrue
\mciteSetBstMidEndSepPunct{\mcitedefaultmidpunct}
{\mcitedefaultendpunct}{\mcitedefaultseppunct}\relax
\EndOfBibitem
\bibitem[Kempe \latin{et~al.}({2006})Kempe, Kitaev, and Regev]{kempe_2006}
Kempe,~J.; Kitaev,~A.; Regev,~O. {The complexity of the local Hamiltonian
  problem}. \emph{{SIAM J. Comp.}} \textbf{{2006}}, \emph{{35}},
  1070--1097\relax
\mciteBstWouldAddEndPuncttrue
\mciteSetBstMidEndSepPunct{\mcitedefaultmidpunct}
{\mcitedefaultendpunct}{\mcitedefaultseppunct}\relax
\EndOfBibitem
\bibitem[Kutzelnigg and Mukherjee(1997)Kutzelnigg, and Mukherjee]{kmgwt}
Kutzelnigg,~W.; Mukherjee,~D. \emph{J. Chem. Phys.} \textbf{1997}, \emph{107},
  432\relax
\mciteBstWouldAddEndPuncttrue
\mciteSetBstMidEndSepPunct{\mcitedefaultmidpunct}
{\mcitedefaultendpunct}{\mcitedefaultseppunct}\relax
\EndOfBibitem
\bibitem[Mukherjee(1997)]{mukherjeecpl}
Mukherjee,~D. Normal ordering and a Wick-like reduction theorem for fermions
  with respect to a multi-determinantal reference state. \emph{Chemical Physics
  Letters} \textbf{1997}, \emph{274}, 561 -- 566\relax
\mciteBstWouldAddEndPuncttrue
\mciteSetBstMidEndSepPunct{\mcitedefaultmidpunct}
{\mcitedefaultendpunct}{\mcitedefaultseppunct}\relax
\EndOfBibitem
\bibitem[Kutzelnigg(2003)]{kutzelnigg2003cumulant}
Kutzelnigg,~W. n-Electron Problem and Its Formulation in Terms of k-Particle
  Density Cumulants. \emph{Int. J. Quant. Chem.} \textbf{2003}, \emph{95},
  404--423\relax
\mciteBstWouldAddEndPuncttrue
\mciteSetBstMidEndSepPunct{\mcitedefaultmidpunct}
{\mcitedefaultendpunct}{\mcitedefaultseppunct}\relax
\EndOfBibitem
\bibitem[Misiewicz \latin{et~al.}(2020)Misiewicz, Turney, and
  III]{schaefer2020cumulant}
Misiewicz,~J.~P.; Turney,~J.~M.; III,~H. F.~S. Reduced Density Matrix
  Cumulants: The Combinatorics of Size-Consistency and Generalized Normal
  Ordering. \emph{J. Chem. Theory Comput.} \textbf{2020}, \emph{16},
  6150--6164\relax
\mciteBstWouldAddEndPuncttrue
\mciteSetBstMidEndSepPunct{\mcitedefaultmidpunct}
{\mcitedefaultendpunct}{\mcitedefaultseppunct}\relax
\EndOfBibitem
\bibitem[Juh\'{a}sh and Mazziotti(2006)Juh\'{a}sh, and
  Mazziotti]{mazziotti2006cumulant}
Juh\'{a}sh,~T.; Mazziotti,~D.~A. The cumulant two-particle reduced density
  matrix as a measure of electron correlation and entanglement. \emph{J. Chem.
  Phys.} \textbf{2006}, \emph{125}, 174105\relax
\mciteBstWouldAddEndPuncttrue
\mciteSetBstMidEndSepPunct{\mcitedefaultmidpunct}
{\mcitedefaultendpunct}{\mcitedefaultseppunct}\relax
\EndOfBibitem
\bibitem[Kong and Valeev(2011)Kong, and Valeev]{kong2011valeev}
Kong,~L.; Valeev,~E.~F. A novel interpretation of reduced density matrix and
  cumulant for electronic structure theories. \emph{J. Chem. Phys.}
  \textbf{2011}, \emph{134}, 214109\relax
\mciteBstWouldAddEndPuncttrue
\mciteSetBstMidEndSepPunct{\mcitedefaultmidpunct}
{\mcitedefaultendpunct}{\mcitedefaultseppunct}\relax
\EndOfBibitem
\bibitem[eva(2025)]{evangelista2025mutual}
Mutual Correlation. \emph{J. Chem. Theory Comput.} \textbf{2025}, \emph{21},
  7471--7484\relax
\mciteBstWouldAddEndPuncttrue
\mciteSetBstMidEndSepPunct{\mcitedefaultmidpunct}
{\mcitedefaultendpunct}{\mcitedefaultseppunct}\relax
\EndOfBibitem
\bibitem[Chilkuri \latin{et~al.}(2017)Chilkuri, DeBeer, and
  Neese]{fecomplexgeom}
Chilkuri,~V.~G.; DeBeer,~S.; Neese,~F. Revisiting the Electronic Structure of
  FeS Monomers Using Ab Initio Ligand Field Theory and the Angular Overlap
  Model. \emph{Inorg. Chem.} \textbf{2017}, \emph{56}, 10418--10436\relax
\mciteBstWouldAddEndPuncttrue
\mciteSetBstMidEndSepPunct{\mcitedefaultmidpunct}
{\mcitedefaultendpunct}{\mcitedefaultseppunct}\relax
\EndOfBibitem
\bibitem[Chilkuri \latin{et~al.}(2020)Chilkuri, DeBeer, and
  Neese]{fe2s2complexgeom}
Chilkuri,~V.~G.; DeBeer,~S.; Neese,~F. Ligand Field Theory and Angular Overlap
  Model Based Analysis of the Electronic Structure of Homovalent Iron-Sulfur
  Dimers. \emph{Inorg. Chem.} \textbf{2020}, \emph{59}, 984--995\relax
\mciteBstWouldAddEndPuncttrue
\mciteSetBstMidEndSepPunct{\mcitedefaultmidpunct}
{\mcitedefaultendpunct}{\mcitedefaultseppunct}\relax
\EndOfBibitem
\bibitem[da~Costa~Gouveia \latin{et~al.}(2024)da~Costa~Gouveia, Maganas, and
  Neese]{NeeseROHF2024}
da~Costa~Gouveia,~T.~L.; Maganas,~D.; Neese,~F. Restricted Open-Shell
  Hartree-Fock Method for a General Configuration State Function Featuring
  Arbitrarily Complex Spin-Couplings. \emph{J. Phys. Chem. A} \textbf{2024},
  \emph{128}, 5041\relax
\mciteBstWouldAddEndPuncttrue
\mciteSetBstMidEndSepPunct{\mcitedefaultmidpunct}
{\mcitedefaultendpunct}{\mcitedefaultseppunct}\relax
\EndOfBibitem
\bibitem[Schollw\"ock(2011)]{Schollwock-2011}
Schollw\"ock,~U. The density-matrix renormalization group in the age of matrix
  product states. \emph{Annals of Physics} \textbf{2011}, \emph{326}, 96 --
  192, January 2011 Special Issue\relax
\mciteBstWouldAddEndPuncttrue
\mciteSetBstMidEndSepPunct{\mcitedefaultmidpunct}
{\mcitedefaultendpunct}{\mcitedefaultseppunct}\relax
\EndOfBibitem
\bibitem[Pittner(2025)]{myspinfree2025}
Pittner,~J. Spin-free orbital entropy, mutual information and correlation
  analysis. \emph{Mol. Phys.} \textbf{2025}, \emph{123}, e2500632\relax
\mciteBstWouldAddEndPuncttrue
\mciteSetBstMidEndSepPunct{\mcitedefaultmidpunct}
{\mcitedefaultendpunct}{\mcitedefaultseppunct}\relax
\EndOfBibitem
\bibitem[Huang and Kais(2005)Huang, and Kais]{kais2005}
Huang,~Z.; Kais,~S. \emph{Chem. Phys. Lett.} \textbf{2005}, \emph{413}, 1\relax
\mciteBstWouldAddEndPuncttrue
\mciteSetBstMidEndSepPunct{\mcitedefaultmidpunct}
{\mcitedefaultendpunct}{\mcitedefaultseppunct}\relax
\EndOfBibitem
\bibitem[Neese(2012)]{Orca}
Neese,~F. \emph{WIREs Comput. Mol. Sci.} \textbf{2012}, \emph{2}, 73--78\relax
\mciteBstWouldAddEndPuncttrue
\mciteSetBstMidEndSepPunct{\mcitedefaultmidpunct}
{\mcitedefaultendpunct}{\mcitedefaultseppunct}\relax
\EndOfBibitem
\bibitem[Brabec \latin{et~al.}(2021)Brabec, Brandejs, Kowalski, Xantheas,
  Legeza, and Veis]{molmps2020}
Brabec,~J.; Brandejs,~J.; Kowalski,~K.; Xantheas,~S.; Legeza,~O.; Veis,~L.
  Massively parallel quantum chemical density matrix renormalization group
  method. \emph{J. Comp. Chem.} \textbf{2021}, \emph{42}, 534--544\relax
\mciteBstWouldAddEndPuncttrue
\mciteSetBstMidEndSepPunct{\mcitedefaultmidpunct}
{\mcitedefaultendpunct}{\mcitedefaultseppunct}\relax
\EndOfBibitem
\bibitem[Zhai \latin{et~al.}(2023)Zhai, Larsson, Lee, Cui, Zhu, Sun, Peng,
  Peng, Liao, T\"{o}lle, Yang, Li, and Chan]{block2code}
Zhai,~H.; Larsson,~H.~R.; Lee,~S.; Cui,~Z.; Zhu,~T.; Sun,~C.; Peng,~L.;
  Peng,~R.; Liao,~K.; T\"{o}lle,~J.; Yang,~J.; Li,~S.; Chan,~G. K.-L. Block2: a
  comprehensive open source framework to develop and apply state-of-the-art
  DMRG algorithms in electronic structure and beyond. \emph{J. Chem. Phys.}
  \textbf{2023}, \emph{159}, 234801\relax
\mciteBstWouldAddEndPuncttrue
\mciteSetBstMidEndSepPunct{\mcitedefaultmidpunct}
{\mcitedefaultendpunct}{\mcitedefaultseppunct}\relax
\EndOfBibitem
\end{mcitethebibliography}

\end{document}